\documentclass[
    12pt,               
    a4paper,            
    bibtotoc,           
    pointlessnumbers,   
    ngerman,            
    titlepage,          
    leqno,              
    ]{scrarticle}       
\usepackage[T1]{fontenc}
\usepackage{float}
\usepackage[utf8]{inputenc}
\usepackage[ngerman,english]{babel}
\usepackage{url}
\usepackage{graphicx}       
\usepackage[a4paper]{geometry}
\usepackage{amssymb}        
\usepackage{amsmath}        
\usepackage{amsthm}         
\usepackage{thmtools}       
\usepackage{csquotes}       
\usepackage{setspace}       
\usepackage{emptypage}      
\usepackage{tabularx}       
\usepackage{makecell}       
\usepackage{caption}        
\usepackage{hyperref}
\usepackage{relsize}        
\usepackage{exscale}        
\hypersetup{colorlinks,allcolors=black}
\usepackage[nodayofweek]{datetime}
\usepackage[figure]{hypcap}
\usepackage[shortcuts]{extdash}
\usepackage{blindtext}
\usepackage{enumitem}
\usepackage{tikz}
\usetikzlibrary{shapes}
\usetikzlibrary{arrows.meta}
\usetikzlibrary{arrows}
\usetikzlibrary{fit}
\usetikzlibrary{decorations.pathmorphing}
\usetikzlibrary{decorations.pathreplacing}
\usetikzlibrary{shapes.geometric}

\usepackage[style=numeric-comp,maxcitenames=2,maxbibnames=99]{biblatex}
\usepackage[normalem]{ulem}
\useunder{\uline}{\ul}{}

\usepackage{chngcntr}
\counterwithout{figure}{section}
\counterwithout{table}{section}

\RedeclareSectionCommand[%
    beforeskip=0.1\baselineskip,
    afterskip=0.01\baselineskip
]{section}
\RedeclareSectionCommand[%
    beforeskip=0.1\baselineskip,
    afterskip=0.01\baselineskip
]{subsection}
\RedeclareSectionCommand[%
    beforeskip=0.1\baselineskip,
    afterskip=0.01\baselineskip
]{subsubsection}
\RedeclareSectionCommand[%
    beforeskip=0.01\baselineskip,
    afterskip=0.01\baselineskip
]{paragraph}

\addtokomafont{caption}{\small}

\include{hyphenation}

\KOMAoptions{parskip=full}

\addtokomafont{disposition}{\rmfamily}

\newcommand\secref{Section~\ref}
\newcommand\figref{Figure~\ref}
\newcommand\tableref{Table~\ref}
\newcommand\thmref{Theorem~\ref}

\newcommand\corref{Corollary~\ref}
\newcommand\obsref{Observation~\ref}

\newcommand\exref{Example~\ref}
\newcommand\factref{Fact~\ref}
\newcommand\caseref{Case~\ref}

\newcommand\ttref[1]{[\hyperref[#1]{Th.\thinspace\ref*{#1}}]}
\newcommand\ccref[1]{[\hyperref[#1]{Co.\thinspace\ref*{#1}}]}
\newcommand\ooref[1]{[\hyperref[#1]{Ob.\thinspace\ref*{#1}}]}

\declaretheoremstyle[
    spaceabove=-4pt,
    spacebelow=0pt,
    headfont=\bfseries\itshape,
    postheadspace=1em,
    qed=\qedsymbol,
    headpunct={}
]{pfstyle}
\declaretheorem[name={Proof},style=pfstyle,unnumbered]{Proof}

\declaretheoremstyle[
    spaceabove=0pt,
    spacebelow=0pt,
    headfont=\bfseries,
    postheadspace=1em,
    headpunct={}
]{mystyle}
\declaretheorem[name={Theorem},style=mystyle]{Theorem}

\declaretheorem[name={Corollary},style=mystyle]{Corollary}
\declaretheorem[name={Observation},style=mystyle]{Observation}

\declaretheorem[name={Example},style=mystyle]{Example}

\newcommand{\topic}{Broadening the Complexity-theoretic Analysis of Manipulative Attacks in Group Identification}
\newcommand{\type}{Masterarbeit}
\newcommand{\authorName}{Emil Junker}
\newcommand{\authorBdate}{15.12.1997}
\newcommand{\authorBplace}{Berlin}
\newcommand{\supervisorA}{Prof.\@ Dr.\@ Robert Bredereck}
\newcommand{\supervisorB}{Prof.\@ Dr.\@  Stefan Kratsch}

\newcommand{\variablegadget}{
\node (o) {};

\node[main node] (aa)   [left of = o, node distance = 2.5cm] {$a_\ast$};
\node[main node] (ax2)  [above of = o, node distance = 3.5cm] {$a_{\bar{x}}$};
\node[main node] (ax1)  [above of = ax2] {$a_x$};

\node[main node] (c1)   [left of = ax1, node distance = 5cm, yshift = 0.9cm] {$a_{c_1}$};
\node[main node] (c2)   [below of = c1] {$a_{c_2}$};
\node[main node] (c3)   [below of = c2] {$a_{c_3}$};

\node (i1) [above left of = aa] {};
\node (i2) [above right of = aa] {};
\node (i3) [left of = i1, node distance = 1cm] {};
\node (i4) [right of = i2, node distance = 1cm] {};

\path[->]
(aa)    edge[loop below]        node {} (aa)
        edge[bend left=10]      node {} (ax1)
        edge[bend left=0]       node {} (ax2)
(ax1)   edge[bend right=30]     node {} (c1)
(ax2)   edge[bend left=30]      node {} (c3)
(i1)    edge[bend right=10]     node {} (aa)
(i2)    edge[bend left=10]      node {} (aa)
(i3)    edge[bend right=10]     node {} (aa)
(i4)    edge[bend left=10]      node {} (aa)
;

\draw [very thick, decorate, decoration = {brace, raise=5pt, amplitude=5pt}]
(-5.8, 7.2) -- (-1.8, 7.2)
node[pos = 0.5, above = 10pt, black] {$A^+$};
}

\newcommand{\variablegadgetextras}{
\node[main node] (qx)   [left of = ax2, node distance = 2.5cm, yshift = 0.9cm] {$q_x$};
\node[main node] (dx1)  [right of = ax1, node distance = 2.5cm, yshift = 1.5cm] {$d_x^1$};
\node[main node] (dx2)  [right of = ax1, node distance = 2.5cm] {$d_x^2$};
\node[main node] (dx3)  [right of = ax2, node distance = 2.5cm] {$d_x^3$};
\node[main node] (dx4)  [right of = ax2, node distance = 2.5cm, yshift = -1.5cm] {$d_x^4$};

\path[->]
(ax1)   edge                    node {} (qx)
        edge                    node {} (dx1)
        edge                    node {} (dx2)
(ax2)   edge                    node {} (qx)
        edge                    node {} (dx3)
        edge                    node {} (dx4)
;
}

\newcommand{\setcoverbriberyreductiongrid}{
\node[main node] (aa_k1)                                {$a^{k+1}_\ast$};
\node[main node] (aa_k)     [below of = aa_k1, node distance = 2cm]  {$a^k_\ast$};
\node            (aa_d)     [below of = aa_k, node distance = 1.5cm] {$\vdots$};
\node[main node] (aa_1)     [below of = aa_d, node distance = 1.5cm] {$a^1_\ast$};

\node[main node] (a_k1_f1)  [right of = aa_k1, node distance=3cm]  {$a^{k+1}_{F_1}$};
\node[main node] (a_k_f1)   [right of = aa_k, node distance = 3cm] {$a^k_{F_1}$};
\node            (a_d_f1)   [right of = aa_d, node distance = 3cm] {$\vdots$};
\node[main node] (a_1_f1)   [right of = aa_1, node distance = 3cm] {$a^1_{F_1}$};

\node[main node] (a_k1_f2)  [right of = a_k1_f1]        {$a^{k+1}_{F_2}$};
\node[main node] (a_k_f2)   [right of = a_k_f1]         {$a^k_{F_2}$};
\node            (a_d_f2)   [right of = a_d_f1]         {$\vdots$};
\node[main node] (a_1_f2)   [right of = a_1_f1]         {$a^1_{F_2}$};

\node            (a_k1_f3)  [right of = a_k1_f2]        {$\ldots$};
\node            (a_k_f3)   [right of = a_k_f2]         {$\ldots$};
\node            (a_d_f3)   [right of = a_d_f2]         {};
\node            (a_1_f3)   [right of = a_1_f2]         {$\ldots$};

\node[main node] (a_k1_fm1) [right of = a_k1_f3]        {$a^{k+1}_{F_{m-1}}$};
\node[main node] (a_k_fm1)  [right of = a_k_f3]         {$a^k_{F_{m-1}}$};
\node            (a_d_fm1)  [right of = a_d_f3]         {$\vdots$};
\node[main node] (a_1_fm1)  [right of = a_1_f3]         {$a^1_{F_{m-1}}$};

\node[main node] (a_k1_fm)  [right of = a_k1_fm1]       {$a^{k+1}_{F_m}$};
\node[main node] (a_k_fm)   [right of = a_k_fm1]        {$a^k_{F_m}$};
\node            (a_d_fm)   [right of = a_d_fm1]        {$\vdots$};
\node[main node] (a_1_fm)   [right of = a_1_fm1]        {$a^1_{F_m}$};

\node (x1)      [below of = a_1_f1, xshift = -2cm]      {};
\node (x2)      [right of = x1, node distance = 1.4cm]  {};
\node (x3)      [right of = x2, node distance = 1.4cm]  {};
\node (x4)      [right of = x3, node distance = 1.4cm]  {};
\node (x5)      [right of = x4, node distance = 1.4cm]  {};
\node (x6)      [right of = x5, node distance = 1.4cm]  {};
\node (x7)      [right of = x6, node distance = 1.4cm]  {};
\node (x8)      [right of = x7, node distance = 1.4cm]  {};
\node (x9)      [right of = x8, node distance = 1.4cm]  {};
\node (x10)     [right of = x9, node distance = 1.4cm]  {};
\node (nxn)     [right of = x5, node distance = 0.7cm, yshift = 0.25cm] {};
\node (nxs)     [right of = x5, node distance = 0.7cm, yshift = -0.5cm] {{$N_X$}};
\node[ellipse, draw, fit={(x3)(nxn)(nxs)(x8)}, inner sep=2mm] (x) {};

\path[->]
(aa_k1)     edge                    node {} (a_k1_f1)
            edge                    node {} (a_k_f1)
            edge                    node {} (a_1_f1)
(aa_k)      edge                    node {} (a_k1_f1)
            edge                    node {} (a_k_f1)
            edge                    node {} (a_1_f1)
(aa_1)      edge[bend left = 8]     node {} (a_k1_f1)
            edge                    node {} (a_k_f1)
            edge                    node {} (a_1_f1)

(a_k1_f1)   edge                    node {} (a_k1_f2)
            edge                    node {} (a_k_f2)
            edge                    node {} (a_1_f2)
(a_k_f1)    edge                    node {} (a_k1_f2)
            edge                    node {} (a_k_f2)
            edge                    node {} (a_1_f2)
(a_1_f1)    edge[bend left = 8]     node {} (a_k1_f2)
            edge                    node {} (a_k_f2)
            edge                    node {} (a_1_f2)

(a_k1_f2)   edge[shorten > = 0pt]   node {} (a_k1_f3)
            edge[shorten > = 0pt]   node {} (a_k_f3)
            edge[shorten > = 0pt]   node {} (a_1_f3)
(a_k_f2)    edge[shorten > = 0pt]   node {} (a_k1_f3.south west)
            edge[shorten > = 0pt]   node {} (a_k_f3)
            edge[shorten > = 0pt]   node {} (a_1_f3.north west)
(a_1_f2)    edge[shorten > = 0pt]   node {} (a_k1_f3)
            edge[shorten > = 0pt]   node {} (a_k_f3)
            edge[shorten > = 0pt]   node {} (a_1_f3)

(a_k1_f3)   edge                    node {} (a_k1_fm1)
            edge                    node {} (a_k_fm1)
            edge                    node {} (a_1_fm1)
(a_k_f3)    edge                    node {} (a_k1_fm1)
            edge                    node {} (a_k_fm1)
            edge                    node {} (a_1_fm1)
(a_1_f3)    edge[bend left = 8]     node {} (a_k1_fm1)
            edge                    node {} (a_k_fm1)
            edge                    node {} (a_1_fm1)

(a_k1_fm1)  edge                    node {} (a_k1_fm)
            edge                    node {} (a_k_fm)
            edge                    node {} (a_1_fm)
(a_k_fm1)   edge                    node {} (a_k1_fm)
            edge                    node {} (a_k_fm)
            edge                    node {} (a_1_fm)
(a_1_fm1)   edge[bend left = 8]     node {} (a_k1_fm)
            edge                    node {} (a_k_fm)
            edge                    node {} (a_1_fm)
;

\path[->]
(a_1_f1)    edge[bend right = 8]    node {} (x2.south east)
            edge[bend left = 8]     node {} (x3.south west)
(a_k_f1)    edge[bend right = 22]   node {} (x2.south west)
            edge[bend left = 14]    node {} (x3.south east)
(a_1_f2)    edge[bend right = 10]   node {} (x4.south east)
            edge[bend left = 8]     node {} (x5.south west)
            edge[bend left = 10]    node {} (x6.south west)
(a_k_f2)    edge[bend right = 25]   node {} (x4.south west)
            edge[bend left = 12]    node {} (x5.south east)
            edge[bend left = 6]     node {} (x6.south east)
(a_1_fm1)   edge[bend right = 10]   node {} (x7.south east)
            edge[bend left = 8]     node {} (x8.south west)
            edge[bend left = 10]    node {} (x9.south west)
(a_k_fm1)   edge[bend right = 14]   node {} (x7.south west)
            edge[bend left = 22]    node {} (x8.south east)
            edge[bend left = 14]    node {} (x9.south east)
;

\draw [very thick, decorate, decoration = {brace, raise=5pt, amplitude=5pt}]
(13.8, 0.9) -- (13.8, -0.9)
node[pos = 0.5, right = 12pt, black] {$\tilde{N}_\mathcal{F}$};

\draw [very thick, decorate, decoration = {brace, raise=5pt, amplitude=5pt}]
(13.8, -1.1) -- (13.8, -6)
node[pos = 0.5, right = 12pt, black] {$N_\mathcal{F}$};

\draw [very thick, decorate, decoration = {brace, raise=5pt, amplitude=5pt}]
(-0.8, 0.9) -- (0.8, 0.9)
node[pos = 0.5, above = 12pt, black] {$Q^{\text{IC}}_0(N, \varphi)$};

\draw [very thick, decorate, decoration = {brace, raise=5pt, amplitude=5pt}]
(2.2, 0.9) -- (3.8, 0.9)
node[pos = 0.5, above = 12pt, black] {$Q^{\text{IC}}_1(N, \varphi)$};

\draw [very thick, decorate, decoration = {brace, raise=5pt, amplitude=5pt}]
(4.7, 0.9) -- (6.3, 0.9)
node[pos = 0.5, above = 12pt, black] {$Q^{\text{IC}}_2(N, \varphi)$};

\draw [very thick, decorate, decoration = {brace, raise=5pt, amplitude=5pt}]
(9.7, 0.9) -- (11.3, 0.9)
node[pos = 0.5, above = 12pt, black] {$Q^{\text{IC}}_{m-1}(N, \varphi)$};

\draw [very thick, decorate, decoration = {brace, raise=5pt, amplitude=5pt}]
(12.2, 0.9) -- (13.8, 0.9)
node[pos = 0.5, above = 12pt, black] {$Q^{\text{IC}}_m(N, \varphi)$};
}

\begin{document}


\pagestyle{empty}

\begin{titlepage}

\begin{minipage}[c][3cm][c]{12cm}
\textsc{
    \hspace{-1.2em}
    \Large{
        Humboldt-Universität zu Berlin \\
    }
    \normalsize{
        Mathematisch-Naturwissenschaftliche Fakultät \\
        Institut für Informatik
    }
}
\end{minipage}
\hfill

\vfill
\linespread{1.5}

\begin{center}
\huge\topic
\vspace{1em}
\Large{
    \type \\
    \vspace{1em}
    zur Erlangung des akademischen Grades \\
    Master of Science (M. Sc.)
    \vspace{\baselineskip}
}
\end{center}

\vfill
\linespread{1}

\large{
    \begin{tabularx}{\textwidth}{ll}
    eingereicht von:    & \authorName \\
    geboren am:         & \authorBdate \\
    geboren in:         & \authorBplace \\
    \vspace{0.5\baselineskip} \\
    Gutachter:          & \supervisorA \\
                        & \supervisorB \\
    \vspace{\baselineskip} \\
    eingereicht am:     & \raisebox{-\height}{\parbox{0.3\linewidth}{\dotfill}} \\
    \vspace{\baselineskip} \\
    verteidigt am:      & \raisebox{-\height}{\parbox{0.3\linewidth}{\dotfill}} \\
    \vspace{\baselineskip} \\
    \end{tabularx}
}

\end{titlepage}

\cleardoublepage

\newgeometry{tmargin=2cm, bmargin=2cm, lmargin=2.5cm, rmargin=2.5cm}

\onehalfspacing


\section*{Abstract}

In the Group Identification problem, we are given a set of individuals and are asked to identify a socially qualified subset among them. Each individual in the set has an opinion about who should be considered socially qualified. There are several different rules that can be used to determine the socially qualified subset based on these mutual opinions. In a manipulative attack, an outsider attempts to exploit the way the used rule works, with the goal of changing the outcome of the selection process to their liking.
\\
In recent years, the complexity of group control and bribery based manipulative attacks in Group Identification has been the subject of intense research. However, the picture is far from complete, and there remain many open questions related to what exactly makes certain problems hard, or certain rules immune to some attacks.
\\
Supplementing previous results, we examine the complexity of group microbribery on so-called protective problem instances; that is, instances where all individuals from the constructive target set are already socially qualified initially. In addition, we study a relaxed variant of group control by deleting individuals for the consent rules, the consensus-start-respecting rule, and the liberal-start-respecting rule. Based on existing literature, we also formalize three new social rules of the iterative consensus type, and we provide a comprehensive complexity-theoretic analysis of group control and bribery problems for these rules.

\cleardoublepage


\begin{otherlanguage*}{ngerman}

\section*{Kurzzusammenfassung}

Beim \emph{Group Identification} Problem ist eine Menge von Individuen gegeben, und die Aufgabe besteht darin, eine sozial qualifizierte Teilmenge von ihnen zu identifizieren. Jedes Individuum in der Menge hat eine Meinung dazu, wer als sozial qualifiziert angesehen werden sollte. Es gibt verschiedene Regeln, die verwendet werden können, um die sozial qualifizierte Teilmenge auf der Grundlage dieser gegenseitigen Meinungen zu bestimmen. Bei einem manipulativen Angriff versucht eine außenstehende Person, die Funktionsweise der verwendeten Regel auszunutzen, um das Ergebnis des Auswahlprozesses nach ihren Vorstellungen zu verändern.
\\
In den letzten Jahren wurde die Komplexität von auf \emph{group control} und \emph{bribery} basierenden manipulativen Angriffen in \emph{Group Identification} intensiv erforscht. Das Bild ist jedoch bei weitem nicht vollständig, und es gibt viele offene Fragen in Bezug darauf, was genau bestimmte Probleme schwer oder bestimmte Regeln immun gegen manche Angriffe macht.
\\
Ergänzend zu früheren Ergebnissen untersuchen wir die Komplexität von \emph{group microbribery} auf sogenannten \emph{protective} Probleminstanzen, d.h.\@ Instanzen, bei denen alle Individuen aus der konstruktiven Zielmenge bereits zu Beginn sozial qualifiziert sind. Außerdem untersuchen wir eine relaxierte Variante des Problems \emph{group control by deleting individuals} für die \emph{consent rules}, die \emph{consensus-start-respecting rule}, sowie die \emph{liberal-start-respecting rule}. Basierend auf bestehender Literatur formalisieren wir zudem drei neue Regeln des Typs \emph{iterative consensus} und führen eine umfassende komplexitätstheoretische Analyse von \emph{group control} und \emph{bribery} Problemen für diese Regeln durch.

\end{otherlanguage*}

\cleardoublepage


\tableofcontents{}

\cleardoublepage

\pagestyle{plain}
\setcounter{page}{1}
\pagenumbering{arabic}

\renewcommand\cellalign{lt}
\setlength\tabcolsep{4pt}


\section{Introduction}
\label{sec:intro}

In many social and economic contexts, it is necessary to identify social groupings among individuals. For example, every person in a society belongs to certain communities, such as a family, a circle of friends, or a political association. Many of these social groups do not have clearly defined extents. Instead, the decision whether an individual belongs to a particular social group depends on various social factors, manifested in the views of the people in the society. Therefore, to determine who the members of the social group are, the societal views must be aggregated and a collective decision extracted from them. In the literature, the problem of deciding who in a given society belongs to a particular social group is commonly referred to as \emph{Group Identification}.


The study of Group Identification was initiated by \textcite{K93}. Based on this groundwork, \textcite{KR97} developed and formalized a problem setting where the individuals in a society try to determine who among them belongs to a certain social group. Thereby, every individual in the society publicly expresses an opinion about which subset of individuals the social group should consist of. For example, consider four students called A, B, C, and D who want to form a learning group for an upcoming exam, but disagree on who should be part of the group. Student A thinks that the group should consist of A, B, and D. Student B wants the group to be just B and D. According to student C, the group members should be B, C, and D. Student D actually prefers to learn alone and thinks that the learning group should comprise only A, B, and C. In this situation, what would be the ideal composition of the group? Clearly, student D does not want to be a member of the group, but all other students want D to be a member. Does the majority overrule liberalism? And student A wants the group to include student B, but B does not want A to be in the group. How should this be resolved? Group Identification deals with exactly these questions.

In addition to identifying social groups, the Group Identification model can also be applied in other contexts. For instance, consider a set of autonomous agents who are presented with a new task and want to determine who among them are in the best position to take on this task. Another possible use case is when a group of people want to form a committee by conducting a survey for the most suitable members. However, it should be noted that there are some important differences between Group Identification and voting/election systems. We address these differences in \secref{sec:intro_related}.

Throughout this thesis, we use $N$ to denote the given set of individuals. Each individual has an opinion about who should be considered as the members of the social group. The individuals express their opinions by either \emph{qualifying} or \emph{disqualifying} one another. To determine which individuals are members of the social group, a special function is applied which aggregates the views of all individuals in $N$ and extrapolates a collective opinion from them. In the original work on Group Identification by \textcite{KR97}, these aggregator functions are called \emph{collective identity functions} (CIFs). However, in the more recent literature on manipulative attacks in Group Identification, they are referred to as \emph{social rules} \cite{YD18,ERY20,BBKL20,J22}. We use the latter name throughout this thesis, and we say that the individuals who are designated by these functions are \emph{socially qualified}. In the following section, we present the social rules that will be studied in this thesis. Formal definitions of these rules will be provided in \secref{sec:social_rules}.

\subsection{Relevant social rules}
\label{sec:intro_social_rules}

The social rules that have received the most attention in the literature are the consent rules, the consensus-start-respecting rule, and the liberal-start-respecting rule \cite{YD18,ERY20,BBKL20,J22}. The behavior of these rules has been studied from a social and economic perspective, and it has been shown that they satisfy certain axioms \cite{SS03,DSX07,M08,N07,D11}.

The consent rules were introduced by \textcite{SS03} and were later slightly revised and re-formalized by \textcite{YD18} in the context of manipulative attacks in Group Identification. In this thesis, we use the definitions and notation from \textcite{YD18}. A consent rule is characterized by two positive integers $s$ and $t$. The social qualification status of an individual depends on their\footnote{Throughout this thesis, we use the singular ``they'' as a gender-neutral pronoun.} own assessment of themselves, as well as the opinions of the other individuals: An individual who qualifies themselves is socially qualified if and only if they are qualified by at least $s$ individuals in total. Likewise, an individual who disqualifies themselves is socially disqualified if and only if they are disqualified by at least $t$ individuals in total.

The consensus-start-respecting (CSR) rule and the liberal-start-respecting (LSR) rule were introduced by \textcite{DSX07}. They determine the set of socially qualified individuals iteratively: With the CSR rule, the initial set of socially qualified individuals consists of all individuals who are qualified by everyone. With the LSR rule, the initial set consists of all individuals who qualify themselves. Subsequently, both rules proceed in an iterative manner where in each iteration, all individuals who are qualified by any of the currently socially qualified individuals are added to the set of socially qualified individuals. This iterative process continues until no new individuals can be added.

The CSR and LSR rules have the disadvantage that they often result in the subset of socially qualified individuals being either empty or very large. An empty subset occurs when no individual is qualified by everyone (in case of the CSR rule), or when no individual qualifies themselves (in case of the LSR rule). On the other hand, as soon as at least some individuals are considered socially qualified, the subset of socially qualified individuals can become very large. This is because in the iterative step, for an individual to be considered socially qualified, it suffices that they are qualified by a single already socially qualified individual. As an alternative, \textcite{KR97} proposed an iterative consensus rule under which an individual only becomes socially qualified if there is a consensus about it among all individuals in the iterative step, and if the individual additionally qualifies themselves. They describe the rule as follows (note that they refer to socially qualified individuals as Js):

\begin{quote}
``Let J(0) be the set of all individuals for which there is a consensus that they are Js (possibly an empty set). Expand the set inductively by adding, at the $t$-th stage, those members of $N$ who consider themselves as Js and for whom there is a consensus among J(t-1) that they are Js.'' \cite{KR97}
\end{quote}

This definition does leave room for interpretation in two important aspects:

\begin{enumerate}
\item
For an individual to become socially qualified in the iterative step, is it necessary that they are qualified by \textit{all socially qualified individuals}, or does it suffice to be qualified only by those individuals who became socially qualified \textit{in the previous iteration}? In the former case, the iterative process would naturally be limited to at most two stages (because any individual who is not socially qualified after the first iterative step can never become socially qualified). We therefore refer to social rules that use the former behavior as \emph{2-stage}.

\item
For a given set $N$ of individuals, let $N_\ast$ denote the subset of individuals who are qualified by everyone. If the set $N_\ast$ is empty (i.e.\@ no individual is qualified by everyone), who should be considered socially qualified? On the one hand, it would be reasonable to consider no one socially qualified in this case. On the other hand, for any individual $a \in N$, if the set $N_\ast$ is empty, it could be argued that there is a consensus among everyone in $N_\ast$ that $a$ should be qualified. Thus, all individuals who qualify themselves would then automatically become socially qualified in the iterative step. We refer to social rules that use the latter behavior as \emph{liberal}.
\end{enumerate}

Based on these considerations, we define three different variants of the iterative consensus social rule: Our first variant is not 2-stage and not liberal. We refer to it simply as the \emph{iterative-consensus rule}. Our second variant is 2-stage, but not liberal. We refer to it as \emph{2-stage-iterative-consensus rule}. Finally, our third variant is both 2-stage and liberal. We refer to it as \emph{2-stage-liberal-iterative-consensus rule}. One could also define a variant that is liberal and not 2-stage, or create even more variants that function slightly differently. However, in this thesis we limit ourselves to the three rules mentioned above. We provide formal definitions for them in \secref{sec:social_rules}.

\subsection{Manipulative attacks}
\label{sec:intro_manipulative_attacks}

Apart from analyzing the axiomatic properties of different social rules, the study of Group Identification has also been focused on so-called manipulative attacks. In a manipulative attack, an \emph{attacker} attempts to change the outcome of the Group Identification process by using some manipulative means. These means include adding, deleting, and partitioning individuals, as well as bribing individuals to change their opinions. By using these means, the attacker tries to achieve a certain strategic objective. We distinguish four kinds of objectives: When the attacker tries to make certain individuals socially qualified, we speak of a \emph{constructive} objective. When the attacker tries to make certain individuals socially disqualified, we speak of a \emph{destructive} objective. If the attacker wants to precisely specify for each individual whether they should be socially qualified or socially disqualified, we speak of an \emph{exact} objective. For any other combination of simultaneously making some individuals socially qualified and some other individuals socially disqualified, we speak of a \emph{general} objective. 
To denote the subset of individuals whom the attacker wants to make socially qualified [resp.\@ socially disqualified] through the attack, we use the term constructive [resp.\@ destructive] \emph{target set}.

Depending on which social rule is used to determine the socially qualified individuals, carrying out a manipulative attack is sometimes easy and sometimes difficult or even impossible. When it is impossible for the attacker to achieve the objective with the given means, we say that the social rule is \emph{immune} to this type of attack. Otherwise, we say that the social rule is \emph{susceptible} to this type of attack, and we then try to determine the computational complexity of calculating the required steps which the attacker must take for a successful attack. If we find that a particular manipulative attack problem is NP-hard, it implies that the used social rule offers some protection against this type of attack, albeit not being completely immune to it.

We now give a brief overview of the manipulative means which the attacker may use to achieve their objective. More formal definitions of the different manipulative attack problems will be provided in \secref{sec:problem_definitions}.

\textbf{Adding individuals}: In this scenario, the attacker expands the set of participants by motivating additional individuals to take part in the process who otherwise would not have done so. In our model, this is realized by initially only considering a subset of individuals $T \subseteq N$ for the Group Identification process. The attacker can then select a certain number of individuals from $N \setminus T$ to be added to $T$.

\textbf{Deleting individuals}: In this scenario, the attacker shrinks the set of participants by preventing some individuals from taking part in the process. In our model, this means that the attacker deletes a certain number of individuals from the set $N$.

\textbf{Bribery}: In this attack scenario, the attacker bribes individuals to change their opinion. Thereby, the attacker can arbitrarily alter the opinions of every bribed individual. Each individual has a price of being bribed, and the attacker has a certain budget which they can spend. There is also an unpriced version where the attacker is allowed to bribe a certain number of individuals.

\textbf{Microbribery}: In the microbribery model, the attacker also bribes individuals, but needs to pay for each changed opinion separately. Each opinion of one individual about another has a price of being changed, and the attacker again has a certain budget which they can spend. There is also an unpriced microbribery version where the attacker can make a certain number of bribes.

Some of the available literature also covers attacks where the individuals are partitioned into subsets. However, we do not study the group control by partitioning problem in this thesis; we only mention it here for the sake of completeness.

\subsection{Related work}
\label{sec:intro_related}

In this section, we give an overview of the existing literature on Group Identification and manipulative attacks, as well as some related problems.

There are several variants of Group Identification that have been studied to date. The original Group Identification problem was introduced by \textcite{K93,KR97}. \textcite{ERY17,R18} studied a variant where not all opinions of the individuals about each other are known. In this variant, one tries to determine which individuals have the possibility of being socially qualified once the missing information is filled, and which individuals are definitely socially qualified regardless of how the missing information is filled. Recently, \textcite{FT20} also studied a version of Group Identification that operates on infinite sets, i.e.\@ with an infinite number of individuals. Moreover, \textcite{FT22}, studied Group Identification in fuzzy environments where the individuals express their opinions not in a binary way, but rather in terms of the degree to which they think someone is socially qualified.

Group Identification has similarities to the peer selection problem. In peer selection, one is given a set of individuals who have pairwise opinions about each other, and the task is to select a subset from them based on these opinions \cite{AFPT11,ALMRW16}. However, in peer selection, it is assumed that each individual wants to be selected. Therefore, each individual only expresses their opinion about the other individuals, not about themselves. Furthermore, the size of the subset to be selected in peer selection is fixed.

Group Identification is also formally related to voting systems, and in particular to multi-winner elections \cite{FSST17,EFSS17}. A voting system involves a set of voters and a set of candidates, and the voters express preference for certain candidates by submitting votes. Based on an aggregation rule, a subset of candidates is then declared winners. However, unlike Group Identification where the size of the socially qualified subset is variable, voting systems usually have a fixed number of winners, and therefore require some sort of tie-breaking mechanism. There also exist multi-winner voting systems that have a variable number of winners \cite{BBMN04,K16,FST20,YW18,DPZ16,LM21}, but they use aggregation rules different from the ones in Group Identification.

The voting systems that are most similar to Group Identification are approval-based systems. In these systems, the voters can either approve or disapprove of the candidates, and a certain number of candidates are then declared winners. There exists extensive literature on approval voting \cite{BEHHR10,FB81,LS10}, including works on the feasibility and complexity of manipulative attacks in these systems \cite{HHR07,L11,YG17,Y19,Y20,BFNT21,BKN21}.

Computational studies of manipulative attacks originated in the context of elections. This includes control-based attacks such as adding and deleting voters \cite{BTT92,MPRZ08,YG14,YG15}, and also bribery and microbribery \cite{FHH09,FHHR09}. See \textcite{FR16} for a survey on manipulative attacks in elections. In particular, microbribery in elections has also been studied in the context of lobbying where a group of legislators is voting on a set of yes/no issues, and a lobbyist tries to influence the outcome \cite{BEFGMR14,CFRS07}. Additionally, bribery and microbribery are related to the concept of \emph{margin of victory} where one tries to determine the number of votes that would need to be altered to change the outcome for a certain candidate \cite{C11,MRSW11,BBFN21,FST17,X12}.

The idea to apply the concept of manipulative attacks to Group Identification came from \textcite{YD18}, although they initially only considered the constructive group control problems. \textcite{ERY20} then extended the scope to the destructive case and to unpriced bribery. Subsequently, \textcite{EY20} also studied several unpriced microbribery problems. \textcite{BBKL20} were the first to consider the general and exact objectives for group bribery, including priced variants of bribery and microbribery. Furthermore, \textcite{BBKL21} obtained some parameterized complexity results for constructive group bribery. Also, \textcite{YD22} recently examined the complexity of group control problems restricted to consecutive domains, including several parameterized complexity results. For a comprehensive overview of manipulative attacks in Group Identification, see \textcite{J22}.

\subsection{Our contributions}
\label{sec:intro_contributions}

In this thesis, we examine various group control and bribery problems from perspectives that have not previously been considered. This way, we hope to get a better understanding of what exactly makes certain manipulative attack problems hard, or certain social rules immune to some attacks.

We introduce the concept of protective instances, i.e.\@ instances where all individuals from the constructive target set are already socially qualified initially, and the attacker only has to take care of the individuals in the destructive target set. Our results show that some problems retain the complexity of the general case when restricted to such instances, but we also obtain a result where such a restriction changes the complexity of the problem.

In the classical definition of group control by deleting individuals, the attacker is not allowed to delete individuals who are in the constructive or destructive target set. In this thesis, we relax that restriction and allow the attacker to also delete individuals from the destructive target set. We show that some social rules that were previously immune to group control by deleting individuals are susceptible to the relaxed variant. On the other hand, some problems that were previously polynomial-time solvable become NP-complete through the relaxation.

Based on the iterative consensus rule proposed by \textcite{KR97}, we formalize three new social rules that have not yet been considered in the context of manipulative attacks. These rules are similar to the consensus-start-respecting rule, but operate in a slightly more stringent way, often leading to more reasonable results. We conduct a comprehensive complexity-theoretic analysis of group control and bribery problems for these rules, obtaining many NP-hardness results, but also some polynomial-time algorithms. Most notably, we find that for certain bribery problems, the priced version has a different complexity than the unpriced one. This is surprising, considering that for all social rules studied previously, the priced and unpriced bribery versions have the same complexity.

\subsection{Organisation}
\label{sec:intro_organization}

The remainder of this thesis is organized as follows: In \secref{sec:basic_definitions}, we introduce our basic notation and define some important concepts and problems. In \secref{sec:social_rules} and \secref{sec:problem_definitions}, we provide formal definitions of the social rules and Group Identification problems that will be studied in this thesis. In \secref{sec:protective_instances}, we examine the complexity of group microbribery on protective instances when using the CSR or LSR social rule. \secref{sec:relaxed_group_control_deleting} is devoted to a relaxed variant of group control by deleting individuals. In \secref{sec:manipulative_attacks_iterative_consensus}, we study manipulative attacks for the previously unstudied iterative consensus rules. \secref{sec:conclusion} concludes this thesis with a summary of our results and ideas for future research.

\clearpage

\section{Basic definitions and notation}
\label{sec:basic_definitions}

In this section, we introduce some basic definitions, notation, and concepts that will be needed to understand the proofs in this thesis.

\subsection{Graphs and operations on graphs}

A graph is a tuple $(V, E)$ where $V$ is the vertex set and $E$ is the edge set. In undirected graphs, the set of edges is defined as $E \subseteq \binom{V}{2}$. For an edge $\{u, v\} \in E$, we call $u$ and $v$ the \emph{endpoints} of the edge, and we say that $u$ and $v$ are \emph{adjacent}. In directed graphs, the set of edges is defined as $E \subseteq V \times V$. For a directed edge $(u, v) \in E$ (also referred to as \emph{arc} from $u$ to $v$), we again call $u$ and $v$ the endpoints and say that they are adjacent. In this case, $v$ is an \emph{out-neighbor} of $u$, and $u$ is an \emph{in-neighbor} of $v$. For a vertex $v \in V$, its degree $\operatorname{deg}(v)$ is defined as the number of vertices it is adjacent to.

Let $G = (V, E)$ be any graph (directed or undirected). For a subset of vertices $V^\prime \subseteq V$, we use $G[V^\prime]$ to denote the subgraph of $G$ induced by the vertices in $V^\prime$. Precisely, $G[V^\prime]$ consists of the vertices in $V^\prime$ along with any edges whose endpoints are both in $V^\prime$. Furthermore, for a subset of vertices $V^\prime \subseteq V$, we use $G - V^\prime$ to denote the graph $G[V \setminus V^\prime]$. In other words, $G - V^\prime$ is the subgraph obtained from $G$ by deleting the vertices in $V^\prime$ and all edges with at least one endpoint in $V^\prime$.

Let $G = (V, E)$ be a directed graph. For a vertex subset $V^\prime \subseteq V$, the operation of \emph{merging} the vertices in $V^\prime$ into a single vertex $w$ is defined as follows:
\begin{enumerate}[nolistsep]
\item
We create one new vertex $w$.
\item
For each vertex $v \in V \setminus V^\prime$ that is an in-neighbor of some vertex in $V^\prime$, we create an edge from $v$ to $w$.
\item
For each vertex $v \in V \setminus V^\prime$ that is an out-neighbor of some vertex in $V^\prime$, we create an edge from $w$ to $v$.
\item
We delete all vertices in $V^\prime$ and all edges with at least one endpoint in $V^\prime$.
\end{enumerate}

For further information about graph theory, we refer to \textcite{W01}.

\subsection{NP-complete problems}

We now provide the definitions of several NP-complete problems which will be used in reductions throughout this thesis.

\clearpage

A \emph{boolean variable} $x$ is either assigned the value $1$ or $0$. Let $X$ be a set of boolean variables. For a variable $x \in X$, we say that $x$ and $\bar{x}$ are \emph{literals} over $X$. A set of literals over $X$ is called a \emph{clause}. A \emph{truth assignment} is a function $\varrho : X \rightarrow \{0, 1\}$. We say that a clause $c$ is satisfied under a truth assignment $\varrho$ if and only if $c$ contains a literal $x$ such that $\varrho(x) = 1$, or $c$ contains a literal $\bar{x}$ such that $\varrho(x) = 0$. Let $C$ be a collection of clauses over $X$. We can interpret $C$ as a formula in \emph{conjunctive normal form} (CNF). A truth assignment $\varrho$ under which all clauses in $C$ are satisfied is called a \emph{satisfying assignment} for $C$. The CNF-SAT problem defined below is well known to be NP-complete \cite{K72}.

\begin{tabularx}{\textwidth}{lX}
\hline
\multicolumn{2}{l}{
\textsc{CNF-Satisfiability} (CNF-SAT)
} \\
\hline
\textbf{Given:} &
A set $X$ of boolean variables, and a collection $C$ of clauses over $X$. \\
\textbf{Question:} &
Is there a satisfying assignment $\varrho : X \rightarrow \{0, 1\}$ for $C$? \\
\hline
\end{tabularx}

\vspace{\baselineskip}

An \emph{independent set} of an undirected graph $G = (V, E)$ is a vertex subset $V^\prime \subseteq V$ where no two vertices in $V^\prime$ are adjacent. The \textsc{Independent Set} problem is NP-complete \cite{K72}.

\begin{tabularx}{\textwidth}{lX}
\hline
\multicolumn{2}{l}{
\textsc{Independent Set}
} \\
\hline
\textbf{Given:} &
An undirected graph $G = (V, E)$ and a positive integer $k$. \\
\textbf{Question:} &
Is there an independent set of size at least $k$ in $G$? \\
\hline
\end{tabularx}

\vspace{\baselineskip}

A \emph{vertex cover} of an undirected graph $G = (V, E)$ is a vertex subset $V^\prime \subseteq V$ such that each edge has at least one endpoint in $V^\prime$. In other words, $V \setminus V^\prime$ is an independent set. The \textsc{Vertex Cover} problem is a known NP-complete problem \cite{K72}.

\begin{tabularx}{\textwidth}{lX}
\hline
\multicolumn{2}{l}{
\textsc{Vertex Cover}
} \\
\hline
\textbf{Given:} &
An undirected graph $G = (V, E)$ and a positive integer $k$. \\
\textbf{Question:} &
Is there a vertex cover of size at most $k$ in $G$? \\
\hline
\end{tabularx}

\vspace{\baselineskip}

A \emph{set cover} for a set $X$ and a family $\mathcal{F}$ of subsets of $X$ is defined as a subfamily $\mathcal{F}^\prime \subseteq \mathcal{F}$ such that each element in $X$ is contained in at least one subset in $\mathcal{F}^\prime$. The \textsc{Set Cover} problem is known to be NP-complete \cite{K72}. Also, it is known to be W[2]-hard with respect to the requested size $k$ of the cover \cite{DF95}.

\begin{tabularx}{\textwidth}{lX}
\hline
\multicolumn{2}{l}{
\textsc{Set Cover}
} \\
\hline
\textbf{Given:} &
A set $X$, a family $\mathcal{F}$ of subsets of $X$, and a positive integer $k$. \\
\textbf{Question:} &
Is there a set cover of cardinality at most $k$ for $X$ in $\mathcal{F}$? \\
\hline
\end{tabularx}

\vspace{\baselineskip}

The \textsc{Restricted Exact Cover by 3-sets} (RX3C) problem is a special case of a set cover. The problem instances have several properties that make them useful for reductions. It has been shown that the RX3C problem is also NP-complete \cite[Theorem A.1]{G85}.

\begin{tabularx}{\textwidth}{lX}
\hline
\multicolumn{2}{l}{
\textsc{Restricted Exact Cover by 3-sets} (RX3C)
} \\
\hline
\textbf{Given:} &
A set $X$ of size $3m$ and a family $\mathcal{F}$ of subsets of $X$. \newline
Each subset $F \in \mathcal{F}$ contains exactly three elements from $X$, and each element in $X$ is contained in exactly three subsets $F \in \mathcal{F}$. \\
\textbf{Question:} &
Is there a subfamily $\mathcal{F}^\prime \subseteq \mathcal{F}$ such that every element in $X$ is contained in exactly one subset in $\mathcal{F}^\prime$? \\
\hline
\end{tabularx}

\vspace{\baselineskip}

\subsection{Group Identification notation}

Let $N$ be a set of individuals. Each individual in $N$ has an opinion about who from $N$ is qualified in a certain way and who is not. We represent these pairwise opinions by a \emph{binary profile} over $N$ which is defined as a function $\varphi : N \times N \rightarrow \{ -1, 1 \}$. For individuals $a, b \in N$, we write $\varphi(a, b) = 1$ when $a$ thinks $b$ is qualified and say \emph{$a$ qualifies $b$}. We write $\varphi(a, b) = -1$ when $a$ thinks $b$ is not qualified and say \emph{$a$ disqualifies $b$}.

Throughout this thesis we always use $n$ to denote the number of individuals in the given instance, i.e.\@ $n = |N|$. In particular, all running time specifications in this thesis use $n$ as the instance size measure.

Let $\varphi$ be a binary profile over a set $N$ of individuals. We sometimes refer to the function values $\varphi(\cdot, \cdot)$ as \emph{entries} of $\varphi$. Naturally, $\varphi$ has $n^2$ entries in total. For an individual $a \in N$, the entries of the form $\varphi(a, \cdot)$ are called \emph{outgoing qualifications of $a$}, and the entries of the form $\varphi(\cdot, a)$ are called \emph{incoming qualifications of $a$}. Furthermore, for any subset $T \subseteq N$ and $x \in \{ -1, 1 \}$, we sometimes write $T^x_\varphi(a)$ to denote the set $\{ a^\prime \in T : \varphi(a^\prime, a) = x \}$. In other words, $T^1_\varphi(a)$ denotes the set of all individuals in $T$ who qualify $a$, and $T^{-1}_\varphi(a)$ denotes the set of all individuals in $T$ who disqualify $a$.

It is often useful to interpret a given Group Identification instance as a directed graph where the vertices represent the individuals and the edges represent the qualifications. Formally, the graph representation of a set $N$ of individuals and a binary profile $\varphi$ over $N$ is defined as $G_{N, \varphi} = (N, E)$ with $E = \{ (a, b) : a, b \in N \text{ and } \varphi(a, b) = 1 \}$. We refer to the graph $G_{N, \varphi}$ as \emph{qualification graph} of $N$ and $\varphi$.

When considering bribery problems, each individual has a certain cost of being bribed. We use a so-called \emph{bribery price function} $\rho : N \rightarrow \mathbb{N}$ to assign a positive integer price to each individual. For a subset of individuals $N^\prime \subseteq N$, we sometimes write $\rho(N^\prime)$ to denote $\sum_{a \in N^\prime} \rho(a)$.

In the microbribery model, the attacker needs to pay for each changed entry of $\varphi$ separately. A \emph{microbribery price function} $\rho : N \times N \rightarrow \mathbb{N}$ assigns every pair $(a, b) \in N \times N$ of individuals a positive integer price, corresponding to the cost of bribing $a$ to change their opinion about $b$. For a set of pairs of individuals $M \subseteq N \times N$, we sometimes write $\rho(M)$ to denote $\sum_{m \in M} \rho(m)$.

\clearpage

\section{Social rules}
\label{sec:social_rules}

Given a set $N$ of individuals and a profile $\varphi$ over $N$, we determine the socially qualified individuals by applying a social rule. A social rule is a function $f$ that assigns a subset $f(T, \varphi) \subseteq T$ to each pair $(T, \varphi)$ where $T \subseteq N$. We refer to the individuals in $f(T, \varphi)$ as the \emph{socially qualified individuals of \thinspace$T\thinspace$and $\varphi$ under $f$}. Analogously, we refer to the individuals in $T \setminus f(T, \varphi)$ as the \emph{socially disqualified individuals of \thinspace$T\thinspace$and $\varphi$ under $f$}.

Below, we give formal definitions of the social rules that will be studied in this thesis. For each social rule, we also state the subset of socially qualified individuals with respect to the following example instance.

\begin{Example} \label{ex:social_rules_example}
Let $N = \{ a_1, a_2, a_3, a_4, a_5, a_6 \}$ be a set of six individuals and let $\varphi$ be defined as follows ($\varphi(a_i, a_j)$ is given by the matrix entry in the $i$-th row and $j$-th column):

\capstartfalse 
\begin{figure}[!h]
\centering
\begin{minipage}[b]{.4\textwidth}

\renewcommand{\arraystretch}{1.2}
\begin{equation*}
\begin{array}{rrrrrrr}
    & a_1 & a_2 & a_3 & a_4 & a_5 & a_6  \\
a_1 &  1 &  1 &  1 &  1 & -1 & -1 \\
a_2 &  1 &  1 & -1 &  1 & -1 & -1 \\
a_3 &  1 &  1 & -1 & -1 & -1 & -1 \\
a_4 &  1 &  1 & -1 &  1 &  1 & -1 \\
a_5 &  1 &  1 &  1 & -1 &  1 & -1 \\
a_6 &  1 &  1 &  1 & -1 & -1 &  1
\end{array}
\end{equation*}
\caption*{Definition matrix of $\varphi$.}

\end{minipage}\hfill
\begin{minipage}[b]{.6\textwidth}
\centering

\raisebox{-0.05\height}{
\begin{tikzpicture}[
    > = Stealth,
    shorten > = 1pt,
    auto,
    node distance = 2cm,
    main node/.style = {circle,draw}
]

\node[main node] (1)                                        {$a_1$};
\node[main node] (2) [right of = 1]                         {$a_2$};
\node[main node] (3) [below left of = 1, yshift = -1cm]     {$a_3$};
\node[main node] (4) [below right of = 2, yshift = -1cm]    {$a_4$};
\node[main node] (5) [below right of = 3]                   {$a_5$};
\node[main node] (6) [below left of = 4]                    {$a_6$};

\path[->]
(1)     edge[loop left]         node {} (1)
        edge[bend right=0]      node {} (2)
        edge[bend right=20]     node {} (3)
        edge[bend right=15]     node {} (4)
(2)     edge[bend right=20]     node {} (1)
        edge[loop right]        node {} (2)
        edge[bend right=0]      node {} (4)
(3)     edge[bend right=0]      node {} (1)
        edge[bend right=5]      node {} (2)
(4)     edge[bend right=0]      node {} (1)
        edge[bend right=20]     node {} (2)
        edge[loop right]        node {} (4)
        edge[bend right=5]      node {} (5)
(5)     edge[bend left=5]       node {} (1)
        edge[bend right=10]     node {} (2)
        edge[bend right=5]      node {} (3)
        edge[loop left]         node {} (5)
(6)     edge[bend left=10]      node {} (1)
        edge[bend right=5]      node {} (2)
        edge[bend right=10]     node {} (3)
        edge[loop right]        node {} (6)
;

\end{tikzpicture}
}
\caption*{Corresponding qualification graph $G_{N, \varphi}$.}

\end{minipage}
\end{figure}
\capstarttrue 
\end{Example}

We now provide the definitions of the social rules that are relevant for this thesis. Let $N$ be a set of individuals and $\varphi$ a binary profile over $N$.

\textbf{Consent rules} $f^{(s, t)}$: The consent rules are a family of social rules. Each consent rule is specified by two parameters $s, t \in \mathbb{N}$. For every subset $T \subseteq N$ and every individual $a \in T$, it holds:
\setlist[itemize,1]{label=\normalfont\bfseries\textendash}
\begin{itemize}[nolistsep]
\item
If $\varphi(a, a) = 1$
then $a \in f^{(s, t)}(T, \varphi)$ if and only if
$|T^1_\varphi(a)| \geq s$.
\item
If $\varphi(a, a) = -1$
then $a \not\in f^{(s, t)}(T, \varphi)$ if and only if
$|T^{-1}_\varphi(a)| \geq t$.
\end{itemize}
\setlist[itemize,1]{label=\textbullet}

In other words, an individual who qualifies themselves is socially qualified if and only if at least $s-1$ other individuals also qualify them. An individual who disqualifies themselves is socially disqualified if and only if at least $t-1$ other individuals also disqualify them.

A special case of the consent rule is when $s = t = 1$. Under the $f^{(1, 1)}$ rule (also referred to as \emph{liberal rule} in the literature \cite{SS03}), an individual is socially qualified if and only if they qualify themselves.

If we apply the liberal rule to the instance depicted in \exref{ex:social_rules_example}, the set of socially qualified individuals would be $f^{(1, 1)}(N, \varphi) = \{ a_1, a_2, a_4, a_5, a_6 \}$ because those and only those five individuals qualify themselves.

If we apply the consent rule $f^{(3, 4)}$ with $s=3$ and $t=4$ to the instance depicted in \exref{ex:social_rules_example}, the individuals $a_1$ and $a_2$ would be socially qualified because they qualify themselves and are qualified by all other individuals. The individual $a_3$ would be socially qualified because they disqualify themselves but are disqualified by less than three other individuals (only $a_2$ and $a_4$ also disqualify $a_3$). The individual $a_4$ would be socially qualified because they qualify themselves and are qualified by two other individuals (namely $a_1$ and $a_2$). Finally, the individuals $a_5$ and $a_6$ would be socially disqualified because they qualify themselves but are each qualified by less than two other individuals. Thus, the set of socially qualified individuals would be $f^{(3, 4)}(N, \varphi) = \{ a_1, a_2, a_3, a_4 \}$.

All the remaining social rules defined below determine the socially qualified individuals in an iterative manner. They start with an initial set of socially qualified individuals and then extend it iteratively based on certain conditions. The iterative process stops when no new individuals can be added.

\textbf{Liberal-start-respecting rule} $f^{\text{LSR}}$: With this rule, the initial set of socially qualified individuals consists of all individuals who qualify themselves. In each iteration, any individual who is qualified by any of the currently socially qualified individuals is added to the set of socially qualified individuals.

More formally, for every subset $T \subseteq N$, the initial set of socially qualified individuals is defined as
$$
K^{\text{L}}_0(T, \varphi) = \{ a \in T : \varphi(a, a) = 1 \}.
$$
And for each positive integer $i$, the set of socially qualified individuals in the $i$-th iteration is defined as
$$
K^{\text{L}}_i(T, \varphi) = K^{\text{L}}_{i-1}(T, \varphi) \cup \{ a \in T : \varphi(a^\prime, a) = 1 \text{ for some } a^\prime \in K^{\text{L}}_{i-1}(T, \varphi) \}.
$$
We define $f^{\text{LSR}}(T, \varphi) = K^{\text{L}}_i(T, \varphi)$ for the smallest $i$ with $K^{\text{L}}_i(T, \varphi) = K^{\text{L}}_{i-1}(T, \varphi)$.

Applying the $f^{\text{LSR}}$ rule to the instance depicted in \exref{ex:social_rules_example}, we initially have $K^{\text{L}}_0(N, \varphi) = \{ a_1, a_2, a_4, a_5, a_6 \}$ since these five individuals each qualify themselves. Because $a_1$ also qualifies $a_3$, we then get $K^{\text{L}}_1(N, \varphi) = \{ a_1, a_2, a_3, a_4, a_5, a_6 \} = f^{\text{LSR}}(N, \varphi)$.

\textbf{Consensus-start-respecting rule} $f^{\text{CSR}}$: With this rule, the initial set of socially qualified individuals consists of all individuals for whom there is a consensus among everyone that they should be qualified. In each iteration, any individual who is qualified by any of the currently socially qualified individuals is added to the set of socially qualified individuals.

More formally, for every subset $T \subseteq N$, the initial set of socially qualified individuals is defined as
$$
K^{\text{C}}_0(T, \varphi) = \{ a \in T : \varphi(a^\prime, a) = 1 \text{ for all } a^\prime \in T \}.
$$
And for each positive integer $i$, the set of socially qualified individuals in the $i$-th iteration is defined as
$$
K^{\text{C}}_i(T, \varphi) = K^{\text{C}}_{i-1}(T, \varphi) \cup \{ a \in T : \varphi(a^\prime, a) = 1 \text{ for some } a^\prime \in K^{\text{C}}_{i-1}(T, \varphi) \}.
$$
We define $f^{\text{CSR}}(T, \varphi) = K^{\text{C}}_i(T, \varphi)$ for the smallest $i$ with $K^{\text{C}}_i(T, \varphi) = K^{\text{C}}_{i-1}(T, \varphi)$.

Applying the $f^{\text{CSR}}$ rule to the instance depicted in \exref{ex:social_rules_example}, we initially have $K^{\text{C}}_0(N, \varphi) = \{ a_1, a_2 \}$ since those two individuals are qualified by everyone. Because $a_1$ also qualifies $a_3$ and $a_4$, we then get $K^{\text{C}}_1(N, \varphi) = \{ a_1, a_2, a_3, a_4 \}$. The individual $a_4$ also qualifies $a_5$, so we get $K^{\text{C}}_2(N, \varphi) = \{ a_1, a_2, a_3, a_4, a_5 \}$. Because $a_5$ qualifies no further individuals, we obtain the final result $f^{\text{CSR}}(N, \varphi) = \{ a_1, a_2, a_3, a_4, a_5 \}$.

\textbf{Iterative-consensus rule} $f^{\text{IC}}$: This rule begins with the same initial set as the consensus-start-respecting rule. Then, in each iteration, the set of socially qualified individuals is extended to include any individual who qualifies themselves and for whom there is a consensus among the individuals added in the previous iteration that they should be qualified. Crucially, this requires that the set of individuals added in the previous iteration is nonempty.

More formally, for every $T \subseteq N$ and each integer $i \geq 0$, we use $Q^{\text{IC}}_i(T, \varphi)$ to denote the set of newly socially qualified individuals in the $i$-th iteration; and we let
$$
K^{\text{IC}}_i(T, \varphi) = \mathsmaller{\bigcup}_{j \in \{ 0, \ldots, i \}} Q^{\text{IC}}_j(T, \varphi)
$$
denote the set of all individuals who are socially qualified after the $i$-th iteration.

The initial set of socially qualified individuals is defined as
$$
Q^{\text{IC}}_0(T, \varphi) = \{ a \in T : \varphi(a^\prime, a) = 1 \text{ for all } a^\prime \in T \}.
$$
And for all $i \geq 1$, we define
\begin{multline*}
Q^{\text{IC}}_i(T, \varphi) = \{ a \in \big(T \setminus K^{\text{IC}}_{i-1}(T, \varphi)\big) : \\
\varphi(a, a) = 1 \text{ and }
Q^{\text{IC}}_{i-1}(T, \varphi) \neq \emptyset \text{ and }
\varphi(a^\prime, a) = 1 \text{ for all } a^\prime \in Q^{\text{IC}}_{i-1}(T, \varphi) \}.
\end{multline*}
We define $f^{\text{IC}}(T, \varphi) = K^{\text{IC}}_i(T, \varphi)$ for the smallest $i$ with $K^{\text{IC}}_i(T, \varphi) = K^{\text{IC}}_{i-1}(T, \varphi)$.

Applying the $f^{\text{IC}}$ rule to the instance depicted in \exref{ex:social_rules_example}, we initially have $K^{\text{IC}}_0(N, \varphi) = Q^{\text{IC}}_0(N, \varphi) = \{ a_1, a_2 \}$ since those two individuals are qualified by everyone. Among $a_1$ and $a_2$, there is a consensus that $a_4$ should be qualified. Because $a_4$ also qualifies themselves, we get $Q^{\text{IC}}_1(N, \varphi) = \{ a_4 \}$ and thus $K^{\text{IC}}_1(N, \varphi) = \{ a_1, a_2, a_4 \}$. Everyone from $Q^{\text{IC}}_1(N, \varphi)$ qualifies the individual $a_5$. Since $a_5$ also qualifies themselves, we get $Q^{\text{IC}}_2(N, \varphi) = \{ a_5 \}$ and thus $K^{\text{IC}}_2(N, \varphi) = \{ a_1, a_2, a_4, a_5 \}$. Now, everyone from $Q^{\text{IC}}_2(N, \varphi)$ qualifies $a_3$, but $a_3$ does not qualify themselves. Therefore, $Q^{\text{IC}}_3(N, \varphi) = \emptyset$ and we obtain the final result $f^{\text{IC}}(N, \varphi) = K^{\text{IC}}_3(N, \varphi) = \{ a_1, a_2, a_4, a_5 \}$.

\textbf{2-stage-iterative-consensus rule} $f^{\text{2IC}}$: This rule begins with the same initial set as the iterative-consensus rule. However, in the iterative step, an individual is only added if they qualify themselves and there is a consensus among \textit{all socially qualified individuals} that they should be qualified. For this reason, the selection process is naturally limited to at most two stages. If no individual is qualified by everyone, then the set of socially qualified individuals remains empty.

More formally, for every subset $T \subseteq N$, the initial set of socially qualified individuals is defined as
$$
K^{\text{2IC}}_0(T, \varphi) = \{ a \in T : \varphi(a^\prime, a) = 1 \text{ for all } a^\prime \in T \}.
$$
And the set of socially qualified individuals after the first iteration is defined as
\begin{multline*}
K^{\text{2IC}}_1(T, \varphi) = K^{\text{2IC}}_0(T, \varphi) \cup \{ a \in T : \\
\varphi(a, a) = 1 \text{ and }
K^{\text{2IC}}_0(T, \varphi) \neq \emptyset \text{ and }
\varphi(a^\prime, a) = 1 \text{ for all } a^\prime \in K^{\text{2IC}}_0(T, \varphi) \}.
\end{multline*}
We define $f^{\text{2IC}}(T, \varphi) = K^{\text{2IC}}_1(T, \varphi)$.

Applying the $f^{\text{2IC}}$ rule to the instance depicted in \exref{ex:social_rules_example}, we initially have $K^{\text{2IC}}_0(N, \varphi) = \{ a_1, a_2 \}$. Among $a_1$ and $a_2$, there is a consensus that $a_4$ should be qualified. Since $a_4$ also qualifies themselves, we get $K^{\text{2IC}}_1(N, \varphi) = \{ a_1, a_2, a_4 \} = f^{\text{2IC}}(N, \varphi)$.

\textbf{2-stage-liberal-iterative-consensus rule} $f^{\text{2LIC}}$: This rule is similar to the 2-stage-iterative-consensus rule, but it relaxes the condition in the iterative step: It still requires that an individual qualifies themselves and that there is a consensus among the individuals who are qualified by everyone. However, if no one is qualified by everyone, then it suffices that an individual qualifies themselves to be considered socially qualified. This way, the $f^{\text{2LIC}}$ rule satisfies the Liberal Principle axiom defined by \textcite{KR97} which requires that the set of socially qualified individuals is nonempty if there exists at least one individual who qualifies themselves.

More formally, for every subset $T \subseteq N$, the initial set of socially qualified individuals is defined as
$$
K^{\text{2LIC}}_0(T, \varphi) = \{ a \in T : \varphi(a^\prime, a) = 1 \text{ for all } a^\prime \in T \}.
$$
And the set of socially qualified individuals after the first iteration is defined as
\begin{multline*}
K^{\text{2LIC}}_1(T, \varphi) = K^{\text{2LIC}}_0(T, \varphi) \cup \{ a \in T : \\
\varphi(a, a) = 1 \text{ and }
\varphi(a^\prime, a) = 1 \text{ for all } a^\prime \in K^{\text{2LIC}}_0(T, \varphi) \}.
\end{multline*}
We define $f^{\text{2LIC}}(T, \varphi) = K^{\text{2LIC}}_1(T, \varphi)$.

If we apply the $f^{\text{2LIC}}$ rule to the instance depicted in \exref{ex:social_rules_example}, we get the same result as with the $f^{\text{2IC}}$ rule. However, if we use a profile $\varphi^\prime$ that is identical to $\varphi$ except that $a_1$ and $a_2$ each disqualify themselves, then we get a different result: The initial set of socially qualified individuals is now empty, i.e.\@ $K^{\text{2LIC}}_0(N, \varphi^\prime) = \emptyset$. Thus, in the iterative step, any individual who qualifies themselves is considered socially qualified, i.e.\@ $K^{\text{2LIC}}_1(N, \varphi^\prime) = \{ a_4, a_5, a_6 \} = f^{\text{2LIC}}(N, \varphi^\prime)$.

\clearpage

\section{Problem definitions}
\label{sec:problem_definitions}

In this section, we provide formal definitions of the manipulative attack problems outlined in \secref{sec:intro_manipulative_attacks}. Let $f$ be a social rule.

\bigskip

\begin{tabularx}{\textwidth}{lX}
\hline
\multicolumn{2}{l}{
$f$-\textsc{Group Control by Adding Individuals} (GCAI)
} \\
\hline
\textbf{Given:} &
A 6-tuple $(N, \varphi, A^+, A^-, T, \ell)$ of a set $N$ of individuals, a profile $\varphi$ over $N$, three subsets $A^+, A^-, T \subseteq N$ with $A^+ \cap A^- = \emptyset$ and $A^+, A^- \subseteq T$, and a positive integer $\ell$. \\
\textbf{Question:} &
Is there a subset $U \subseteq N \setminus T$ such that $|U| \leq \ell$ and $A^+ \subseteq f(T \cup U, \varphi)$ and $A^- \cap f(T \cup U, \varphi) = \emptyset$? \\
\hline
\end{tabularx}

In addition to the general $f$-GCAI problem, there are the three special variants
\begin{itemize}
\item
$f$-\textsc{Constructive Group Control by Adding Individuals} (CGCAI) where the set $A^-$ is dropped,
\item
$f$-\textsc{Destructive Group Control by Adding Individuals} (DGCAI) where the set $A^+$ is dropped,
\item
$f$-\textsc{Exact Group Control by Adding Individuals} (EGCAI) where we add the premise $A^+ \cup A^- = T$.
\end{itemize}

\bigskip

\begin{tabularx}{\textwidth}{lX}
\hline
\multicolumn{2}{l}{
$f$-\textsc{Group Control by Deleting Individuals} (GCDI)
} \\
\hline
\textbf{Given:} &
A 5-tuple $(N, \varphi, A^+, A^-, \ell)$ of a set $N$ of individuals, a profile $\varphi$ over $N$, two subsets $A^+, A^- \subseteq N$ with $A^+ \cap A^- = \emptyset$, and a positive integer $\ell$. \\
\textbf{Question:} &
Is there a subset $U \subseteq N \setminus (A^+ \cup A^-)$ such that $|U| \leq \ell$ and $A^+ \subseteq f(N \setminus U, \varphi)$ and $A^- \cap f(N \setminus U, \varphi) = \emptyset$? \\
\hline
\end{tabularx}

In addition to the general $f$-GCDI problem, there are the two special variants
\begin{itemize}
\item
$f$-\textsc{Constructive Group Control by Deleting Individuals} (CGCDI) where the set $A^-$ is dropped,
\item
$f$-\textsc{Destructive Group Control by Deleting Individuals} (DGCDI) where the set $A^+$ is dropped.
\end{itemize}

The $f$-\textsc{Exact Group Control by Deleting Individuals} problem is not defined because we only allow the attacker to delete individuals from $N \setminus (A^+ \cup A^-)$. Thus, when $A^+ \cup A^- = N$, the attacker would not be allowed to delete anyone. In \secref{sec:relaxed_group_control_deleting}, we consider a relaxed variant of the problem where we also allow the attacker to delete individuals from $A^-$.




\bigskip

\begin{tabularx}{\textwidth}{lX}
\hline
\multicolumn{2}{l}{
$f$-\textsc{\$Group Bribery} (\$GB)
} \\
\hline
\textbf{Given:} &
A 6-tuple $(N, \varphi, A^+, A^-, \rho, \ell)$ of a set $N$ of individuals, a profile $\varphi$ over $N$, two subsets $A^+, A^- \subseteq N$ with $A^+ \cap A^- = \emptyset$, a bribery price function $\rho : N \rightarrow \mathbb{N}$, and a positive integer $\ell$. \\
\textbf{Question:} &
Is there a way to obtain a profile $\varphi^\prime$ over $N$ by changing the outgoing qualifications in $\varphi$ of some individuals $U \subseteq N$ such that $\rho(U) \leq \ell$ and $A^+ \subseteq f(N, \varphi^\prime)$ and $A^- \cap f(N, \varphi^\prime) = \emptyset$? \\
\hline
\end{tabularx}

In addition to the general $f$-\$GB problem, there are the three special variants
\begin{itemize}
\item
$f$-\textsc{\$Constructive Group Bribery} (\$CGB) where the set $A^-$ is dropped,
\item
$f$-\textsc{\$Destructive Group Bribery} (\$DGB) where the set $A^+$ is dropped,
\item
$f$-\textsc{\$Exact Group Bribery} (\$EGB) where we add the premise $A^+ \cup A^- = N$.
\end{itemize}

Also, there are the four unpriced versions $f$-\textsc{Group Bribery} (GB), $f$-\textsc{Constructive Group Bribery} (CGB), $f$-\textsc{Destructive Group Bribery} (DGB), and $f$-\textsc{Exact Group Bribery} (EGB) where $\rho$ assigns the price $1$ to every individual, i.e.\@ $\rho(a) = 1$ for all $a \in N$.

\bigskip

\begin{tabularx}{\textwidth}{lX}
\hline
\multicolumn{2}{l}{
$f$-\textsc{\$Group Microbribery} (\$GMB)
} \\
\hline
\textbf{Given:} &
A 6-tuple $(N, \varphi, A^+, A^-, \rho, \ell)$ of a set $N$ of individuals, a profile $\varphi$ over $N$, two subsets $A^+, A^- \subseteq N$ with $A^+ \cap A^- = \emptyset$, a microbribery price function $\rho : N \times N \rightarrow \mathbb{N}$, and a positive integer $\ell$. \\
\textbf{Question:} &
Is there a way to obtain a profile $\varphi^\prime$ over $N$ by changing the entries of $\varphi$ for some pairs $M \subseteq N \times N$ of individuals such that $\rho(M) \leq \ell$ and $A^+ \subseteq f(N, \varphi^\prime)$ and $A^- \cap f(N, \varphi^\prime) = \emptyset$? \\
\hline
\end{tabularx}

In addition to the general $f$-\$GMB problem, there are the three special variants
\begin{itemize}
\item
$f$-\textsc{\$Constructive Group Microbribery} (\$CGMB) where the set $A^-$ is dropped,
\item
$f$-\textsc{\$Destructive Group Microbribery} (\$DGMB) where the set $A^+$ is dropped,
\item
$f$-\textsc{\$Exact Group Microbribery} (\$EGMB) where we add the premise \\ $A^+ \cup A^- = N$.
\end{itemize}

Furthermore, there are the four unpriced versions $f$-\textsc{Group Microbribery} (GMB), $f$-\textsc{Constructive Group Microbribery} (CGMB), $f$-\textsc{Destructive Group Microbribery} (DGMB), and $f$-\textsc{Exact Group Microbribery} (EGMB) where $\rho$ assigns the price $1$ to every pair of individuals, i.e.\@ $\rho\big((a, b)\big) = 1$ for all $(a, b) \in N \times N$.

\subsection{Some remarks on problem instances and complexity}

In this thesis, we only consider problem instances where at least one of the target sets $A^+$ or $A^-$ is nonempty. Obviously, for the constructive and destructive problems (which only have one target set), this means that the target set must always be nonempty.

For most problem instances considered in this thesis, if $A^+$ is nonempty, we require that $A^+ \not\subseteq f(N, \varphi)$. In other words, at least one individual from $A^+$ must be socially disqualified initially. Likewise, if $A^-$ is nonempty, we require that $A^- \cap f(N, \varphi) \neq \emptyset$. In other words, at least one individual from $A^-$ must be socially qualified initially. Sometimes, we also consider instances where $A^+ \subseteq f(N, \varphi)$ or $A^- \cap f(N, \varphi) = \emptyset$ already holds initially. However, this is only the case if we explicitly state it in the corresponding theorem.

Throughout this thesis, whenever we claim that a problem is NP-complete, we only show NP-hardness. It is easy to verify that for each of the social rules defined in \secref{sec:social_rules}, the set of socially qualified individuals for a profile $\varphi$ over a given set $N$ can be computed in polynomial time. Hence, all problems considered in this thesis are in NP because the solutions can be verified in polynomial time.

\clearpage

\section{Protective group microbribery is still NP-complete}
\label{sec:protective_instances}

\textcite[Theorem 2]{BBKL20} showed that DGMB and EGMB are polynomial-time solvable for both the $f^{\text{CSR}}$ rule and the $f^{\text{LSR}}$ rule. On the other hand, they showed that CGMB is NP-complete for these rules by a reduction from \textsc{Set Cover} \cite[Theorem 3]{BBKL20}. One direct implication of this is that the general $f^{\text{CSR}} / f^{\text{LSR}}$-GMB problem is also NP-complete, provided that $A^+ \neq \emptyset$ and $A^+ \not\subseteq f^{\text{CSR}} / f^{\text{LSR}}(N, \varphi)$ holds initially. However, it is not immediately clear what would happen if we consider instances where everyone in $A^+$ is already socially qualified. In such instances, the attacker just needs to ``protect'' the individuals in $A^+$ from turning socially disqualified, and meanwhile take care of the individuals in $A^-$. Below, we answer this question and show that $f^{\text{CSR}} / f^{\text{LSR}}$-GMB is still NP-complete even when restricted to such \emph{protective} instances.

\begin{Theorem} \label{thm:gmb_protective_instances_npc}
$f^{\text{CSR}} / f^{\text{LSR}}$-\textsc{Group Microbribery} where both $A^+ \neq \emptyset$ and $A^- \neq \emptyset$ is NP-complete even if $A^+ \subseteq f^{\text{CSR}} / f^{\text{LSR}}(N, \varphi)$ already holds initially.
\end{Theorem}

\begin{Proof}
Given a CNF-SAT instance $(X, C)$ with $|X|=m$, we construct an instance of $f^{\text{CSR}} / f^{\text{LSR}}$-GMB as follows (the reduction is the same for both $f^{\text{CSR}}$ and $f^{\text{LSR}}$):

For each clause $c \in C$, we introduce an individual $a_c$. For each variable $x \in X$, we introduce a total of six individuals: two individuals $a_x$ and $a_{\bar{x}}$ who represent the literals $x$ and $\bar{x}$, two individuals $q_x^1$ and $q_x^2$ whom we want to make socially qualified, and two individuals $d_x^1$ and $d_x^2$ whom we want to make socially disqualified. Finally, we introduce one more individual $a_\ast$. We then set $A^+ = \{a_\ast\} \cup \{ a_c : c \in C \} \cup \{ q_x^1, q_x^2 : x \in X \}$ and $A^- = \{ d_x^1, d_x^2 : x \in X \}$. We define the profile $\varphi$ as follows:

\begin{itemize}[nolistsep]
\item
All individuals qualify $a_\ast$.
\item
The individual $a_\ast$ qualifies all individuals in $\{ a_x, a_{\bar{x}} : x \in X \}$.
\item
For each variable $x \in X$, the individuals $a_x$ and $a_{\bar{x}}$ each qualify the four individuals $q_x^1$, $q_x^2$, $d_x^1$ and $d_x^2$.
\item
For each variable $x \in X$, and each clause $c \in C$:
    \begin{itemize}[nolistsep]
    \item
    If $x \in c$ then $a_x$ qualifies $a_c$.
    \item
    If $\bar{x} \in c$ then $a_{\bar{x}}$ qualifies $a_c$.
    \end{itemize}
\item
All remaining entries of $\varphi$ not explicitly defined above are set to $-1$.
\end{itemize}

See \figref{fig:gmb_variable_gadget} for an example illustration of one ``variable gadget'' for some variable $x \in X$ created in this reduction. To conclude the reduction, we let $\ell = 3m$, i.e.\@ the budget is three times the number of variables. Clearly, the reduction can be completed in polynomial time.

\begin{figure}[!tbh]
\centering
\raisebox{0.04\height}{
\begin{tikzpicture}[
    > = Stealth,
    shorten > = 1pt,
    auto,
    node distance = 1.8cm,
    main node/.style = {circle,draw,minimum size = 1cm}
]

\variablegadget

\node[main node] (qx1)  [left of = ax1, node distance = 2.5cm] {$q_x^1$};
\node[main node] (qx2)  [below of = qx1] {$q_x^2$};
\node[main node] (dx1)  [right of = ax1, node distance = 2.5cm] {$d_x^1$};
\node[main node] (dx2)  [below of = dx1] {$d_x^2$};

\path[->]
(ax1)   edge                    node {} (qx1)
        edge                    node {} (qx2)
        edge                    node {} (dx1)
        edge                    node {} (dx2)
(ax2)   edge                    node {} (qx1)
        edge                    node {} (qx2)
        edge                    node {} (dx1)
        edge                    node {} (dx2)
;

\draw [very thick, decorate, decoration = {brace, raise=5pt, amplitude=5pt}]
(1.8, 6.2) -- (3.2, 6.2)
node[pos = 0.5, above = 10pt, black] {$A^-$};

\end{tikzpicture}
}
\caption{An illustration of one ``variable gadget'' for some variable $x \in X$ created in the reduction in \thmref{thm:gmb_protective_instances_npc}. Here, we assume that there are three clauses $c_1$, $c_2$, $c_3$, and that $c_1$ contains the literal $x$, $c_3$ contains the literal $\bar{x}$, and $c_2$ contains neither $x$ nor $\bar{x}$.}
\label{fig:gmb_variable_gadget}
\end{figure}

It is easy to see that initially all individuals in the created instance are socially qualified. This is because $a_\ast$ is qualified by everyone, i.e.\@ $a_\ast \in K^{\text{C}}_0(N, \varphi)$ (for the $f^{\text{CSR}}$ rule) and $a_\ast \in K^{\text{L}}_0(N, \varphi)$ (for the $f^{\text{LSR}}$ rule). For each variable $x \in X$, the individual $a_\ast$ then qualifies $a_x$ and $a_{\bar{x}}$ who in turn qualify $q_x^1$, $q_x^2$, $d_x^1$ and $d_x^2$. Also, each individual $a_c$ that represents a clause $c \in C$ is qualified by some variable individual because the clause contains at least one literal. In particular, this means that all individuals in $A^+$ are already socially qualified initially.

Before we show the correctness of the above reduction, it will be useful to observe a few simple facts:

\begin{enumerate}[
    topsep=-4pt,
    leftmargin=*,
    itemindent=4.4em,
    label={\textit{Fact \ref*{thm:gmb_protective_instances_npc}.\arabic*}:},
    ref={\ref*{thm:gmb_protective_instances_npc}.\arabic*}
]
\item \label{fact:gmb_at_least_three}
For each variable $x \in X$, we need to modify at least three entries of $\varphi$ to turn $d_x^1$ and $d_x^2$ socially disqualified while keeping $q_x^1$ and $q_x^2$ socially qualified. Assume for the sake of contradiction that we would turn $d_x^1$ and $d_x^2$ socially disqualified by changing only two entries of $\varphi$. Since $a_\ast$ is in $A^+$ and must therefore remain socially qualified, the only way to do that would be by bribing $a_\ast$ to disqualify $a_x$ and $a_{\bar{x}}$. But in this case, $q_x^1$ and $q_x^2$ would also turn socially disqualified, and bribing someone else to qualify them would require at least two more modifications in $\varphi$, contradicting our assumption.

\item \label{fact:gmb_at_most_three}
For each variable $x \in X$, we can modify at most three entries of $\varphi$. Assume for the sake of contradiction that we would change more than three entries of $\varphi$ in the variable gadget of some $x \in X$. Since we can only make at most $\ell = 3m$ modifications in total, this would imply that we can only make at most two modifications for some other variable $x^\prime \in X$, contradicting \factref{fact:gmb_at_least_three}.

\item \label{fact:gmb_exactly_three}
From Facts \ref{fact:gmb_at_least_three} and \ref{fact:gmb_at_most_three}, it follows that we need to modify exactly three entries of $\varphi$ for each variable $x \in X$. Thus, for each variable $x \in X$, we only have two options to achieve the objective: (1) bribe $a_\ast$ to disqualify $a_x$ and bribe $a_{\bar{x}}$ to disqualify $d_x^1$ and $d_x^2$, or (2) bribe $a_\ast$ to disqualify $a_{\bar{x}}$ and bribe $a_x$ to disqualify $d_x^1$ and $d_x^2$. In both scenarios, we make exactly three modifications in $\varphi$ and we successfully turn $d_x^1$ and $d_x^2$ socially disqualified while keeping $q_x^1$ and $q_x^2$ socially qualified.
\end{enumerate}

Equipped with these facts, we now show that the constructed instance is a YES-instance if and only if there exists a satisfying assignment for $C$.

($\Rightarrow$) Assume that $\varrho : X \rightarrow \{0, 1\}$ is a satisfying assignment for $C$. For each $x \in X$ we do the following:
\begin{itemize}[nolistsep, topsep=-1.25em]
\item
If $\varrho(x) = 1$ then we bribe $a_\ast$ to disqualify $a_{\bar{x}}$ and bribe $a_x$ to disqualify $d_x^1$ and $d_x^2$.
\item
If $\varrho(x) = 0$ then we bribe $a_\ast$ to disqualify $a_x$ and bribe $a_{\bar{x}}$ to disqualify $d_x^1$ and $d_x^2$.
\end{itemize}
Clearly, we are modifying exactly three entries of $\varphi$ per variable, so the budget of $\ell = 3m$ is sufficient. Also, after the bribery, all individuals $d_x^1$ and $d_x^2$ for $x \in X$ are socially disqualified and all individuals $q_x^1$ and $q_x^2$ for $x \in X$ are still socially qualified. Moreover, because $\varrho$ is a satisfying assignment for $C$, each clause $c \in C$ contains at least one literal of the form $x$ with $\varrho(x) = 1$ or one literal of the form $\bar{x}$ with $\varrho(x) = 0$. In the former case, the corresponding individual $a_c$ is qualified by the socially qualified individual $a_x$ and is therefore also socially qualified. In the latter case, the corresponding individual $a_c$ is qualified by the socially qualified individual $a_{\bar{x}}$ and is therefore also socially qualified. Hence, all individuals $a_c$ for $c \in C$ are socially qualified.

($\Leftarrow$) Assume we are given a successful microbribery $M \subseteq N \times N$ consisting of at most $\ell = 3m$ pairs of individuals. As we established in \factref{fact:gmb_exactly_three}, for each variable $x \in X$, there are only two possible ways how the individuals in the corresponding variable gadget may have been bribed:
\begin{itemize}[nolistsep, topsep=-1.25em]
\item
If $a_\ast$ was bribed to disqualify $a_{\bar{x}}$, and $a_x$ was bribed to disqualify $d_x^1$ and $d_x^2$, then we set $\varrho(x) = 1$.
\item
If $a_\ast$ was bribed to disqualify $a_x$, and $a_{\bar{x}}$ was bribed to disqualify $d_x^1$ and $d_x^2$, then we set $\varrho(x) = 0$.
\end{itemize}
We know that all individuals in $\{ a_c : c \in C \} \subseteq A^+$ are still socially qualified after the bribery. Therefore, for each clause $c \in C$, there either is a literal $x \in c$ for which the individual $a_x$ is still socially qualified after the bribery, or there is a literal $\bar{x} \in c$ for which the individual $a_{\bar{x}}$ is still socially qualified after the bribery. In the former case, we have set $\varrho(x) = 1$, so the clause will be satisfied under the truth assignment $\varrho$. In the latter case, we have set $\varrho(x) = 0$, so the clause will also be satisfied under the truth assignment $\varrho$. Hence, $\varrho$ is a satisfying assignment for $C$.
\end{Proof}

Because we can reduce from GMB to \$GMB, the NP-completeness also extends to the priced problem versions:

\begin{Corollary} \label{cor:gmb_protective_instances_npc}
$f^{\text{CSR}} / f^{\text{LSR}}$-\textsc{\$Group Microbribery} where both $A^+ \neq \emptyset$ and $A^- \neq \emptyset$ is NP-complete even if $A^+ \subseteq f^{\text{CSR}} / f^{\text{LSR}}(N, \varphi)$ already holds initially.
\end{Corollary}

\clearpage

\section{Relaxed group control by deleting individuals}
\label{sec:relaxed_group_control_deleting}

In the standard definition of the group control by deleting individuals problems, we do not allow the attacker to delete individuals that are in $A^+$ or $A^-$. For the set $A^+$, this restriction makes intuitive sense because the attacker cannot make a deleted individual socially qualified. However, regarding the individuals in $A^-$, it is conceivable that the attacker would also be satisfied if an individual whom they want to make socially disqualified is simply deleted instead. In this section, we therefore consider a relaxed variant of group control by deleting individuals and study how this relaxation affects the complexity of the problems. For this, we consider the rules $f^{(s, t)}$, $f^{\text{CSR}}$ and $f^{\text{LSR}}$.

The relaxed problem is defined as follows. The input is the same as for the non-relaxed variant, but instead of requiring the solution to be a subset of $N \setminus (A^+ \cup A^-)$, we extend the solution space to $N \setminus A^+$. In other words, we allow the attacker to delete individuals from $A^-$.

\bigskip

\begin{tabularx}{\textwidth}{lX}
\hline
\multicolumn{2}{l}{
$f$-\textsc{Relaxed Group Control by Deleting Individuals} (R-GCDI)
} \\
\hline
\textbf{Given:} &
A 5-tuple $(N, \varphi, A^+, A^-, \ell)$ of a set $N$ of individuals, a profile $\varphi$ over $N$, two subsets $A^+, A^- \subseteq N$ with $A^+ \cap A^- = \emptyset$, and a positive integer $\ell$. \\
\textbf{Question:} &
Is there a subset $U \subseteq N \setminus A^+$ such that $|U| \leq \ell$ and $A^+ \subseteq f(N \setminus U, \varphi)$ and $A^- \cap f(N \setminus U, \varphi) = \emptyset$? \\
\hline
\end{tabularx}

In addition to the general $f$-R-GCDI problem, there are the three special variants
\begin{itemize}
\item
$f$-\textsc{Relaxed Constructive Group Control by Deleting Individuals} \\ (R-CGCDI) where the set $A^-$ is dropped,
\item
$f$-\textsc{Relaxed Destructive Group Control by Deleting Individuals} \\ (R-DGCDI) where the set $A^+$ is dropped, and
\item
$f$-\textsc{Relaxed Exact Group Control by Deleting Individuals} \\ (R-EGCDI) where we add the premise $A^+ \cup A^- = N$.
\end{itemize}

Note that the relaxed constructive problem is exactly the same as its non-relaxed counterpart. This is because the relaxation only affects the destructive target set $A^-$ while the rules regarding the constructive target set $A^+$ remain unchanged. Also note that, unlike the non-relaxed variant, the relaxed exact problem is defined.

\tableref{tab:group_control_deleting} shows the known complexity results for the non-relaxed group control by deleting individuals problems. See \textcite{J22} for a survey of these results. \tableref{tab:relaxed_group_control_deleting} provides an overview of our results for the relaxed problem variants.

\renewcommand{\arraystretch}{1.8}

\begin{table}[!htb]
\begin{tabularx}{\textwidth}{
p{0.092\textwidth}
X
X
X
X
X
X
X
X
X
p{0.108\textwidth}
X
}
\hline & \multicolumn{9}{l}{Consent rules $f^{(s, t)}$} & $f^{\text{CSR}}$ & $f^{\text{LSR}}$ \\
\cline { 2 - 10 } & \multicolumn{3}{l}{$s=1$} & \multicolumn{3}{l}{$s=2$} & \multicolumn{3}{l}{$s \geq 3$} & & \\
\cline { 2 - 10 } & $t=1$ & $t=2$ & $t \geq 3$ & $t=1$ & $t=2$ & $t \geq 3$ & $t=1$ & $t=2$ & $t \geq 3$ & & \\
\hline

\multicolumn{1}{p{0.092\textwidth}|}{CGCDI} &
I &
P &
\multicolumn{1}{X|}{NP-c} &
I &
P &
\multicolumn{1}{X|}{NP-c} &
I &
P &
\multicolumn{1}{X|}{NP-c} &
P &
I \\

\multicolumn{1}{p{0.092\textwidth}|}{DGCDI} &
I &
I &
\multicolumn{1}{X|}{I} &
P &
P &
\multicolumn{1}{X|}{P} &
NP-c &
NP-c &
\multicolumn{1}{X|}{NP-c} &
\textit{P} \ttref{thm:csr_dgcdi_p} &
P \\

\multicolumn{1}{p{0.092\textwidth}|}{EGCDI} &
- &
- &
\multicolumn{1}{X|}{-} &
- &
- &
\multicolumn{1}{X|}{-} &
- &
- &
\multicolumn{1}{X|}{-} &
- &
- \\

\multicolumn{1}{p{0.092\textwidth}|}{GCDI} &
I &
I &
\multicolumn{1}{X|}{I} &
I &
P &
\multicolumn{1}{X|}{NP-c} &
I &
NP-c &
\multicolumn{1}{X|}{NP-c} &
\textit{I} \ooref{obs:csr_immune_gcdi} &
I \\

\hline
\end{tabularx}
\caption{
A summary of the complexity results for group control by deleting individuals.
In the table, ``P'' stands for ``polynomial-time solvable'', ``NP-c'' stands for ``NP-complete'', and ``I'' stands for ``immune''. The symbol ``-'' indicates that the problem is not defined.
The \textit{italicized} entries are new results for which we present the proofs below.
}
\label{tab:group_control_deleting}
\end{table}

\begin{table}[!htb]
\begin{tabularx}{\textwidth}{
p{0.116\textwidth}
p{0.125\textwidth}
X
X
p{0.125\textwidth}
X
X
l
p{0.125\textwidth}
}
\hline & \multicolumn{6}{l}{Consent rules $f^{(s, t)}$} & $f^{\text{CSR}}$ & $f^{\text{LSR}}$ \\
\cline { 2 - 7 } & \multicolumn{3}{l}{$s=1$} & \multicolumn{3}{l}{$s \geq 2$} & & \\
\cline { 2 - 7 } & $t=1$ & $t=2$ & $t \geq 3$ & $t=1$ & $t=2$ & $t \geq 3$ & & \\
\hline

\multicolumn{1}{p{0.116\textwidth}|}{\makecell{R-CGCDI \\ ~}} &
\makecell{I \\ \ooref{obs:f_rcgcdi}} &
\makecell{P \\ \ooref{obs:f_rcgcdi}} &
\multicolumn{1}{X|}{ \makecell{NP-c \\ \ooref{obs:f_rcgcdi}} } &
\makecell{I \\ \ooref{obs:f_rcgcdi}} &
\makecell{P \\ \ooref{obs:f_rcgcdi}} &
\multicolumn{1}{X|}{ \makecell{NP-c \\ \ooref{obs:f_rcgcdi}} } &
\makecell{P \\ \ooref{obs:f_rcgcdi}} &
\makecell{I \\ \ooref{obs:f_rcgcdi}} \\

\multicolumn{1}{p{0.116\textwidth}|}{\makecell{R-DGCDI \\ ~}} &
\makecell{P \\ \ttref{thm:fst_rdgcdi_p}} &
\makecell{P \\ \ttref{thm:fst_rdgcdi_p}} &
\multicolumn{1}{X|}{ \makecell{P \\ \ttref{thm:fst_rdgcdi_p}} } &
\makecell{NP-c \\ \ttref{thm:fst_rdgcdi_npc}} &
\makecell{NP-c \\ \ttref{thm:fst_rdgcdi_npc}} &
\multicolumn{1}{X|}{ \makecell{NP-c \\ \ttref{thm:fst_rdgcdi_npc}} } &
\makecell{P \\ \ccref{cor:csr_rdgcdi_p}} &
\makecell{P \\ \ttref{thm:lsr_rdgcdi_p}} \\

\multicolumn{1}{p{0.116\textwidth}|}{\makecell{R-EGCDI \\ ~}} &
\makecell{I if $A^+ \neq \emptyset$ \\ P else \\ \ooref{obs:f11_regcdi_i_p}} &
\makecell{P \\ \ooref{obs:f12_regcdi_p}} &
\multicolumn{1}{X|}{ \makecell{NP-c \\ \ttref{thm:f1t_regcdi_npc}} } &
\makecell{I if $A^+ \neq \emptyset$ \\ NP-c else \\ \ooref{obs:f21_regcdi_i_npc}} &
\makecell{NP-c \\ \ttref{thm:fst_regcdi_npc}} &
\multicolumn{1}{X|}{ \makecell{NP-c \\ \ttref{thm:fst_regcdi_npc}} } &
\makecell{P \\ \ttref{thm:csr_regcdi_p}} &
\makecell{I if $A^+ \neq \emptyset$ \\ P else \\ \ooref{obs:lsr_regcdi_i_p}} \\

\multicolumn{1}{p{0.116\textwidth}|}{\makecell{R-GCDI \\ ~}} &
\makecell{I \\ \ooref{obs:f11_rgcdi_i}} &
\makecell{P \\ \ooref{obs:f12_rgcdi_p}} &
\multicolumn{1}{X|}{ \makecell{NP-c \\ \ccref{cor:f1t_rgcdi_npc}} } &
\makecell{I \\ \ooref{obs:f21_rgcdi_i}} &
\makecell{NP-c \\ \ccref{cor:fst_rgcdi_npc}} &
\multicolumn{1}{X|}{ \makecell{NP-c \\ \ccref{cor:fst_rgcdi_npc}} } &
\makecell{NP-c \\ \ttref{thm:csr_rgcdi_npc}} &
\makecell{I \\ \ooref{obs:lsr_rgcdi_i}} \\

\hline
\end{tabularx}
\caption{
A summary of the complexity results for relaxed group control by deleting individuals.
In the table, ``P'' stands for ``polynomial-time solvable'', ``NP-c'' stands for ``NP-complete'', and ``I'' stands for ``immune''.
}
\label{tab:relaxed_group_control_deleting}
\end{table}

\renewcommand{\arraystretch}{1.0}

\clearpage

\subsection{Constructive objective}

Because the relaxed constructive problems have the same definition as their non-relaxed counterparts, the complexity of R-CGDCI is identical to that of CGCDI.

\begin{Observation} \label{obs:f_rcgcdi}
For all social rules $f \in \{ f^{(s, t)}, f^{\text{CSR}}, f^{\text{LSR}} \}$, the problem $f$-\textsc{Relaxed Constructive Group Control by Deleting Individuals} has the same complexity as $f$-\textsc{Constructive Group Control by Deleting Individuals}.
\end{Observation}

\subsection{Destructive objective}

The complexity of the relaxed destructive problems does differ from the non-relaxed variants. While all consent rules with $s=1$ are immune to DGCDI, the $f^{(s, t)}$-R-DGCDI problem is polynomial-time solvable when $s=1$ for all $t \geq 1$:

\begin{Theorem} \label{thm:fst_rdgcdi_p}
$f^{(s, t)}$-\textsc{Relaxed Destructive Group Control by Deleting Individuals} can be solved in time $\mathcal{O}(n^2)$ when $s=1$ for all $t \geq 1$.
\end{Theorem}

\begin{Proof}
Given the $f^{(s, t)}$-R-DGCDI instance $(N, \varphi, A^-, \ell)$, we first compute the two sets $A^-_{1} = \{ a \in A^- : \varphi(a, a) = 1 \}$, i.e.\@ the subset of individuals in $A^-$ who qualify themselves, and $A^-_{-1} = \{ a \in A^- : \varphi(a, a) = -1 \}$, i.e.\@ the subset of individuals in $A^-$ who disqualify themselves.

Since $s=1$, all individuals in $A^-_{1}$ will end up socially qualified if we do not interfere. The only way to prevent this is by deleting them. Therefore, if $|A^-_{1}| > \ell$ we immediately conclude that the instance is a NO-instance. Otherwise, we delete the individuals in $A^-_{1}$ and update $\ell := \ell - |A^-_{1}|$.

For each individual $a \in A^-_{-1}$, we initialize $d_a := |\{ a^\prime \in N \setminus A^-_{1} : \varphi(a^\prime, a) = -1 \}|$, i.e.\@ $d_a$ stores the number of individuals in $N \setminus A^-_{1}$ who disqualify $a$. Any individual in $A^-_{-1}$ who is disqualified by less than $t$ individuals will end up socially qualified and therefore needs to be deleted. To do this, we prepare a queue $Q$ consisting of all individuals from $A^-_{-1}$ in an arbitrary order. We then iterate over the individuals in the queue and delete the ones who are disqualified by less than $t$ individuals.

More precisely, for each individual $a \leftarrow \operatorname{pop}(Q)$, we check whether $q_a < t$. If this is the case, we delete $a$ and update $\ell := \ell - 1$. Additionally, we then go over all individuals in $A^-_{-1}$ who were disqualified by $a$, i.e.\@ for each $a^\prime \in A^-_{-1}$ with $\varphi(a, a^\prime) = -1$ we update $q_{a^\prime} := q_{a^\prime} - 1$, and if the new value of $q_{a^\prime}$ is less than $t$, we add $a^\prime$ to $Q$ (unless it is already in the queue) because $a^\prime$ now also needs to be deleted.

If we successfully emptied the queue and still have $\ell \geq 0$, the instance is a YES-instance. This is because all remaining individuals in $A^-$ disqualify themselves and are disqualified by at least $t-1$ other individuals, making them socially disqualified.

It remains to analyze the running time of the algorithm: The sets $A^-_{1}$ and $A^-_{-1}$ can be computed in time $\mathcal{O}(n)$. Deleting the individuals in $A^-_{1}$ can be done in time $\mathcal{O}(n^2)$. Initializing $d_a$ for each $a \in A^-_{-1}$ also takes time $\mathcal{O}(n^2)$. Iterating over the individuals in $Q$ takes time $\mathcal{O}(n)$ because each $a \in A^-_{-1}$ can be added to the queue at most twice before being deleted. When processing the queue, the algorithm spends only $\mathcal{O}(n)$ time per iteration. Overall, the running time is bounded by $\mathcal{O}(n^2)$.
\end{Proof}

For all consent rules with $s \geq 2$, the $f^{(s, t)}$-R-DGCDI problem becomes NP-complete. This can be shown via a reduction from \textsc{Vertex Cover}:

\begin{Theorem} \label{thm:fst_rdgcdi_npc}
$f^{(s, t)}$-\textsc{Relaxed Destructive Group Control by Deleting Individuals} is NP-complete for all $s \geq 2$ and $t \geq 1$.
\end{Theorem}

\begin{Proof}
We first consider the case where $s=2$. Given a \textsc{Vertex Cover} instance $(G, k)$ with $G = (V, E)$, we construct an instance of $f^{(s, t)}$-R-DGCDI as follows:

We introduce one individual $a_v$ for each vertex $v \in V$. We then let each individual qualify themselves, i.e.\@ for all $v \in V$ we set $\varphi(a_v, a_v) = 1$. For each pair of individuals, we let them qualify each other if and only if they are adjacent in $G$, i.e.\@ for all $u, v \in V$ with $u \neq v$ we set
$$
\varphi(a_u, a_v) = \varphi(a_v, a_u) = \left\{\begin{array}{cl}
1 & \text { if }\{u, v\} \in E \\
-1 & \text { else. }
\end{array}\right.
$$

Finally, we set $A^- = N = \{ a_v : v \in V \}$, $s=2$, and $\ell = k$. The value of $t$ can be set arbitrarily (thus, the theorem holds for any $t \geq 1$).

Clearly, the reduction only takes polynomial time. We now show that the constructed instance is a YES-instance if and only if there is a vertex cover for $G$ of size at most $k$.

($\Rightarrow$) Assume that $V^\prime \subseteq V$ is a vertex cover for $G$ with $|V^\prime| \leq k$. We delete all individuals $a_v$ where $v \in V^\prime$. Afterwards, no two remaining individuals qualify one another since $V \setminus V^\prime$ is an independent set. Thus, all remaining individuals are only qualified by themselves and are therefore socially disqualified (recall that $s=2$).

($\Leftarrow$) Assume we are given a set $U \subseteq N$ of size at most $\ell$ such that after deleting the individuals in $U$, all remaining individuals in $A^-$ are socially disqualified. Because $s=2$ and all individuals in $A^-$ qualify themselves, no two remaining individuals from $A^-$ can qualify one another. By construction of $\varphi$, this implies that $V^\prime = \{ v \in V : a_v \in U \}$ is a vertex cover for $G$ of size at most $\ell = k$.

Next, we show how to extend this proof to work for larger values of $s$. For each individual in $\{ a_v : v \in V \}$, we add further $s-2$ dummy individuals to $A^-$ who qualify only themselves and $a_v$. For each $v \in V$, let $D_v = \{ d_v^1, d_v^2, \ldots, d_v^{s-2} \}$ denote these dummy individuals. Furthermore, let $D = \cup_{v \in V} D_v$ denote the set of all dummy individuals. The destructive target set is now defined as $A^- = N = \{ a_v : v \in V \} \cup D$. For all $u, v \in V$ and $i \in \{ 1, \ldots, s-2 \}$, we have $\varphi(d_v^i, d_v^i) = 1$, $\varphi(d_v^i, a_v) = 1$, $\varphi(a_v, d_v^i) = -1$, and for $u \neq v$ $\varphi(d_u^i, d_v^i) = -1$, $\varphi(d_u^i, a_v) = -1$, $\varphi(a_v, d_u^i) = -1$.

We now show that the correspondence between a vertex cover for the graph and a solution for the constructed instance still holds. Specifically, there exists a group control solution of size at most $\ell$ if and only if the graph $G$ has a vertex cover of size at most $k$:

($\Rightarrow$) Assume that $V^\prime \subseteq V$ is a vertex cover for $G$ with $|V^\prime| \leq k$. We delete all individuals $a_v$ where $v \in V^\prime$. Afterwards, for all $u, v \in V \setminus V^\prime$ with $u \neq v$, the two corresponding individuals $a_u$ and $a_v$ do not qualify one another because $V \setminus V^\prime$ is an independent set. Thus, for each remaining individual in $\{ a_v : v \in V \setminus V^\prime \}$, the only individuals who still qualify $a_v$ are $a_v$ themselves and the $s-2$ dummy individuals from $D_v$. Furthermore, for each dummy individual $d \in D$, the only individual who qualifies $d$ is $d$ themselves. Hence, all remaining individuals are qualified by less than $s$ individuals and are therefore socially disqualified.

($\Leftarrow$) Assume we are given a minimal R-DGCDI solution $U \subseteq N$ with $|U| \leq \ell$ such that after deleting the individuals in $U$, all remaining individuals in $A^-$ are socially disqualified. Without loss of generality, we can assume that $U \cap D = \emptyset$, i.e.\@ no dummy individuals are being deleted. This is because deleting a dummy individual $d_v^i$ for some $v \in V$ and $i \in \{ 1, \ldots, s-2 \}$ only has the effect of taking one qualification away from $a_v$. Thus, instead of deleting the dummy individual $d_v^i$, we can always delete an individual $a_u$ with $\{u, v\} \in E$ since this also takes one qualification away from $a_v$. Clearly, such a neighbor $u$ of $v$ must exist because else the individual $a_v$ would already have had at most $s-1$ qualifications and there would have been no need to delete $d_v^i$ in the first place (recall that $U$ is a minimal solution). Thus, after deleting the individuals in $U$, each remaining individual $a_v$ for $v \in V$ is still qualified by themselves and by the $s-2$ dummy individuals in $D_v$. This implies that no two remaining individuals $a_u$ and $a_v$ for $u, v \in V$ qualify one another because otherwise they would have at least $s$ qualifications and would be socially qualified. By construction of $\varphi$, it follows that $V^\prime = \{ v \in V : a_v \in U \}$ is a vertex cover for $G$ of size at most $\ell = k$.
\end{Proof}

Next, we turn to the $f^{\text{LSR}}$ rule. We show that $f^{\text{LSR}}$-R-DGCDI can be solved in polynomial time by reducing the problem to finding a minimum $(v^\ast, w)$-separator in an auxiliary graph. This proof is very similar to the proof for the non-relaxed variant by \textcite[Theorem 7]{ERY20}.

\begin{Theorem} \label{thm:lsr_rdgcdi_p}
$f^{\text{LSR}}$-\textsc{Relaxed Destructive Group Control by Deleting Individuals} can be solved in time $\mathcal{O}(n^{2.5})$.
\end{Theorem}

\begin{Proof}
Let $(N, \varphi, A^-, \ell)$ be an instance of $f^{\text{LSR}}$-R-DGCDI. In a first step, we construct a directed graph from the given instance. That is, for each individual $a \in N$ we create a vertex $v_a$, and there is an edge from $v_a$ to $v_b$ for $a, b \in N$ if and only if $\varphi(a, b) = 1$. Next, we insert two additional vertices $v^\ast$ and $w$ into the graph. Let $V^{\text{L}}$ denote the subset of vertices that represent individuals who qualify themselves, i.e.\@ $V^{\text{L}} = \{ v_a : a \in K^{\text{L}}_0(N, \varphi) \}$. Let $V^-$ denote the subset of vertices that represent the individuals from $A^-$, i.e.\@ $V^- = \{ v_a : a \in A^- \}$. Note that $V^{\text{L}}$ and $V^-$ may overlap. For each $u \in V^{\text{L}}$, we add an edge from $v^\ast$ to $u$, and for each $v \in V^-$, we add an edge from $v$ to $w$. Let $G_{N, \varphi}^{\text{L}} = (V, E)$ denote the resulting auxiliary graph. The only difference between $G_{N, \varphi}^{\text{L}}$ and the auxiliary graph from the proof by \textcite[Theorem 7]{ERY20} is that we do not merge the vertices from $V^-$ into $w$. This is because in the relaxed problem variant, the attacker may also delete the individuals in $A^-$.

Clearly, the socially qualified individuals under the $f^{\text{LSR}}$ rule are exactly the ones represented by vertices in $G_{N, \varphi}^{\text{L}}$ that are reachable from $v^\ast$. We now compute a minimum $(v^\ast, w)$-separator $S \subseteq V$. Since for each $v \in V^-$ there exists an edge $(v, w) \in E$, and no vertices from $V \setminus V^-$ are adjacent to $w$, any $v^\ast$-$w$-path in $G_{N, \varphi}^{\text{L}}$ ends with an edge from a vertex in $V^-$ to the vertex $w$. Therefore, if we delete all individuals who are represented by the vertices in the $(v^\ast, w)$-separator $S$, none of the individuals in $A^-$ is socially qualified. Also, because $S$ is a minimum separator, there is no way to achieve the objective by deleting less than $|S|$ individuals. Hence, if $|S| > \ell$, we conclude that the given instance is a NO-instance. Otherwise, it is a YES-instance.

It remains to analyze the running time of the algorithm: The auxiliary graph $G_{N, \varphi}^{\text{L}}$ can be constructed in time $\mathcal{O}(n^2)$. Computing a minimum separator can be done in time $\mathcal{O}(n^{2.5})$ \cite{E75,ET75}. The running time is therefore bounded by $\mathcal{O}(n^{2.5})$.
\end{Proof}

Finally, we now consider the $f^{\text{CSR}}$ rule. We begin by showing a polynomial-time reduction from $f^{\text{CSR}}$-R-DGCDI to $f^{\text{CSR}}$-DGCDI. This reduction implies that, if the non-relaxed variant can be solved efficiently, then the same is true for the relaxed one.

\begin{Observation} \label{obs:csr_rdgcdi_reduction_dgcdi}
Any instance of $f^{\text{CSR}}$-\textsc{Relaxed Destructive Group Control by Deleting Individuals} can be turned into an equivalent instance of $f^{\text{CSR}}$-\textsc{Destructive Group Control by Deleting Individuals} in time $\mathcal{O}(n^2)$.
\end{Observation}

\begin{Proof}
Given an instance $(N, \varphi, A^-, \ell)$ of $f^{\text{CSR}}$-R-DGCDI, we construct an equivalent instance $(N^\prime, \varphi^\prime, {A^-}^\prime, \ell^\prime)$ of $f^{\text{CSR}}$-DGCDI as follows:

We insert one additional individual $w$, i.e.\@ $N^\prime = N \cup \{w\}$. We then let

\begin{itemize}[nolistsep, topsep=-1.25em]
\item
$\varphi^\prime(a, b) = \varphi(a, b)$ for all $a, b \in N$,
\item
$\varphi^\prime(a, w) = 1$ for all $a \in A^-$,
\item
$\varphi^\prime(a, w) = -1$ for all $a \in N \setminus A^-$,
\item
$\varphi^\prime(w, a) = -1$ for all $a \in N$, and
\item
$\varphi^\prime(w, w) = -1$.
\end{itemize}

In other words, $w$ is qualified by exactly those individuals from the original instance who are in $A^-$, and $w$ disqualifies everyone. Finally, we let ${A^-}^\prime = \{w\}$ and $\ell^\prime = \ell$.

We now show that the created instance $(N^\prime, \varphi^\prime, {A^-}^\prime, \ell^\prime)$ is a YES-instance if and only if the original instance $(N, \varphi, A^-, \ell)$ was a YES-instance:

($\Rightarrow$) Assume that the original instance $(N, \varphi, A^-, \ell)$ was a YES-instance. Then there is a subset $U \subseteq N$ with $|U| \leq \ell$ such that after deleting the individuals in $U$, all remaining individuals in $A^-$ are socially disqualified. Because the individuals in $A^-$ are exactly the ones who qualify $w$ in the created instance, it follows that $U$ is also a solution for the created instance. That is, after deleting the individuals in $U$, the individual $w$ becomes socially disqualified.

($\Leftarrow$) Assume that the created instance $(N^\prime, \varphi^\prime, {A^-}^\prime, \ell^\prime)$ is a YES-instance. Then there is a subset $U^\prime \subseteq N^\prime \setminus {A^-}^\prime$ with $|U^\prime| \leq \ell^\prime$ such that after deleting the individuals in $U^\prime$, all individuals in ${A^-}^\prime$ are socially disqualified. Since ${A^-}^\prime = \{w\}$ and the individual $w$ is qualified by exactly the individuals in $A^-$ from the original instance, it follows that after deleting the individuals in $U^\prime$, none of the individuals in $A^-$ is socially qualified (because otherwise $w$ would also be socially qualified). Thus, $U^\prime$ is also a solution for the original instance.
\end{Proof}

It remains to show a polynomial-time algorithm for the non-relaxed variant. \textcite[Theorem 7]{ERY20} claim that $f^{\text{CSR}}$-DGCDI can be solved in time $\mathcal{O}(n^{3.5})$ by reducing the problem to finding a minimum $(v^\ast, w)$-separator in an auxiliary graph. If this result were correct, we would obtain an equivalent result for the relaxed variant $f^{\text{CSR}}$-R-DGCDI. However, there is, in fact, an error in the proof by \textcite{ERY20}. To demonstrate this, we now present a counterexample for which their algorithm fails:

\clearpage

\begin{Example}
Let $N = \{ a_1, a_2, a_3, a_4, a_5, a_6 \}$ be a set of six individuals and let $\varphi$ be defined as follows ($\varphi(a_i, a_j)$ is given by the matrix entry in the $i$-th row and $j$-th column):

\capstartfalse 
\begin{figure}[!h]
\centering
\begin{minipage}[b]{.4\textwidth}

\renewcommand{\arraystretch}{1.2}
\begin{equation*}
\begin{array}{rrrrrrr}
    & a_1 & a_2 & a_3 & a_4 & a_5 & a_6  \\
a_1 &  1 & -1 & -1 &  1 &  1 & -1 \\
a_2 &  1 &  1 & -1 &  1 &  1 & -1 \\
a_3 &  1 &  1 &  1 &  1 &  1 & -1 \\
a_4 &  1 &  1 &  1 & -1 & -1 &  1 \\
a_5 &  1 &  1 &  1 & -1 & -1 &  1 \\
a_6 &  1 &  1 &  1 & -1 & -1 & -1
\end{array}
\end{equation*}
\caption*{Definition matrix of $\varphi$.}
\renewcommand{\arraystretch}{1.0}

\end{minipage}\hfill
\begin{minipage}[b]{.6\textwidth}
\centering

\raisebox{0.04\height}{
\begin{tikzpicture}[
    > = Stealth,
    shorten > = 1pt,
    auto,
    node distance = 2cm,
    main node/.style = {circle,draw}
]

\node[main node] (1) {$a_1$};
\node[main node] (2) [right of = 1] {$a_2$};
\node[main node] (3) [right of = 2] {$a_3$};
\node[main node] (4) [below of = 1, node distance = 2.6cm, xshift = 0.4cm] {$a_4$};
\node[main node] (5) [below of = 3, node distance = 2.6cm, xshift = -0.4cm] {$a_5$};
\node[main node] (6) [below of = 2, node distance = 3cm] {$a_6$};

\node (l) [above of = 1, node distance = 1.2cm, xshift = -0.8cm] {};
\node (r) [above of = 3, node distance = 1.2cm, xshift = 0.8cm] {};

\path[->]
(1)         edge[loop left]         node {} (1)
(2)         edge[bend left=25]      node {} (1)
            edge[loop left]         node {} (2)
(3)         edge[bend right=30]     node {} (1)
            edge[bend left=25]      node {} (2)
            edge[loop left]         node {} (3)
(4)         edge                    node {} (1)
            edge                    node {} (2)
            edge[bend right=10]     node {} (3)
            edge                    node {} (6)
(5)         edge[bend left=10]      node {} (1)
            edge                    node {} (2)
            edge                    node {} (3)
            edge                    node {} (6)
(6)         edge[bend left=5]       node {} (1)
            edge[bend left=10]      node {} (2)
            edge[bend right=5]      node {} (3)
;

\path[-, shorten > = 0]
(1)         edge[bend right=40]     node {} (l.south)
(2)         edge[bend right=30]     node {} (l.south)
(3)         edge[bend right=40]     node {} (l.south)
;

\path[-, shorten > = 0]
(1)         edge[bend left=40]      node {} (r.south)
(2)         edge[bend left=30]      node {} (r.south)
(3)         edge[bend left=40]      node {} (r.south)
;

\path[->]
(l.south)    edge[bend right=60]     node {} (4)
(r.south)    edge[bend left=60]      node {} (5)
;

\end{tikzpicture}
}
\caption*{Corresponding qualification graph $G_{N, \varphi}$.}

\end{minipage}
\end{figure}
\capstarttrue 

Now let $A^- = \{a_6\}$ and $\ell = 2$. Note that the created $f^{\text{CSR}}$-DGCDI instance $(N, \varphi, A^-, \ell)$ has the following properties:
\begin{itemize}[nolistsep]
\item
Initially, only the individual $a_1$ is qualified by everyone.
\item
If we delete $a_1$, the individual $a_2$ is qualified by all remaining individuals.
\item
If we delete $a_1$ and $a_2$, the individual $a_3$ is qualified by all remaining individuals.
\end{itemize}

It is easy to see that the created instance is a YES-instance. By deleting $a_4$ and $a_5$, we can turn $a_6$ socially disqualified. However, if we follow the steps of the algorithm by \textcite[Theorem 7]{ERY20}, it will conclude that the instance is a NO-instance:

The algorithm begins by constructing an auxiliary graph similar to the one in \thmref{thm:lsr_rdgcdi_p}. For each individual $a \in N$ there is a vertex $v_a$, and there exists an edge from $v_a$ to $v_b$ for $a, b \in N$ if and only if $\varphi(a, b) = 1$. Additionally, there is a vertex $v^\ast$ that has an edge to all vertices that represent individuals who are qualified by everyone. Finally, the vertices that represent the individuals from $A^-$ are merged into a single vertex $w$.

The algorithm now computes a minimum $(v^\ast, w)$-separator $S$. In our example, we would obtain $S = \{v_{a_1}\}$. To make the individuals from $A^-$ socially disqualified, we delete the individuals represented by the vertices in $S$, i.e.\@ we delete $a_1$. Now, we have deleted all individuals who were qualified by everyone, and we therefore need to go through another iteration of the algorithm.

In the second iteration, the individual $a_2$ is qualified by everyone (because $a_1$ has been deleted). Hence, $v^\ast$ now has an edge to $v_{a_2}$; and the new minimum $(v^\ast, w)$-separator is $S = \{v_{a_2}\}$. We therefore delete the individual $a_2$ and continue with another iteration (because we again deleted all individuals who were qualified by everyone).

In the third iteration, the individual $a_3$ is qualified by everyone (because $a_1$ and $a_2$ have been deleted). Thus, $v^\ast$ now has an edge to $v_{a_3}$; and the new minimum $(v^\ast, w)$-separator is $S = \{v_{a_3}\}$. This implies that we have to delete a third individual, namely $a_3$. But this would exceed our budget of $\ell = 2$, so the algorithm now wrongly concludes that the instance is a NO-instance.
\end{Example}

\vspace{1em}

The above example illustrates that the algorithm by \textcite{ERY20} has a flaw. From the start, it would have been optimal to delete $a_4$ and $a_5$. However, because the algorithm only looks for a minimum $(v^\ast, w)$-separator, it never considers this option and always deletes the single individual who is qualified by everyone (which, unfortunately, only results in a new individual now being qualified by everyone).

To fix this problem, we need to modify the algorithm and introduce some additional checks in each iteration. If we do this carefully, we can even maintain the same upper bound for the running time:

\begin{Theorem} \label{thm:csr_dgcdi_p}
$f^{\text{CSR}}$-\textsc{Destructive Group Control by Deleting Individuals} can be solved in time $\mathcal{O}(n^{3.5})$.
\end{Theorem}

\begin{Proof}
Let $(N, \varphi, A^-, \ell)$ be an instance of $f^{\text{CSR}}$-DGCDI. In a first step, we construct the same auxiliary graph as before. That is, for each individual $a \in N$ we create a vertex $v_a$, and there is an edge from $v_a$ to $v_b$ for $a, b \in N$ if and only if $\varphi(a, b) = 1$. We also create one additional vertex $v^\ast$. Let $V^{\text{C}}$ denote the subset of vertices that represent individuals who are qualified by everyone, i.e.\@ $V^{\text{C}} = \{ v_a : a \in K^{\text{C}}_0(N, \varphi) \}$. Let $V^-$ denote the subset of vertices that represent the individuals from $A^-$, i.e.\@ $V^- = \{ v_a : a \in A^- \}$. If $V^{\text{C}} \cap V^- \neq \emptyset$, we immediately conclude that the instance is a NO-instance since we cannot delete individuals in $A^-$. Otherwise, we add an edge from $v^\ast$ to $v$ for each $v \in V^{\text{C}}$, and we merge the vertices in $V^-$ into a single vertex $w$. Let $G_{N, \varphi}^{\text{C}} = (V, E)$ denote the resulting auxiliary graph.

Clearly, the socially qualified individuals under the $f^{\text{CSR}}$ rule are exactly the ones represented by vertices in $G_{N, \varphi}^{\text{C}}$ that are reachable from $v^\ast$. We now compute a minimum $(v^\ast, w)$-separator $S \subseteq V$ and do the following:

If $|S| > \ell$, we conclude that the given instance is a NO-instance. The reason for this is as follows: Since $S$ is a minimum $(v^\ast, w)$-separator, we have to delete at least $|S|$ individuals to make the individuals in $A^-$ (represented by $w$ in the auxiliary graph) socially disqualified. But the fact that $|S| > \ell$ implies that this would exceed our budget.

If $|S| \leq \ell$ and $V^{\text{C}} \setminus S \neq \emptyset$, we conclude that the instance is a YES-instance. The reason for this is as follows: After deleting the individuals in $\{ a : v_a \in S \}$, there still remains at least one individual who is qualified by everyone, but there no longer exist any $v^\ast$-$w$-paths in the auxiliary graph. Hence, all individuals from $A^-$ are socially disqualified.

If $|S| \leq \ell$ and $V^{\text{C}} \setminus S = \emptyset$, we have to do more work. There are now two possible ways how we could achieve the objective: (a) delete all individuals who are represented by the vertices in $S$, after which we would need to go through another iteration of the algorithm because we have deleted all individuals who were qualified by everyone; or (b) compute a minimum $(v^\ast, w)$-separator $S^\prime \subseteq V$ with $V^{\text{C}} \not\subseteq S^\prime$ (i.e.\@ $S^\prime$ does not include all vertices from $V^{\text{C}}$) and delete the individuals represented by the vertices in $S^\prime$. If we choose the latter option, we may have to delete more individuals than in the former, but it has the advantage that we do not need to go through another iteration.

To compute a minimum $(v^\ast, w)$-separator that does not include all vertices from $V^{\text{C}}$, we need at least one vertex from $V^{\text{C}}$ to survive: For each $v \in V^{\text{C}}$, let $G_{N, \varphi}^{\text{C}, v}$ denote the graph obtained from $G_{N, \varphi}^{\text{C}}$ by removing the vertex $v$ and inserting a direct edge from $v^\ast$ to $v^\prime$ for every out-neighbor $v^\prime$ of $v$ with $v^\prime \neq v$. For each vertex $v \in V^{\text{C}}$, we compute the minimum $(v^\ast, w)$-separator in $G_{N, \varphi}^{\text{C}, v}$. Let $S^\prime$ denote the smallest $(v^\ast, w)$-separator found by this approach. Clearly, $S^\prime$ is also a $(v^\ast, w)$-separator in $G_{N, \varphi}^{\text{C}}$ since for any out-neighbor $v^\prime$ of $v$ there is no path from $v^\prime$ to $w$ in $G_{N, \varphi}^{\text{C}, v} - S^\prime$ (and thus also not in $G_{N, \varphi}^{\text{C}} - S^\prime$).

If $|S^\prime| \leq \ell$, we conclude that the instance is a YES-instance. The reason for this is as follows: After deleting the individuals in $\{ a : v_a \in S^\prime \}$, there still remains at least one individual from $K^{\text{C}}_0(N, \varphi)$ who is qualified by everyone (namely the individual represented by the removed vertex $v \in V^{\text{C}}$), but there no longer exist any $v^\ast$-$w$-paths in the auxiliary graph. Hence, all individuals from $A^-$ are socially disqualified.

If $|S^\prime| > \ell$, we need to go through another iteration of the algorithm to determine whether deleting the individuals represented by the vertices in $S$ leads to a solution. Hence, we delete the individuals in $\{ a : v_a \in S \}$, update $\ell := \ell - |S|$, and call the algorithm again with this new instance as input. Any solution for this smaller instance can be combined with deleting the individuals in $\{ a : v_a \in S \}$ to obtain a solution for the given instance. Therefore, the given instance is a YES-instance if and only if this recursive algorithm call returns YES.

It remains to analyze the running time of the algorithm: The auxiliary graph $G_{N, \varphi}^{\text{C}}$ can be constructed in time $\mathcal{O}(n^2)$, and computing a minimum separator can be done in time $\mathcal{O}(n^{2.5})$ \cite{E75,ET75}. After each iteration, either the algorithm terminates, or some individuals are deleted and a new (smaller) instance is created. In the worst case, the algorithm goes through $n$ iterations, requiring a total time of $n \cdot \mathcal{O}(n^{2.5})$. Furthermore, the algorithm spends additional time in each iteration to construct the auxiliary graph $G_{N, \varphi}^{\text{C}, v}$ for each $v \in V^{\text{C}}$ and to compute a minimum $(v^\ast, w)$-separator in $G_{N, \varphi}^{\text{C}, v}$. Each such computation again requires time $\mathcal{O}(n^{2.5})$. However, because the vertices in $V^{\text{C}}$ are deleted after that (or the algorithm terminates right away), this computation can only happen at most once per individual. Thus, over all iterations of the algorithm, the time required by these additional checks is at most $n \cdot \mathcal{O}(n^{2.5})$. The total running time of the algorithm is therefore bounded by $\mathcal{O}(n^{3.5})$.
\end{Proof}

From \thmref{thm:csr_dgcdi_p} and the reduction in \obsref{obs:csr_rdgcdi_reduction_dgcdi}, we immediately get an equivalent result for the relaxed variant:

\begin{Corollary} \label{cor:csr_rdgcdi_p}
$f^{\text{CSR}}$-\textsc{Relaxed Destructive Group Control by Deleting Individuals} can be solved in time $\mathcal{O}(n^{3.5})$.
\end{Corollary}

\clearpage

\subsection{Exact objective}

Next, we consider the problems where the attacker has an exact objective. For all consent rules with $s=1$, we know from \thmref{thm:fst_rdgcdi_p} that the destructive problem is polynomial-time solvable. Roughly speaking, the exact problem therefore inherits the complexity of the constructive problem for these rules, provided that the constructive target set $A^+$ is nonempty. We first show this for the liberal rule $f^{(1, 1)}$.

\begin{Observation} \label{obs:f11_regcdi_i_p}
If $A^+ \neq \emptyset$, then the $f^{(1, 1)}$ rule is immune to \textsc{Relaxed Exact Group Control by Deleting Individuals}. Otherwise, the $f^{(1, 1)}$-\textsc{Relaxed Exact Group Control by Deleting Individuals} problem can be solved in time $\mathcal{O}(n)$.
\end{Observation}

\begin{Proof}
Because the $f^{(1, 1)}$ rule is immune to CGCDI as shown by \textcite[Theorem 1]{YD18}, it is impossible to achieve the objective if $A^+$ is nonempty.

If $A^+$ is empty, we let $U = \{ a \in N : \varphi(a, a) = 1 \}$ denote the set of individuals in $N = A^-$ who qualify themselves. Clearly, $U$ can be computed in time $\mathcal{O}(n)$. After deleting the individuals in $U$ (which is necessary since $s=1$), all remaining individuals are socially disqualified (since $t=1$). Thus, the instance is a YES-instance if and only if $|U| \leq \ell$.
\end{Proof}

For $s=1$ and $t=2$, both $f^{(s, t)}$-R-CGCDI and $f^{(s, t)}$-R-DGCDI can be solved in polynomial time. This leads us to the following algorithm for solving $f^{(s, t)}$-R-EGCDI:

\begin{Observation} \label{obs:f12_regcdi_p}
$f^{(1, 2)}$-\textsc{Relaxed Exact Group Control by Deleting Individuals} can be solved in time $\mathcal{O}(n^2)$.
\end{Observation}

\begin{Proof}
To solve a given instance of $f^{(1, 2)}$-R-EGCDI, we first take care of the individuals in $A^+$. As shown by \textcite[Theorem 3]{YD18}, to do this, we need to iterate over every individual in $A^+$ who disqualifies themselves and ensure that they are not disqualified by anyone else (since $t=2$). Clearly, this takes time $\mathcal{O}(n^2)$.

After taking care of the individuals in $A^+$, we must delete everyone from $A^-$ who qualifies themselves (because $s=1$), and we must also delete everyone from $A^-$ who disqualifies themselves but is disqualified by no one else (because $t=2$). As shown in the proof of \thmref{thm:fst_rdgcdi_p}, this can be done in time $\mathcal{O}(n^2)$ by using a queue. If the budget suffices for these deletions, the instance is a YES-instance. Else, it is a NO-instance.
\end{Proof}

For $s=1$ and all $t \geq 3$, the $f^{(s, t)}$-R-EGCDI problem is NP-complete. This can be shown via a reduction from \textsc{Restricted Exact Cover by 3-sets}, similar to the one for the constructive case by \textcite[Theorem 4]{YD18}. From the reduction below, it follows that the NP-completeness also applies when the attacker has an exact objective and both $A^+$ and $A^-$ are nonempty.

\begin{Theorem} \label{thm:f1t_regcdi_npc}
$f^{(s, t)}$-\textsc{Relaxed Exact Group Control by Deleting Individuals} is NP-complete when $s=1$ for all $t \geq 3$.
\end{Theorem}

\begin{Proof}
Given a RX3C instance $(X, \mathcal{F})$ with $|X|=3m$ (and thus $|\mathcal{F}|=3m$), we construct an equivalent instance of R-EGCDI as follows:

For each element $x \in X$, we introduce one individual $a_x$. Let $N_X = \{ a_x : x \in X \}$. For each triplet $F \in \mathcal{F}$, we introduce one individual $a_F$. Let $N_\mathcal{F} = \{ a_F : F \in \mathcal{F} \}$. If $t > 3$, we also introduce further $t-3$ dummy individuals $\{ d_1, \ldots, d_{t-3} \} = N_D$. Finally, we introduce one more individual $d$. We set $N = N_X \cup N_\mathcal{F} \cup N_D \cup \{d\}$ and define the profile $\varphi$ over $N$ as follows:

\begin{itemize}[nolistsep]
\item
For each $x \in X$, the individual $a_x$ qualifies everyone except themselves and the individuals in $N_\mathcal{F}$:
    \begin{itemize}[nolistsep]
    \item
    $\varphi(a_x, a_x) = -1$.
    \item
    $\varphi(a_x, a_F) = -1$ for all $F \in \mathcal{F}$.
    \item
    $\varphi(a_x, b) = 1$ for all $b \in N \setminus (\{a_x\} \cup N_\mathcal{F})$.
    \end{itemize}
\item
For each $F \in \mathcal{F}$, the individual $a_F$ qualifies all individuals in $\{ a_x \in N_X : x \not\in F \}$ and disqualifies all other individuals (including themselves):
    \begin{itemize}[nolistsep]
    \item
    $\varphi(a_F, a_x) = 1$ for all individuals in $\{ a_x \in N_X : x \not\in F \}$.
    \item
    $\varphi(a_F, b) = -1$ for all $b \in N \setminus \{ a_x \in N_X : x \not\in F \}$.
    \end{itemize}
\item
For each $i \in \{ 1, \ldots, t-3 \}$, the individual $d_i$ qualifies only themselves:
    \begin{itemize}[nolistsep]
    \item
    $\varphi(d_i, d_i) = 1$.
    \item
    $\varphi(d_i, b) = -1$ for all $b \in N \setminus \{d_i\}$.
    \end{itemize}
\item
The individual $d$ qualifies only themselves and the individuals in $N_X$:
    \begin{itemize}[nolistsep]
    \item
    $\varphi(d, d) = 1$.
    \item
    $\varphi(d, a_x) = 1$ for each $x \in X$.
    \item
    $\varphi(d, b) = -1$ for all $b \in N \setminus (\{d\} \cup N_X)$.
    \end{itemize}
\end{itemize}

To conclude the reduction, we set $A^+ = N_X \cup N_D$, $A^- = N_\mathcal{F} \cup \{d\}$, and $\ell = 2m+1$. Before we show the correctness of the above reduction, it will be useful to observe a few simple facts:

\begin{enumerate}[
    topsep=-4pt,
    leftmargin=*,
    itemindent=4.4em,
    label={\textit{Fact \ref*{thm:f1t_regcdi_npc}.\arabic*}:},
    ref={\ref*{thm:f1t_regcdi_npc}.\arabic*}
]
\item \label{fact:regcdi_nx}
Each individual $a_x \in N_X$ is disqualified by themselves, by the $t-3$ dummy individuals in $N_D$, and by the three individuals $a_F \in N_\mathcal{F}$ with $x \in F$. Since $a_x$ and the dummy individuals are in $A^+$, we are not allowed to delete them. Therefore, to turn $a_x$ socially qualified, we have to make sure that at most one other individual (besides $a_x$ and the dummy individuals) disqualifies $a_x$. The only way to do that is by deleting two of the three individuals from $N_\mathcal{F}$ who disqualify $a_x$.

\item \label{fact:regcdi_d}
Since $s=1$, the individual $d$ is initially socially qualified because they qualify themselves (this is to fulfill the requirement that at least one individual in $A^-$ is initially socially qualified, i.e.\@ $A^- \cap f^{(s, t)}(N, \varphi) \neq \emptyset$). Clearly, the only way to achieve the destructive objective is by deleting $d$ which decreases the budget by $1$. Hence, for achieving the constructive objective, only at most $\ell - 1 = 2m$ deletions are allowed.

\item \label{fact:regcdi_nd_nf}
All dummy individuals from $N_D$ are already socially qualified initially because $s=1$ and they qualify themselves. All individuals from $N_\mathcal{F}$ are already socially disqualified initially because everyone disqualifies them.
\end{enumerate}

Equipped with these facts, we now show that the constructed instance is a YES-instance if and only if there exists an exact 3-set cover for $X$ in $\mathcal{F}$.

($\Rightarrow$) Assume that $\mathcal{F}^\prime \subseteq \mathcal{F}$ is an exact 3-set cover for $X$. Let $U = \{ a_F : F \in \mathcal{F} \setminus \mathcal{F}^\prime \}$ denote the subset of individuals who represent the sets which are not part of the exact 3-set cover. Clearly, $|U| = 2m$. We now delete the individuals in $U$. After the deletion, for each individual $a_x \in N_X$, there remains exactly one individual $a_F \in N_\mathcal{F} \setminus U$ who disqualifies $a_x$. Hence, each individual $a_x \in N_X$ is now disqualified by $t-1$ individuals and is therefore socially qualified. As a last step, we also delete the individual $d$ as discussed in \factref{fact:regcdi_d}.

($\Leftarrow$) Assume we are given a set $U \subseteq N \setminus A^+$ of size at most $\ell = 2m+1$ such that after deleting the individuals in $U$, all individuals in $A^+$ are socially qualified and all remaining individuals in $A^-$ are socially disqualified. From \factref{fact:regcdi_d}, it follows that $d \in U$. Let $U^\prime = U \setminus \{d\}$. From \factref{fact:regcdi_nx}, we know that all individuals in $U^\prime$ are from $N_\mathcal{F}$ and that each individual $a_x \in N_X$ is disqualified by at most one individual in $N_\mathcal{F} \setminus U^\prime$. Let $\tilde{U} = N_\mathcal{F} \setminus U^\prime$. By construction of $\varphi$, each individual $a_F \in \tilde{U}$ disqualifies exactly three individuals in $N_X$. Hence, from $|N_X| = 3m$, we obtain that $|\tilde{U}| \leq m$. Furthermore, from $|U^\prime| \leq 2m$, it follows that $|\tilde{U}| = 3m - |U^\prime| \geq m$. Thus, it must be that $|\tilde{U}| = m$. This implies that each individual $a_x \in N_X$ is disqualified by exactly one individual $a_F \in \tilde{U}$. It follows that $\mathcal{F}^\prime = \{ F \in \mathcal{F} : a_F \in \tilde{U} \}$ is an exact 3-set cover for $X$.
\end{Proof}

We now turn to the consent rules with $s \geq 2$. We begin with an observation for the case $s \geq 2$ and $t=1$:

\begin{Observation} \label{obs:f21_regcdi_i_npc}
If $A^+ \neq \emptyset$, then the $f^{(s, t)}$ rule is immune to \textsc{Relaxed Exact Group Control by Deleting Individuals} for all $s \geq 2$ and $t=1$. Otherwise, $f^{(s, t)}$-\textsc{Relaxed Exact Group Control by Deleting Individuals} is NP-complete for all $s \geq 2$ and $t=1$.
\end{Observation}

\begin{Proof}
Because all $f^{(s, t)}$ rules with $t=1$ are immune to CGCDI as shown by \textcite[Theorem 2]{YD18}, it is impossible to achieve the objective if $A^+$ is nonempty.

If $A^+$ is empty, we have $A^- = N$. In other words, the objective is to make all individuals socially disqualified. As we have shown in \thmref{thm:fst_rdgcdi_npc}, this special case of $f^{(s, t)}$-R-DGCDI is NP-complete for all $s \geq 2$ and $t \geq 1$.
\end{Proof}

For the consent rules with $s \geq 2$ and $t \geq 2$, $f^{(s, t)}$-R-EGCDI is NP-complete. When $A^+$ is empty, this again follows from the reduction in \thmref{thm:fst_rdgcdi_npc} where we show that $f^{(s, t)}$-R-DGCDI with $A^- = N$ is NP-complete for all $s \geq 2$ and $t \geq 1$. Below, we show how to extend this result to the case where both $A^+$ and $A^-$ are nonempty.

\begin{Theorem} \label{thm:fst_regcdi_npc}
For all $s \geq 2$ and $t \geq 2$, the $f^{(s, t)}$-\textsc{Relaxed Exact Group Control by Deleting Individuals} problem with $A^+ \neq \emptyset$ and $A^- \neq \emptyset$ is NP-complete.
\end{Theorem}

\begin{Proof}
We first apply the same reduction from \textsc{Vertex Cover} to $f^{(s, t)}$-R-DGCDI as in \thmref{thm:fst_rdgcdi_npc}. To turn the resulting instance of R-DGCDI into an instance of R-EGCDI with $A^+ \neq \emptyset$, we insert one additional individual $q$ and $t-1$ additional dummy individuals $\{ d_q^1, \ldots, d_q^{t-1} \}$. Both $q$ and the dummy individuals disqualify everyone, including themselves. All individuals from the original instance qualify $q$ and disqualify all the dummy individuals. Finally, we add the dummy individuals to the set $A^-$, define $A^+ = \{q\}$, and increase the budget $\ell$ by $1$.

Note that all individuals from the original instance qualify themselves, so their social qualification status is unaffected by the individuals in $\{q\} \cup \{ d_q^1, \ldots, d_q^{t-1} \}$.

Since $q$ disqualifies themselves and is disqualified by the $t-1$ dummy individuals from $\{ d_q^1, \ldots, d_q^{t-1} \}$, the attacker is always forced to delete one of the dummy individuals to turn $q$ socially qualified. This decreases the budget by $1$, so solving the remaining instance is again equivalent to solving the destructive case.
\end{Proof}

We now consider the consensus-start-respecting rule. We show that $f^{\text{CSR}}$-R-EGCDI is polynomial-time solvable. Interestingly, we get a worse bound for the running time when the constructive target set $A^+$ is nonempty:

\begin{Theorem} \label{thm:csr_regcdi_p}
When $A^+ = \emptyset$, $f^{\text{CSR}}$-\textsc{Relaxed Exact Group Control by Deleting Individuals} can be solved in time $\mathcal{O}(n^2)$. Otherwise, it can be solved in time $\mathcal{O}(n^3)$.
\end{Theorem}

\begin{Proof}
We first consider the case that $A^+ = \emptyset$. In this case, we have $A^- = N$, i.e.\@ the objective is to make all individuals socially disqualified. To do that, we obviously need to ensure that no individual is qualified by everyone.

For each $a \in N$, we initialize $d_a := |\{ a^\prime \in N : \varphi(a^\prime, a) = -1 \}|$, i.e.\@ $d_a$ stores the number of individuals in $N$ who disqualify $a$. We then prepare a queue $Q$ consisting of all individuals in an arbitrary order. To solve the instance, we iterate over the queue and delete all individuals who are qualified by everyone. This is necessary because the only way to get rid of an individual who is qualified by everyone is to delete that individual.

More precisely, for each individual $a \leftarrow \operatorname{pop}(Q)$, we check whether $d_a = 0$. If this is the case, we delete $a$ and update $\ell := \ell - 1$. Additionally, we then go over all individuals who were disqualified by $a$, i.e.\@ for each $a^\prime \in N$ with $\varphi(a, a^\prime) = -1$ we update $d_{a^\prime} := d_{a^\prime} - 1$, and if the new value of $d_{a^\prime}$ is $0$, we add $a^\prime$ to $Q$ (unless it is already in the queue) because $a^\prime$ now also needs to be deleted.

If we successfully emptied the queue and still have $\ell \geq 0$, the instance is a YES-instance. This is because none of the remaining individuals is qualified by everyone. Thus, all the remaining individuals are socially disqualified.

It remains to analyze the running time of the algorithm: Initializing $d_a$ for each $a \in N$ takes time $\mathcal{O}(n^2)$. Iterating over the individuals in $Q$ takes time $\mathcal{O}(n)$ because each $a \in N$ can be added to the queue at most twice before being deleted. When processing the queue, the algorithm spends only $\mathcal{O}(n)$ time per iteration. Overall, the running time is bounded by $\mathcal{O}(n^2)$.

\vspace{1em}

If $A^+ \neq \emptyset$, we need to use a different approach to solve the instance. We begin by deleting everyone from $A^-$ who is qualified by someone in $A^+$. This is always necessary because, if a solution exists, all individuals in $A^+$ will end up socially qualified.

More formally, we iterate over each $a \in A^-$ and check whether there is an $a^\prime \in A^+$ with $\varphi(a^\prime, a) = 1$. Whenever this is the case, we delete the individual $a$ and update $\ell := \ell - 1$.

If we have $\ell < 0$ after this operation, we can conclude that the instance is a NO-instance. Otherwise, all remaining individuals in $A^-$ are now socially disqualified. This is because each remaining individual in $A^-$ is at least disqualified by everyone from $A^+$. Thus, none of the remaining individuals in $A^-$ is in the initial set of socially qualified individuals. Moreover, the individuals in $A^-$ are isolated from the individuals in $A^+$, so they cannot become socially qualified in a later iteration either.

Now we take care of the individuals in $A^+$. First, we make the following observation: If there exists a solution (i.e.\@ if it is possible to make everyone in $A^+$ socially qualified by deleting individuals), then there is an $a \in A^+$ such that it suffices to delete all individuals who disqualify $a$ to achieve the objective. Below, we present a proof of this observation:

Assume for the sake of contradiction that we need to make two individuals $a, b \in A^+$ qualified by everyone to achieve the objective. Let $U_a = \{ a^\prime \in N : \varphi(a^\prime, a) = -1 \}$ denote the subset of individuals we have to delete to make $a$ qualified by all remaining individuals, and let $U_b = \{ a^\prime \in N : \varphi(a^\prime, b) = -1 \}$ denote the subset of individuals we have to delete to make $b$ qualified by all remaining individuals. Clearly, after deleting the individuals in $U_a \cup U_b$, both $a$ and $b$ are qualified by everyone and, thus, are in the initial set of socially qualified individuals. However, this implies that $a$ and $b$ also qualify each other, i.e.\ $\varphi(a, b) = \varphi(b, a) = 1$. Therefore, instead of deleting all individuals in $U_a \cup U_b$, it suffices to only delete the individuals in $U_a$. This way, $a$ is in the initial set of socially qualified individuals, and $b$ is socially qualified after the first iterative step (at the latest). Hence, the final set of socially qualified individuals stays the same. Obviously, it holds that $|U_a| \leq |U_a \cup U_b|$, so deleting $U_a$ is not more expensive than deleting $U_a \cup U_b$.

Equipped with the above observation, we can solve the instance as follows: We iterate over each $a \in A^+$ and compute the set $U_a$ of remaining individuals who disqualify $a$. If $A^+ \cap U_a \neq \emptyset$ or $|U_a| > \ell$, we immediately continue with the next $a \in A^+$. Otherwise, we check if deleting the individuals in $U_a$ makes all individuals from $A^+$ socially qualified, i.e.\@ $A^+ \subseteq f^{\text{CSR}}(N \setminus U_a, \varphi)$. If we find an individual $a \in A^+$ where this is the case, the instance is a YES-instance. Otherwise, it is a NO-instance.

It remains to analyze the running time of the algorithm: Deleting all individuals from $A^-$ who are qualified by someone in $A^+$ takes time $\mathcal{O}(n^2)$. Iterating over the individuals in $A^+$ and performing the necessary checks takes time $\mathcal{O}(n \cdot n^2)$. Overall, the running time is bounded by $\mathcal{O}(n^3)$.
\end{Proof}

Finally, we consider the liberal-start-respecting rule. The complexity of $f^{\text{LSR}}$-R-EGCDI again depends on whether the set $A^+$ is empty or not:

\begin{Observation} \label{obs:lsr_regcdi_i_p}
If $A^+ \neq \emptyset$, then the $f^{\text{LSR}}$ rule is immune to \textsc{Relaxed Exact Group Control by Deleting Individuals}. Otherwise, the $f^{\text{LSR}}$-\textsc{Relaxed Exact Group Control by Deleting Individuals} problem is solvable in time $\mathcal{O}(n)$.
\end{Observation}

\begin{Proof}
Because the $f^{\text{LSR}}$ rule is immune to CGCDI as shown by \textcite[Theorem 7]{YD18}, it is impossible to achieve the objective if $A^+$ is nonempty.

If $A^+$ is empty, we let $U = \{ a \in N : \varphi(a, a) = 1 \}$ denote the set of individuals in $N = A^-$ who qualify themselves. Clearly, the set $U$ can be computed in time $\mathcal{O}(n)$. To achieve the destructive objective, we have to delete all individuals in $U$. Afterwards, none of the remaining individuals is socially qualified since $K^{\text{L}}_0(N \setminus U, \varphi) = \emptyset$. Thus, the instance is a YES-instance if and only if $|U| \leq \ell$.
\end{Proof}

\clearpage

\subsection{General objective}

To conclude this section, we now consider the general problems. Here, the attacker wants to simultaneously make everyone from $A^+$ socially qualified and everyone from $A^-$ socially disqualified. Note that when $A^- = \emptyset$ [resp.\@ $A^+ = \emptyset$], the problems are equivalent to the constructive [resp.\@ destructive] cases. Below, we therefore only consider instances where both $A^+$ and $A^-$ are nonempty.

We begin with an observation for the liberal rule $f^{(1, 1)}$.

\begin{Observation} \label{obs:f11_rgcdi_i}
The $f^{(1, 1)}$ rule is immune to \textsc{Relaxed Group Control by Deleting Individuals}, provided that $A^+ \neq \emptyset$.
\end{Observation}

\begin{Proof}
Because the $f^{(1, 1)}$ rule is immune to CGCDI as shown by \textcite[Theorem 1]{YD18}, it is impossible to achieve the objective if $A^+$ is nonempty.
\end{Proof}

By using the same approach as in \obsref{obs:f12_regcdi_p}, $f^{(1, 2)}$-R-GCDI can be solved in polynomial time:

\begin{Observation} \label{obs:f12_rgcdi_p}
$f^{(1, 2)}$-\textsc{Relaxed Group Control by Deleting Individuals} can be solved in time $\mathcal{O}(n^2)$.
\end{Observation}

\begin{Proof}
As outlined in the proof of \obsref{obs:f12_regcdi_p}, to solve an instance of $f^{(1, 2)}$-R-GCDI, we first take care of the individuals in $A^+$. This can be done in time $\mathcal{O}(n^2)$. After that, we delete everyone from $A^-$ who is not yet socially disqualified. This can also be done in time $\mathcal{O}(n^2)$.
\end{Proof}

When $s=1$ and $t \geq 3$, the $f^{(s, t)}$-R-GCDI problem is NP-complete. This follows directly from the NP-completeness of $f^{(s, t)}$-R-EGCDI shown in \thmref{thm:f1t_regcdi_npc}. Note that the instances created in the proof of \thmref{thm:f1t_regcdi_npc} fulfill the requirement that both $A^+$ and $A^-$ are nonempty.

\begin{Corollary} \label{cor:f1t_rgcdi_npc}
For $s=1$ and all $t \geq 3$, the problem $f^{(s, t)}$-\textsc{Relaxed Group Control by Deleting Individuals} with $A^+ \neq \emptyset$ and $A^- \neq \emptyset$ is NP-complete.
\end{Corollary}

We now turn to the consent rules with $s \geq 2$. When $s \geq 2$ and $t=1$, the $f^{(s, t)}$ rule is immune to R-GCDI:

\begin{Observation} \label{obs:f21_rgcdi_i}
For all $s \geq 2$ and $t=1$, the $f^{(s, t)}$ rule is immune to \textsc{Relaxed Group Control by Deleting Individuals}, provided that $A^+ \neq \emptyset$.
\end{Observation}

\begin{Proof}
Because all $f^{(s, t)}$ rules with $t=1$ are immune to CGCDI as shown by \textcite[Theorem 2]{YD18}, it is impossible to achieve the objective if $A^+$ is nonempty.
\end{Proof}

For the consent rules with $s \geq 2$ and $t \geq 2$, $f^{(s, t)}$-R-GCDI is NP-complete. This follows directly from the NP-completeness of $f^{(s, t)}$-R-EGCDI shown in \thmref{thm:fst_regcdi_npc}. Note that the instances created in the proof of \thmref{thm:fst_regcdi_npc} fulfill the requirement that both $A^+$ and $A^-$ are nonempty.

\begin{Corollary} \label{cor:fst_rgcdi_npc}
For all $s \geq 2$ and $t \geq 2$, the problem $f^{(s, t)}$-\textsc{Relaxed Group Control by Deleting Individuals} with $A^+ \neq \emptyset$ and $A^- \neq \emptyset$ is NP-complete.
\end{Corollary}

We now consider the $f^{\text{CSR}}$ rule. We begin with an observation for the non-relaxed variant of GCDI. The complexity of this problem was left open in the survey by \textcite{J22}. On the class of GCDI instances where at least one individual from $A^+$ is socially disqualified initially, and at least one individual from $A^-$ is socially qualified initially, the $f^{\text{CSR}}$ rule is immune to GCDI:

\begin{Observation} \label{obs:csr_immune_gcdi}
$f^{\text{CSR}}$ is immune to \textsc{Group Control by Deleting Individuals}, provided that $A^+ \neq \emptyset$ and $A^- \neq \emptyset$ and that $A^+ \not\subseteq f^{\text{CSR}}(N, \varphi)$ and $A^- \cap f^{\text{CSR}}(N, \varphi) \neq \emptyset$ holds initially.
\end{Observation}

\begin{Proof}
Assume we are given an instance of $f^{\text{CSR}}$-GCDI where $A^+ \neq \emptyset$ and $A^- \neq \emptyset$. We fix any individual $a^+ \in A^+$ who is initially socially disqualified, i.e.\@ $a^+ \not\in f^{\text{CSR}}(N, \varphi)$. We also fix any individual $a^- \in A^-$ who is initially socially qualified, i.e.\@ $a^- \in f^{\text{CSR}}(N, \varphi)$. 

To turn $a^+$ socially qualified, we have to ensure that some individual $a_\ast \in N$ is qualified by everyone (because else no one is socially qualified). Assume for the sake of contradiction that there is an individual $a_\ast \in N$ such that $a^+$ becomes socially qualified after we delete all individuals who disqualify $a_\ast$. We distinguish between two cases:

\begin{enumerate}[
    leftmargin=*,
    label={\textit{Case \ref*{obs:csr_immune_gcdi}.\arabic*}:},
    nolistsep
]
\item
The individual $a_\ast$ is disqualified by $a^-$. \\
In this case, it is impossible to make $a_\ast$ qualified by everyone because we are not allowed to delete $a^-$.
\item
The individual $a_\ast$ is qualified by $a^-$. \\
In this case, $a_\ast$ was already socially qualified (since $a^-$ is socially qualified). Therefore, making $a_\ast$ qualified by everyone does not change the qualification status of $a^+$.
\end{enumerate}

Both of the above cases result in a contradiction. Hence, it is impossible to make $a^+$ socially qualified by deleting individuals.
\end{Proof}

In the relaxed variant, there are fewer restrictions on the allowed operations. Therefore, the $f^{\text{CSR}}$ rule is susceptible to R-GCDI. However, finding a solution is NP-hard. This can be shown via a reduction from CNF-SAT, similar to the one in \thmref{thm:gmb_protective_instances_npc}:

\begin{Theorem} \label{thm:csr_rgcdi_npc}
$f^{\text{CSR}}$-\textsc{Relaxed Group Control by Deleting Individuals} is NP-complete, provided that $A^+ \neq \emptyset$ and $A^- \neq \emptyset$.
\end{Theorem}

\begin{Proof}
Given a CNF-SAT instance $(X, C)$ with $|X|=m$, we construct an instance of $f^{\text{CSR}}$-R-GCDI as follows:

For each clause $c \in C$, we introduce an individual $a_c$. For each variable $x \in X$, we introduce a total of seven individuals: two individuals $a_x$ and $a_{\bar{x}}$ who represent the literals $x$ and $\bar{x}$, one individual $q_x$ whom we want to make socially qualified, and four individuals $d_x^1$, $d_x^2$, $d_x^3$ and $d_x^4$ whom we want to make socially disqualified. Finally, we introduce two more individuals $a_\ast$ and $a_{\ast\ast}$. We then set $A^+ = \{a_\ast\} \cup \{ a_c : c \in C \} \cup \{ q_x : x \in X \}$ and $A^- =  \{a_{\ast\ast}\} \cup \{ d_x^1, d_x^2, d_x^3, d_x^4 : x \in X \}$. We define the profile $\varphi$ as follows:

\begin{itemize}[nolistsep]
\item
All individuals qualify $a_{\ast\ast}$.
\item
The individual $a_{\ast\ast}$ disqualifies $a_\ast$; all other individuals (including $a_\ast$) qualify $a_\ast$.
\item
The individual $a_\ast$ qualifies all individuals in $\{ a_x, a_{\bar{x}} : x \in X \}$.
\item
For each variable $x \in X$, the individual $a_x$ qualifies $q_x$, $d_x^1$ and $d_x^2$.
\item
For each variable $x \in X$, the individual $a_{\bar{x}}$ qualifies $q_x$, $d_x^3$ and $d_x^4$.
\item
For each variable $x \in X$, and each clause $c \in C$:
    \begin{itemize}[nolistsep]
    \item
    If $x \in c$ then $a_x$ qualifies $a_c$.
    \item
    If $\bar{x} \in c$ then $a_{\bar{x}}$ qualifies $a_c$.
    \end{itemize}
\item
All remaining entries of $\varphi$ not explicitly defined above are set to $-1$.
\end{itemize}

See \figref{fig:rgcdi_variable_gadget} for an example illustration of one ``variable gadget'' for some variable $x \in X$ created in this reduction. To conclude the reduction, we let $\ell = 3m+1$. Clearly, the reduction can be completed in polynomial time.

\begin{figure}[!tbh]
\centering
\raisebox{0.04\height}{
\begin{tikzpicture}[
    > = Stealth,
    shorten > = 1pt,
    auto,
    node distance = 1.8cm,
    main node/.style = {circle,draw,minimum size = 1cm}
]

\variablegadget

\variablegadgetextras

\node[main node] (aaa)  [right of = aa, node distance = 5cm, yshift = -1.5cm] {$a_{\ast\ast}$};

\node (ii1) [above left of = aaa] {};
\node (ii2) [above right of = aaa] {};
\node (ii3) [left of = ii1, node distance = 1cm] {};
\node (ii4) [right of = ii2, node distance = 1cm] {};
\node (ii5) [above of = aaa, node distance = 1.3cm] {};

\path[->]
(aaa)   edge[loop below]        node {} (aaa)
(ii1)   edge[bend right=10]     node {} (aaa)
(ii2)   edge[bend left=10]      node {} (aaa)
(ii3)   edge[bend right=10]     node {} (aaa)
(ii4)   edge[bend left=10]      node {} (aaa)
(ii5)   edge                    node {} (aaa)
;

\draw [very thick, decorate, decoration = {brace, raise=5pt, amplitude=5pt}]
(1.8, 7.6) -- (3.2, 7.6)
node[pos = 0.5, above = 10pt, black] {$A^-$};

\end{tikzpicture}
}
\caption{An illustration of one ``variable gadget'' for some variable $x \in X$ created in the reduction in \thmref{thm:csr_rgcdi_npc}. Here, we assume that there are three clauses $c_1$, $c_2$, $c_3$, and that $c_1$ contains the literal $x$, $c_3$ contains the literal $\bar{x}$, and $c_2$ contains neither $x$ nor $\bar{x}$.}
\label{fig:rgcdi_variable_gadget}
\end{figure}

Before we show the correctness of the above reduction, it will be useful to observe a few simple facts:

\begin{enumerate}[
    topsep=-4pt,
    leftmargin=*,
    itemindent=4.4em,
    label={\textit{Fact \ref*{thm:csr_rgcdi_npc}.\arabic*}:},
    ref={\ref*{thm:csr_rgcdi_npc}.\arabic*}
]
\item \label{fact:rgcdi_aaa}
Initially, $a_{\ast\ast}$ is the only individual who is socially qualified because everyone qualifies $a_{\ast\ast}$, but $a_{\ast\ast}$ qualifies only themselves. This is to ensure that $A^+ \not\subseteq f^{\text{CSR}}(N, \varphi)$ and $A^- \cap f^{\text{CSR}}(N, \varphi) \neq \emptyset$ holds initially. Since $a_{\ast\ast} \in A^-$, the attacker is always forced to delete $a_{\ast\ast}$ which decreases the budget by $1$. Hence, for the remaining work, only at most $\ell - 1 = 3m$ deletions are allowed.

\item \label{fact:rgcdi_deleting_aaa}
After the deletion of $a_{\ast\ast}$, all remaining individuals are socially qualified. This is because $a_\ast$ is now qualified by everyone, i.e.\@ $a_\ast \in K^{\text{C}}_0(N \setminus \{a_{\ast\ast}\}, \varphi)$. For each variable $x \in X$, the individual $a_\ast$ then qualifies $a_x$ and $a_{\bar{x}}$ who in turn qualify $q_x$, $d_x^1$, $d_x^2$, $d_x^3$ and $d_x^4$. Also, each individual $a_c$ that represents a clause $c \in C$ is qualified by some variable individual because the clause contains at least one literal.

\item \label{fact:rgcdi_at_least_three}
For each variable $x \in X$, we need to delete at least three individuals to ensure that $q_x$ stays socially qualified while none of the individuals $d_x^1$, $d_x^2$, $d_x^3$ and $d_x^4$ is socially qualified. Assume for the sake of contradiction that we would achieve this by deleting only two individuals. Since the individual $a_\ast$ is in $A^+$, we are not allowed to delete them. Therefore, the only way to make $d_x^1$, $d_x^2$, $d_x^3$ and $d_x^4$ socially disqualified would be by deleting $a_x$ and $a_{\bar{x}}$. But in this case, the individual $q_x \in A^+$ would also turn socially disqualified, contradicting our assumption.

\item \label{fact:rgcdi_at_most_three}
For each variable $x \in X$, we can delete at most three individuals from the variable gadget of $x$. Assume for the sake of contradiction that we would delete more than three individuals in the variable gadget of some $x \in X$. Since we can only delete at most $\ell = 3m$ individuals in total (see \factref{fact:rgcdi_aaa}), this would imply that we can only delete at most two individuals for some other variable $x^\prime \in X$, contradicting \factref{fact:rgcdi_at_least_three}.

\item \label{fact:rgcdi_exactly_three}
From Facts \ref{fact:rgcdi_at_least_three} and \ref{fact:rgcdi_at_most_three}, it follows that for each variable $x \in X$, we need to delete exactly three individuals from the variable gadget of $x$. Thus, for each $x \in X$, we only have two options to achieve the objective: (1) delete $a_x$ along with $d_x^3$ and $d_x^4$, or (2) delete $a_{\bar{x}}$ along with $d_x^1$ and $d_x^2$.
\end{enumerate}

Equipped with these facts, we now show that the constructed instance is a YES-instance if and only if there exists a satisfying assignment for $C$.

($\Rightarrow$) Assume that $\varrho : X \rightarrow \{0, 1\}$ is a satisfying assignment for $C$. For each $x \in X$ we do the following:
\begin{itemize}[nolistsep, topsep=-1.25em]
\item
If $\varrho(x) = 1$ then we delete $a_{\bar{x}}$ along with $d_x^1$ and $d_x^2$.
\item
If $\varrho(x) = 0$ then we delete $a_x$ along with $d_x^3$ and $d_x^4$.
\end{itemize}
Clearly, we are deleting exactly three individuals per variable, so the budget of $\ell = 3m+1$ is sufficient. As a last step, we also delete the individual $a_{\ast\ast}$ as discussed in \factref{fact:rgcdi_aaa}. After that, it is easy to verify that none of the individuals $d_x^1$, $d_x^2$, $d_x^3$ and $d_x^4$ for $x \in X$ is socially qualified and each individual $q_x$ for $x \in X$ is still socially qualified. Furthermore, because $\varrho$ is a satisfying assignment for $C$, each clause $c \in C$ contains at least one literal of the form $x$ with $\varrho(x) = 1$ or one literal of the form $\bar{x}$ with $\varrho(x) = 0$. In the former case, the corresponding individual $a_c$ is qualified by the socially qualified individual $a_x$ and is therefore also socially qualified. In the latter case, the corresponding individual $a_c$ is qualified by the socially qualified individual $a_{\bar{x}}$ and is therefore also socially qualified. Hence, all individuals $a_c$ for $c \in C$ are socially qualified.

($\Leftarrow$) Assume we are given a set $U \subseteq N \setminus A^+$ of size at most $\ell$ such that after deleting the individuals in $U$, all individuals in $A^+$ are socially qualified and all remaining individuals in $A^-$ are socially disqualified. As we established in \factref{fact:rgcdi_exactly_three}, for each variable $x \in X$, there are only two possibilities which individuals may have been deleted from the corresponding variable gadget:
\begin{itemize}[nolistsep, topsep=-1.25em]
\item
If $a_{\bar{x}}$ was deleted along with $d_x^1$ and $d_x^2$, then we set $\varrho(x) = 1$.
\item
If $a_x$ was deleted along with $d_x^3$ and $d_x^4$, then we set $\varrho(x) = 0$.
\end{itemize}
We know that all individuals in $\{ a_c : c \in C \} \subseteq A^+$ are still socially qualified after deleting the individuals in $U$. Therefore, for each clause $c \in C$, there either is a literal $x \in c$ for which the individual $a_x$ is still socially qualified after the deletion, or there is a literal $\bar{x} \in c$ for which the individual $a_{\bar{x}}$ is still socially qualified after the deletion. In the former case, we have set $\varrho(x) = 1$, so the clause will be satisfied under the truth assignment $\varrho$. In the latter case, we have set $\varrho(x) = 0$, so the clause will also be satisfied under the truth assignment $\varrho$. Hence, $\varrho$ is a satisfying assignment for $C$.
\end{Proof}

By slightly modifying the instances created in the above reduction, we can get a similar result for the non-relaxed variant of GCDI. Although we have shown in \obsref{obs:csr_immune_gcdi} that the $f^{\text{CSR}}$ rule is immune to GCDI, the problem becomes NP-complete if we consider protective instances, i.e.\@ if $A^+ \subseteq f^{\text{CSR}}(N, \varphi)$ already holds initially:

\begin{Corollary} \label{cor:csr_gcdi_npc}
When $A^+ \neq \emptyset$, $A^- \neq \emptyset$, and $A^+ \subseteq f^{\text{CSR}}(N, \varphi)$ already holds initially, then $f^{\text{CSR}}$-\textsc{Group Control by Deleting Individuals} is NP-complete.
\end{Corollary}

\begin{Proof}
We first apply the same reduction from CNF-SAT as in \thmref{thm:csr_rgcdi_npc}. We then make the following minor modifications:

\begin{itemize}[nolistsep]
\item
We drop the individual $a_{\ast\ast}$.
\item
For each variable $x \in X$, we create an additional individual $d_x$ who qualifies only $a_\ast$ and is qualified only by the four individuals $d_x^1$, $d_x^2$, $d_x^3$ and $d_x^4$.
\item
We define $A^- = \{ d_x : x \in X \}$.
\item
We let $\ell = 3m$.
\end{itemize}

See \figref{fig:gcdi_variable_gadget} for an example illustration of one ``variable gadget'' for some variable $x \in X$ created in this reduction.

\begin{figure}[!tbh]
\centering
\raisebox{0.04\height}{
\begin{tikzpicture}[
    > = Stealth,
    shorten > = 1pt,
    auto,
    node distance = 1.8cm,
    main node/.style = {circle,draw,minimum size = 1cm}
]

\variablegadget

\variablegadgetextras

\node[main node] (dx)   [right of = qx, node distance = 7.5cm] {$d_x$};

\path[->]
(dx1)   edge                    node {} (dx)
(dx2)   edge                    node {} (dx)
(dx3)   edge                    node {} (dx)
(dx4)   edge                    node {} (dx)
;

\draw [very thick, decorate, decoration = {brace, raise=5pt, amplitude=5pt}]
(4.3, 5.6) -- (5.7, 5.6)
node[pos = 0.5, above = 10pt, black] {$A^-$};

\end{tikzpicture}
}
\caption{An illustration of one ``variable gadget'' for some variable $x \in X$ created in the reduction in \corref{cor:csr_gcdi_npc}. Here, we assume that there are three clauses $c_1$, $c_2$, $c_3$, and that $c_1$ contains the literal $x$, $c_3$ contains the literal $\bar{x}$, and $c_2$ contains neither $x$ nor $\bar{x}$.}
\label{fig:gcdi_variable_gadget}
\end{figure}

It is easy to see that initially all individuals in the created instance are socially qualified. In particular, this means that all individuals from $A^+$ are already socially qualified, i.e.\@ $A^+ \subseteq f^{\text{CSR}}(N, \varphi)$. For each variable $x \in X$, the only way to turn $d_x$ socially disqualified is to ensure that none of the individuals $d_x^1$, $d_x^2$, $d_x^3$ and $d_x^4$ is socially qualified. Using the same arguments as in \thmref{thm:csr_rgcdi_npc}, one can verify that the constructed instance is a YES-instance if and only if there exists a satisfying assignment for $C$.
\end{Proof}

Finally, we now turn to the liberal-start-respecting rule. It is easy to see that this rule is immune to R-GCDI:

\begin{Observation} \label{obs:lsr_rgcdi_i}
The $f^{\text{LSR}}$ rule is immune to \textsc{Relaxed Group Control by Deleting Individuals}, provided that $A^+ \neq \emptyset$.
\end{Observation}

\begin{Proof}
Because the $f^{\text{LSR}}$ rule is immune to CGCDI as shown by \textcite[Theorem 7]{YD18}, it is impossible to achieve the objective if $A^+$ is nonempty.
\end{Proof}

\clearpage

\section{Manipulative attacks on iterative consensus rules}
\label{sec:manipulative_attacks_iterative_consensus}

Until now, the study of manipulative attacks in Group Identification has been focused on the consent rules, the consensus-start-respecting rule, and the liberal-start-respecting rule \cite{YD18,ERY20,BBKL20,J22}. In this section, we extend the analysis to three previously unstudied social rules: We consider the iterative-consensus rule $f^{\text{IC}}$, the 2-stage-iterative-consensus rule $f^{\text{2IC}}$, and the 2-stage-liberal-iterative-consensus rule $f^{\text{2LIC}}$, each paired with every problem listed in \secref{sec:problem_definitions}. See \secref{sec:social_rules} for the formal definitions of these social rules. \tableref{tab:manipulative_attacks_iterative_consensus} shows an overview of our results.

\renewcommand{\arraystretch}{1.8}

\begin{table}[!htb]
\begin{tabularx}{\textwidth}{
p{0.25\textwidth}
p{0.07\textwidth}
X
p{0.07\textwidth}
X
p{0.07\textwidth}
X
}
\hline & \multicolumn{2}{l}{$f^{\text{IC}}$} & \multicolumn{2}{l}{$f^{\text{2IC}}$} & \multicolumn{2}{l}{$f^{\text{2LIC}}$} \\
\hline

CGCAI &
NP-c                            & \ccref{cor:ic_cgcai_npc} &
NP-c                            & \ccref{cor:ic_cgcai_npc} &
NP-c                            & \ccref{cor:ic_cgcai_npc} \\

DGCAI &
NP-c                            & \ttref{thm:ic_dgcai_npc} &
NP-c                            & \ttref{thm:ic_dgcai_npc} &
NP-c                            & \ttref{thm:ic_dgcai_npc} \\

EGCAI &
NP-c                            & \ccref{cor:ic_egcai_gcai_npc} &
NP-c                            & \ccref{cor:2ic_egcai_gcai_npc} &
NP-c                            & \ccref{cor:2ic_egcai_gcai_npc} \\

GCAI &
NP-c                            & \ccref{cor:ic_egcai_gcai_npc} &
NP-c                            & \ccref{cor:2ic_egcai_gcai_npc} &
NP-c                            & \ccref{cor:2ic_egcai_gcai_npc} \\

\hline

CGCDI &
NP-c                            & \ttref{thm:ic_cgcdi_npc} &
P                               & \ttref{thm:2ic_cgcdi_p} &
P                               & \ccref{cor:2lic_cgcdi_p} \\

DGCDI &
NP-c                            & \ttref{thm:ic_dgcdi_npc} &
NP-c                            & \ttref{thm:ic_dgcdi_npc} &
NP-c                            & \ttref{thm:ic_dgcdi_npc} \\

GCDI &
NP-c                            & \ccref{cor:ic_gcdi_npc} &
NP-c                            & \ccref{cor:2ic_gcdi_npc} &
NP-c                            & \ccref{cor:2ic_gcdi_npc} \\

\hline

CGB / \$CGB &
NP-c                            & \ttref{thm:ic_cgb_npc} &
P                               & \ttref{thm:2ic_cgb_p} &
P                               & \ttref{thm:2lic_cgb_p} \\

\makecell{ DGB / \\ \$DGB }     &
\makecell{ P \\ P }             &
\makecell{ \ttref{thm:2ic_dgb_p} \\ \ttref{thm:2ic_dgb_p} } &
\makecell{ P \\ P }             &
\makecell{ \ttref{thm:2ic_dgb_p} \\ \ttref{thm:2ic_dgb_p} } &
\makecell{ P \\ NP-c }          &
\makecell{ \ttref{thm:2lic_gb_p} \\ \ttref{thm:2lic_dgb_npc} } \\

\makecell{ EGB / \\ \$EGB }     &
\makecell{ NP-c \\ NP-c }       &
\makecell{ \ccref{cor:ic_gb_npc} \\ \ccref{cor:ic_gb_npc} } &
\makecell{ P \\ NP-c }          &
\makecell{ \ccref{cor:2ic_gb_p} \\ \ccref{cor:2ic_egb_gb_npc} } &
\makecell{ P \\ NP-c }          &
\makecell{ \ttref{thm:2lic_gb_p} \\ \ccref{cor:2ic_egb_gb_npc} } \\

\makecell{ GB / \\ \$GB }       &
\makecell{ NP-c \\ NP-c }       &
\makecell{ \ccref{cor:ic_gb_npc} \\ \ccref{cor:ic_gb_npc} } &
\makecell{ P \\ NP-c }          &
\makecell{ \ccref{cor:2ic_gb_p} \\ \ccref{cor:2ic_egb_gb_npc} } &
\makecell{ P \\ NP-c }          &
\makecell{ \ttref{thm:2lic_gb_p} \\ \ccref{cor:2ic_egb_gb_npc} } \\[1.5em]

\hline

\makecell{ CGMB / \\ \$CGMB }   &
\makecell{ NP-c \\ NP-c }       &
\makecell{ \ttref{thm:ic_cgmb_npc} \\ \ttref{thm:ic_cgmb_npc} } &
\makecell{ P \\ NP-c }          &
\makecell{ \ttref{thm:2ic_cgmb_p} \\ \ttref{thm:2ic_cgmb_npc} } &
\makecell{ P \\ NP-c }          &
\makecell{ \ttref{thm:2lic_cgmb_p} \\ \ttref{thm:2ic_cgmb_npc} } \\

DGMB / \$DGMB &
NP-c                            & \ttref{thm:ic_dgmb_npc} &
NP-c                            & \ttref{thm:ic_dgmb_npc} &
NP-c                            & \ttref{thm:ic_dgmb_npc} \\

EGMB / \$EGMB &
NP-c                            & \ccref{cor:ic_egmb_gmb_npc} &
NP-c                            & \ccref{cor:ic_egmb_gmb_npc} &
NP-c                            & \ccref{cor:ic_egmb_gmb_npc} \\

GMB / \$GMB &
NP-c                            & \ccref{cor:ic_egmb_gmb_npc} &
NP-c                            & \ccref{cor:ic_egmb_gmb_npc} &
NP-c                            & \ccref{cor:ic_egmb_gmb_npc} \\

\hline
\end{tabularx}
\caption{
A summary of the complexity results for manipulative attacks on iterative consensus rules.
In the table, ``P'' stands for ``polynomial-time solvable'' and ``NP-c'' stands for ``NP-complete''.
Note that for some of the bribery and microbribery problems, the priced version has a different complexity than the unpriced one.
}
\label{tab:manipulative_attacks_iterative_consensus}
\end{table}

\renewcommand{\arraystretch}{1.0}

In many proofs throughout this section, the set of socially qualified individuals in the $i$-th iteration is the same for all considered social rules. In this case, we sometimes write $K_i(N, \varphi)$ instead of $K^{\text{x}}_i(N, \varphi)$ with $\text{x} \in \{ \text{IC}, \text{2IC}, \text{2LIC} \}$. Likewise, we then write $f(N, \varphi)$ instead of $f^{\text{x}}(N, \varphi)$ for the final set of socially qualified individuals.

\subsection{Group control by adding individuals}

We begin with the group control by adding individuals problems. When the attacker has a destructive objective, the problem is NP-complete for all three considered social rules. This can be shown via a reduction from \textsc{Set Cover}. Because \textsc{Set Cover} is W[2]-hard with respect to $k$, this also implies that DGCAI is W[2]-hard with respect to $\ell$:

\begin{Theorem} \label{thm:ic_dgcai_npc}
For all $f \in \{ f^{\text{IC}}, f^{\text{2IC}}, f^{\text{2LIC}} \}$, the problem $f$-\textsc{Destructive Group Control by Adding Individuals} is NP-complete and W[2]-hard with respect to $\ell$.
\end{Theorem}

\begin{Proof}
Given a \textsc{Set Cover} instance $(X, \mathcal{F}, k)$, we construct an equivalent instance of DGCAI as follows:

For each element $x \in X$, we introduce one individual $a_x$. Let $N_X = \{ a_x : x \in X \}$. For each set $F \in \mathcal{F}$, we introduce one individual $a_F$. Let $N_\mathcal{F} = \{ a_F : F \in \mathcal{F} \}$. We set $N = N_X \cup N_\mathcal{F}$ and define the profile $\varphi$ over $N$ as follows:

\begin{itemize}[nolistsep]
\item
Each individual $a_x$ where $x \in X$ qualifies everyone (including themselves).
\item
Each individual $a_F$ where $F \in \mathcal{F}$ disqualifies all $a_x \in N_X$ with $x \in F$ and qualifies all other individuals (including themselves).
\end{itemize}

Finally, we define $A^- = T = N_X$ and $\ell = k$. Clearly, the above reduction only takes polynomial time.

Initially, each individual in $A^- = T = N_X$ is qualified by everyone from $A^-$. Therefore, all individuals in $A^-$ are socially qualified. To turn them socially disqualified, we need to ensure that (a) none of them is qualified by everyone and (b) among the individuals who are qualified by everyone, there is no consensus that any individual in $A^-$ should be qualified. We now show that the constructed instance is a YES-instance if and only if there is a set cover for $X$ in $\mathcal{F}$ of cardinality at most $k$.

($\Rightarrow$) Assume that $\mathcal{F}^\prime \subseteq \mathcal{F}$ is a set cover for $X$ with $|\mathcal{F}^\prime| \leq k$. We let $U = \{ a_F : F \in \mathcal{F}^\prime \}$. Clearly, it holds that $|U| \leq k = \ell$ and that each individual in $U$ is qualified by everyone. Furthermore, since $\mathcal{F}^\prime$ is a set cover for $X$, each individual $a_x$ where $x \in X$ is disqualified by at least one individual in $U$. Therefore, we have $K_0(T \cup U, \varphi) = U$, $K_1(T \cup U, \varphi) = U$, and thus $f(T \cup U, \varphi) = U$. In other words, all individuals in $A^-$ are socially disqualified after adding $U$ to $T$.

($\Leftarrow$) Assume we are given a set $U \subseteq N \setminus T$ of size at most $\ell$ such that all individuals in $A^-$ are socially disqualified after adding $U$ to $T$. This implies that each individual in $A^-$ is disqualified by at least one individual from $U$. By construction of $\varphi$, it follows that $\mathcal{F}^\prime = \{ F \in \mathcal{F} : a_F \in U \}$ is a set cover for $X$ of size at most $\ell = k$.
\end{Proof}

Next, we consider the constructive problems. By extending the above reduction, we can show that CGCAI is also NP-complete and W[2]-hard for all three social rules:

\begin{Corollary} \label{cor:ic_cgcai_npc}
For all $f \in \{ f^{\text{IC}}, f^{\text{2IC}}, f^{\text{2LIC}} \}$, the problem $f$-\textsc{Constructive Group Control by Adding Individuals} is NP-complete and W[2]-hard with respect to $\ell$.
\end{Corollary}

\begin{Proof}
We first apply the same reduction from \textsc{Set Cover} as in \thmref{thm:ic_dgcai_npc}. We then make the following minor modifications:

\begin{itemize}[nolistsep]
\item
We add one further individual $q_1$ to $T$.
\item
We let $q_1$ qualify everyone (including themselves).
\item
All individuals from $N_\mathcal{F}$ qualify $q_1$, and all individuals from $N_X$ disqualify $q_1$.
\item
We define $A^+ = \{q_1\}$ (the set $A^-$ is dropped for the constructive problem).
\end{itemize}

As before, each individual in $N_X$ is initially qualified by everyone and therefore socially qualified, i.e.\@ $K_0(T, \varphi) = N_X$. Everyone in $N_X$ disqualifies $q_1$, so $q_1$ is socially disqualified. To turn $q_1$ socially qualified, we need to ensure that none of the individuals from $N_X$ is qualified by everyone (because this way we create a consensus among the individuals who are qualified by everyone that $q_1$ should be qualified). Using the same arguments as in \thmref{thm:ic_dgcai_npc}, one can verify that the constructed instance is a YES-instance if and only if there exists a set cover for $X$ in $\mathcal{F}$ of cardinality at most $k$.
\end{Proof}

For the rules $f^{\text{2IC}}$ and $f^{\text{2LIC}}$, we can also obtain an NP-completeness proof for EGCAI and GCAI by again modifying the above reduction:

\begin{Corollary} \label{cor:2ic_egcai_gcai_npc}
For $f \in \{ f^{\text{2IC}}, f^{\text{2LIC}} \}$, the problems $f$-\textsc{Exact Group Control by Adding Individuals} and $f$-\textsc{Group Control by Adding Individuals} with $A^+ \neq \emptyset$ and $A^- \neq \emptyset$ are NP-complete and W[2]-hard with respect to $\ell$.
\end{Corollary}

\begin{Proof}
We first apply the same reduction from \textsc{Set Cover} as in \corref{cor:ic_cgcai_npc}. We then additionally define $A^- = T \setminus \{q_1\} = N_X$. We keep the definition of $A^+ = \{q_1\}$.

As before, the attacker needs to ensure that none of the individuals in $N_X$ is qualified by everyone to turn $q_1$ socially qualified. This also has the effect that none of the individuals in $A^- = N_X$ is socially qualified after the first iterative step (since each of them is disqualified by someone in $U$). Because the social rules $f^{\text{2IC}}$ and $f^{\text{2LIC}}$ terminate after the first iterative step, this implies that $A^- \cap f(T \cup U, \varphi) = \emptyset$. In other words, if the constructive objective is achieved, the destructive objective follows automatically.
\end{Proof}

For the $f^{\text{IC}}$ rule, the above argument from \corref{cor:2ic_egcai_gcai_npc} does not work. This is because the $f^{\text{IC}}$ rule does not terminate after only one iteration, so we would get $Q^{\text{IC}}_0(T \cup U, \varphi) = U$, $Q^{\text{IC}}_1(T \cup U, \varphi) = \{q_1\}$, and then $Q^{\text{IC}}_2(T \cup U, \varphi) = N_X$ because $q_1$ qualifies everyone in $N_X$. Hence, $A^- \cap f^{\text{IC}}(T \cup U, \varphi) = \emptyset$ would not hold. We fix this problem by further extending the reduction from \corref{cor:2ic_egcai_gcai_npc}:

\begin{Corollary} \label{cor:ic_egcai_gcai_npc}
The problems $f^{\text{IC}}$-\textsc{Exact Group Control by Adding Individuals} and $f^{\text{IC}}$-\textsc{Group Control by Adding Individuals} with $A^+ \neq \emptyset$ and $A^- \neq \emptyset$ both are NP-complete and W[2]-hard with respect to $\ell$.
\end{Corollary}

\begin{Proof}
We first apply the same reduction from \textsc{Set Cover} as in \corref{cor:2ic_egcai_gcai_npc}. We then make the following additional modifications:

\begin{itemize}[nolistsep]
\item
We add two further individuals $q_2$ and $q_3$ to $T$.
\item
The individual $q_1$ now disqualifies everyone in $N_X$, but still qualifies all remaining individuals (including themselves).
\item
The individual $q_2$ also disqualifies everyone in $N_X$, and qualifies all remaining individuals (including themselves).
\item
The individual $q_3$ disqualifies $q_1$ and $q_2$, and qualifies all remaining individuals (including themselves).
\item
All individuals from $N_\mathcal{F}$ qualify $q_1$, but disqualify $q_2$ and $q_3$.
\item
All individuals from $N_X$ qualify $q_3$, but disqualify $q_1$ and $q_2$.
\item
We define $A^+ = \{ q_1, q_2, q_3 \}$ and keep the definition of $A^- = N_X$.
\end{itemize}

Initially, the only individual who is qualified by everyone is $q_3$, i.e.\@ $Q^{\text{IC}}_0(T, \varphi) = \{q_3\}$. Since $q_3$ qualifies everyone in $N_X$ and disqualifies $q_1$ and $q_2$, we get $Q^{\text{IC}}_1(T, \varphi) = N_X$. The individuals in $N_X$ disqualify $q_1$ and $q_2$, thus $Q^{\text{IC}}_2(T, \varphi) = \emptyset$ and $f^{\text{IC}}(T, \varphi) = \{q_3\} \cup N_X$.

To turn the individuals from $A^-$ socially disqualified while making $q_1$, $q_2$ and $q_3$ socially qualified, the attacker obviously needs to add at least one individual from $N \setminus T$ to $T$. As soon as that happens, the situation changes: Since $N \setminus T = N_\mathcal{F}$ and all individuals in $N_\mathcal{F}$ disqualify $q_3$, the individual $q_3$ is no longer qualified by everyone. The only individuals who are qualified by everyone are the ones in the subset $U \subseteq N_\mathcal{F}$ which the attacker decides to add to $T$. Hence, we now have $Q^{\text{IC}}_0(T \cup U, \varphi) = U$. Among the individuals in $U$, there must not be a consensus that any $a_x \in N_X$ should be qualified (since otherwise this $a_x$ would be socially qualified after the first iterative step). In other words, as before, $U$ must represent a set cover for $X$ of cardinality at most $\ell = k$, making the problem NP-hard. But among the individuals in $U$, there does exist a consensus that $q_1$ should be qualified, i.e.\@ $Q^{\text{IC}}_1(T \cup U, \varphi) = \{q_1\}$. The individual $q_1$ qualifies $q_2$ and $q_3$, and disqualifies everyone in $N_X$, i.e.\@ $Q^{\text{IC}}_2(T \cup U, \varphi) = \{ q_2, q_3 \}$. Among $q_2$ and $q_3$, there is no consensus that anyone in $N_X$ should be qualified, i.e.\@ $Q^{\text{IC}}_3(T \cup U, \varphi) = \emptyset$. We therefore obtain $f^{\text{IC}}(T \cup U, \varphi) = U \cup A^+$, and $A^- \cap f^{\text{IC}}(T \cup U, \varphi) = \emptyset$ now holds.
\end{Proof}

\clearpage

\subsection{Group control by deleting individuals}

Next, we turn to the group control by deleting individuals problems. When the iterative-consensus rule is used, the constructive problem CGCDI is NP-complete and W[2]-hard with respect to $\ell$. This can be shown via another reduction from \textsc{Set Cover}:

\begin{Theorem} \label{thm:ic_cgcdi_npc}
The problem $f^{\text{IC}}$-\textsc{Constructive Group Control by Deleting Individuals} is NP-complete and W[2]-hard with respect to $\ell$.
\end{Theorem}

\begin{Proof}
Let $(X, \mathcal{F}, k)$ be an instance of \textsc{Set Cover}. Without loss of generality, we can assume that the set family $\mathcal{F}$ is of the form $\{ F_1, F_2, \ldots, F_m \}$ where $m = |\mathcal{F}| > k$. We now construct an equivalent instance of $f^{\text{IC}}$-CGCDI as follows:

For each element $x \in X$, we introduce one individual $a_x$. Let $N_X = \{ a_x : x \in X \}$. For each set $F_i$ with $i \in \{ 1, \ldots, m \}$, we introduce two individuals $a_{F_i}$ and $\tilde{a}_{F_i}$. Let $N_\mathcal{F} = \big\{ a_{F_i} : i \in \{ 1, \ldots, m \} \big\}$ and $\tilde{N}_\mathcal{F} = \big\{ \tilde{a}_{F_i} : i \in \{ 1, \ldots, m \} \big\}$. Finally, we introduce two more individuals $a_\ast$ and $z_\ast$. We then set $N = N_X \cup N_\mathcal{F} \cup \tilde{N}_\mathcal{F} \cup \{ a_\ast, z_\ast \}$ and define the profile $\varphi$ over $N$ as follows:

\begin{itemize}[nolistsep]
\item
All individuals qualify themselves and the individual $a_\ast$.
\item
Each individual $a_x$ where $x \in X$ also qualifies all other individuals.
\item
The individual $a_\ast$ qualifies the two individuals $a_{F_1}$ and $\tilde{a}_{F_1}$.
\item
The two individuals $a_{F_m}$ and $\tilde{a}_{F_m}$ each qualify $z_\ast$.
\item
For every $i \in \{ 1, \ldots, m-1 \}$, the individuals $a_{F_i}$ and $\tilde{a}_{F_i}$ each qualify $a_{F_{i+1}}$ and $\tilde{a}_{F_{i+1}}$.
\item
For every $i \in \{ 1, \ldots, m \}$, the individual $a_{F_i}$ additionally qualifies all individuals $a_x \in N_X$ with $x \in F_i$.
\item
All remaining entries of $\varphi$ not explicitly defined above are set to $-1$.
\end{itemize}

Finally, we set $A^+ = N \setminus \tilde{N}_\mathcal{F} = N_X \cup N_\mathcal{F} \cup \{ a_\ast, z_\ast \}$ and $\ell = k$. See \figref{fig:ic_cgcdi_npc} for an example illustration of the CGCDI instance created in this reduction. Clearly, the reduction can be completed in polynomial time. Before we show the correctness of the reduction, it will be useful to observe a few simple facts:

\begin{figure}[!tbh]
\centering
\raisebox{0.04\height}{
\begin{tikzpicture}[
    > = Stealth,
    shorten > = 1pt,
    auto,
    node distance = 2.5cm,
    main node/.style = {circle, draw, minimum size = 1.45cm}
]

\node[main node] (aa)                                   {$a_\ast$};
\node[main node] (ft1)  [right of = aa, yshift = 2.5cm] {$\tilde{a}_{F_1}$};
\node[main node] (f1)   [right of = aa]                 {$a_{F_1}$};
\node[main node] (ft2)  [right of = ft1]                {$\tilde{a}_{F_2}$};
\node[main node] (f2)   [right of = f1]                 {$a_{F_2}$};
\node            (ft3)  [right of = ft2]                {$\ldots$};
\node            (f3)   [right of = f2]                 {$\ldots$};
\node[main node] (ftm)  [right of = ft3]                {$\tilde{a}_{F_m}$};
\node[main node] (fm)   [right of = f3]                 {$a_{F_m}$};
\node[main node] (za)   [right of = fm]                 {$z_\ast$};

\node (x1)      [below of = f1, xshift = -1.8cm]        {};
\node (x2)      [right of = x1, node distance = 1cm]    {};
\node (x3)      [right of = x2, node distance = 1cm]    {};
\node (x4)      [right of = x3, node distance = 1cm]    {};
\node (x5)      [right of = x4, node distance = 1cm]    {};
\node (x6)      [right of = x5, node distance = 1cm]    {};
\node (x7)      [right of = x6, node distance = 1cm]    {};
\node (x8)      [right of = x7, node distance = 1cm]    {};
\node (x9)      [right of = x8, node distance = 1cm]    {};
\node (x10)     [right of = x9, node distance = 1cm]    {};
\node (x11)     [right of = x10, node distance = 1cm]   {};
\node (x12)     [right of = x11, node distance = 1cm]   {};
\node (nxn)     [right of = x6, node distance = 0.5cm, yshift = 0.25cm] {};
\node (nxs)     [right of = x6, node distance = 0.5cm, yshift = -0.5cm] {$N_X$};
\node[ellipse, draw, fit={(x3)(nxn)(nxs)(x10)}, inner sep=1mm] (x) {};

\path[->]
(aa)    edge                    node {} (ft1)
        edge                    node {} (f1)
(ft1)   edge                    node {} (ft2)
        edge                    node {} (f2)
(f1)    edge                    node {} (ft2)
        edge                    node {} (f2)
(ft2)   edge[shorten > = 0pt]   node {} (ft3)
        edge[shorten > = 0pt]   node {} (f3)
(f2)    edge[shorten > = 0pt]   node {} (ft3)
        edge[shorten > = 0pt]   node {} (f3)
(ft3)   edge                    node {} (ftm)
        edge                    node {} (fm)
(f3)    edge                    node {} (ftm)
        edge                    node {} (fm)
(ftm)   edge                    node {} (za)
(fm)    edge                    node {} (za)
;

\path[->]
(f1)    edge                    node {} (x2.east)
        edge                    node {} (x4.west)
(f2)    edge                    node {} (x4.east)
        edge                    node {} (x5.east)
        edge                    node {} (x7.west)
        edge                    node {} (x8.west)
(fm)    edge                    node {} (x8.east)
        edge                    node {} (x9.east)
        edge                    node {} (x11.west)
;

\draw [very thick, decorate, decoration = {brace, raise=5pt, amplitude=5pt}]
(-1, -3.8) -- (-1, 1)
node[pos = 0.5, left = 12pt, black] {$A^+$};

\draw [very thick, decorate, decoration = {brace, raise=5pt, amplitude=5pt}]
(-0.8, 3.3) -- (0.8, 3.3)
node[pos = 0.5, above = 12pt, black] {$Q^{\text{IC}}_0(N, \varphi)$};

\draw [very thick, decorate, decoration = {brace, raise=5pt, amplitude=5pt}]
(1.7, 3.3) -- (3.3, 3.3)
node[pos = 0.5, above = 12pt, black] {$Q^{\text{IC}}_1(N, \varphi)$};

\draw [very thick, decorate, decoration = {brace, raise=5pt, amplitude=5pt}]
(4.2, 3.3) -- (5.8, 3.3)
node[pos = 0.5, above = 12pt, black] {$Q^{\text{IC}}_2(N, \varphi)$};

\draw [very thick, decorate, decoration = {brace, raise=5pt, amplitude=5pt}]
(9.2, 3.3) -- (10.8, 3.3)
node[pos = 0.5, above = 12pt, black] {$Q^{\text{IC}}_m(N, \varphi)$};

\draw [very thick, decorate, decoration = {brace, raise=5pt, amplitude=5pt}]
(11.7, 3.3) -- (13.3, 3.3)
node[pos = 0.5, above = 12pt, black] {$Q^{\text{IC}}_{m+1}(N, \varphi)$};
\end{tikzpicture}
}
\caption{An example illustration of the CGCDI instance created in the reduction in \thmref{thm:ic_cgcdi_npc}. In addition to the drawn qualifications, all individuals qualify themselves and $a_\ast$, and every individual in $N_X$ qualifies all other individuals.}
\label{fig:ic_cgcdi_npc}
\end{figure}
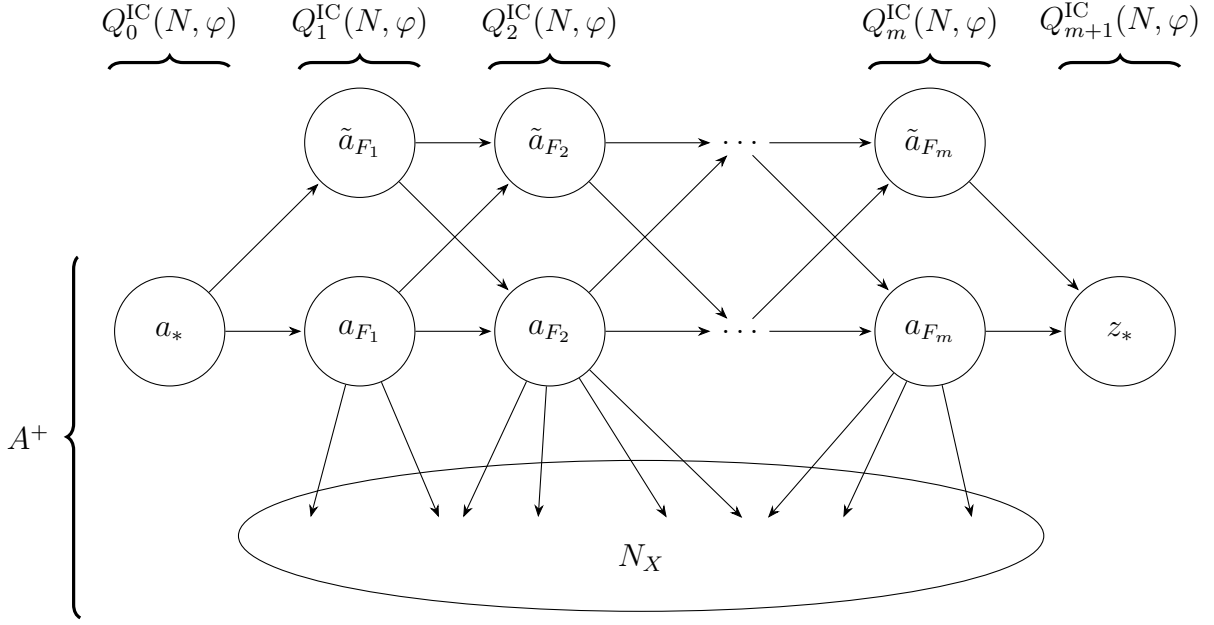

\begin{enumerate}[
    topsep=-4pt,
    leftmargin=*,
    itemindent=4.96em,
    label={\textit{Fact \ref*{thm:ic_cgcdi_npc}.\arabic*}:},
    ref={\ref*{thm:ic_cgcdi_npc}.\arabic*}
]
\item \label{fact:ic_cgcdi_aa}
Initially, only the individual $a_\ast$ is qualified by everyone, i.e.\@ $Q^{\text{IC}}_0(N, \varphi) = \{a_\ast\}$. The individual $a_\ast$ only qualifies themselves, $a_{F_1}$ and $\tilde{a}_{F_1}$. However, the individual $z_\ast$ (who is in $A^+$ and therefore cannot be deleted) does not qualify $a_{F_1}$ and $\tilde{a}_{F_1}$. Hence, no matter which individuals are deleted, $a_\ast$ will always be the only individual who is qualified by everyone (also note that $a_\ast$ themself is in $A^+$ and therefore cannot be deleted).

\item \label{fact:ic_cgcdi_chain}
Among the individuals in $Q^{\text{IC}}_0(N, \varphi) = \{a_\ast\}$, there is a consensus that $a_{F_1}$ and $\tilde{a}_{F_1}$ should be qualified. Furthermore, for any $i \in \{ 1, \ldots, m-1 \}$, there is a consensus among $a_{F_i}$ and $\tilde{a}_{F_i}$ that $a_{F_{i+1}}$ and $\tilde{a}_{F_{i+1}}$ should be qualified. This way, all individuals in $N_\mathcal{F} \cup \tilde{N}_\mathcal{F}$ are socially qualified. Meanwhile, for any $i \in \{ 1, \ldots, m \}$, there is no consensus among $a_{F_i}$ and $\tilde{a}_{F_i}$ that someone in $N_X$ should be qualified ($a_{F_i}$ qualifies some individuals in $N_X$, but $\tilde{a}_{F_i}$ does not). Finally, there is a consensus among $a_{F_m}$ and $\tilde{a}_{F_m}$ that $z_\ast$ should be qualified. Thus, we initially have $f^{\text{IC}}(N, \varphi) = \{a_\ast\} \cup N_\mathcal{F} \cup \tilde{N}_\mathcal{F} \cup \{z_\ast\}$.

\item \label{fact:ic_cgcdi_must_delete_from_n_f}
To turn the individuals from $N_X \subseteq A^+$ socially qualified, the attacker needs to ensure that, for each $a_x \in N_X$, there is at some point a consensus among the individuals in a set $Q^{\text{IC}}_i(N, \varphi)$ that $a_x$ should be qualified. Clearly, the only way to do that is by deleting certain individuals from $\tilde{N}_\mathcal{F}$ because all individuals outside of $\tilde{N}_\mathcal{F}$ are in $A^+$ and therefore cannot be deleted.

\item \label{fact:ic_cgcdi_delete_some_n_f}
Once the attacker deletes an individual $\tilde{a}_{F_i}$ with $i \in \{ 1, \ldots, m \}$, there is a consensus among everyone in $Q^{\text{IC}}_i(N \setminus \{\tilde{a}_{F_i}\}, \varphi)$ that the individuals in $\{ a_x : x \in F_i \}$ should be qualified. More formally, $\{ a_x : x \in F_i \} \subseteq Q^{\text{IC}}_{i+1}(N \setminus \{\tilde{a}_{F_i}\}, \varphi)$. Because the individuals from $N_X$ qualify everyone, this does not cause any changes in subsequent iterations: Among the individuals in $Q^{\text{IC}}_{i+1}(N \setminus \{\tilde{a}_{F_i}\}, \varphi)$, there still exists a consensus that $a_{F_{i+2}}$ and $\tilde{a}_{F_{i+2}}$ should be qualified. Thus, the argument also applies if multiple individuals from $\tilde{N}_\mathcal{F}$ are deleted: For every individual $\tilde{a}_{F_i}$ that is deleted from $\tilde{N}_\mathcal{F}$, the individuals in $\{ a_x : x \in F_i \}$ become socially qualified.

\item \label{fact:ic_cgcdi_chain_ends}
In any solution for the created instance, the iterative process ends after the $(m+1)$-th iteration. This is because $z_\ast \in Q^{\text{IC}}_{m+1}(N, \varphi)$, and the individual $z_\ast$ only qualifies themselves and $a_\ast$.
\end{enumerate}

Equipped with these facts, we now show that the constructed instance is a YES-instance if and only if there exists a set cover for $X$ in $\mathcal{F}$ of cardinality at most $k$.

($\Rightarrow$) Assume that $\mathcal{F}^\prime \subseteq \mathcal{F}$ is a set cover for $X$ with $|\mathcal{F}^\prime| \leq k$. We let $U = \{ \tilde{a}_{F_i} : {F_i} \in \mathcal{F}^\prime \}$. Clearly, $|U| \leq k = \ell$. We delete the individuals in $U$. From \factref{fact:ic_cgcdi_delete_some_n_f} and the fact that $\mathcal{F}^\prime$ is a set cover for $X$, we know that all individuals from $N_X$ are now socially qualified. From \factref{fact:ic_cgcdi_chain}, we also know that all individuals in $\{ a_\ast, z_\ast \} \cup N_\mathcal{F}$ are socially qualified too. Thus, $U$ is a solution for the constructed instance.

($\Leftarrow$) Assume we are given a set $U \subseteq N \setminus A^+$ of size at most $\ell$ such that all individuals in $A^+$ are socially qualified after deleting the individuals in $U$. From \factref{fact:ic_cgcdi_must_delete_from_n_f}, we know that $U \subseteq \tilde{N}_\mathcal{F}$. Moreover, from \factref{fact:ic_cgcdi_chain_ends}, we know that $Q^{\text{IC}}_{m+2}(N \setminus U, \varphi) = \emptyset$. In conjunction with \factref{fact:ic_cgcdi_delete_some_n_f} and the fact that $N_X \subseteq f^{\text{IC}}(N \setminus U, \varphi)$, this implies that for each $x \in X$, there exists a set $F_i \in \mathcal{F}$ with $x \in F_i$ for which the individual $\tilde{a}_{F_i}$ is in $U$. From this, it follows that $\mathcal{F}^\prime = \{ F_i \in \mathcal{F} : \tilde{a}_{F_i} \in U \}$ is a set cover for $X$ of size at most $\ell = k$.
\end{Proof}

By extending the above reduction, we can also show that the general $f^{\text{IC}}$-GCDI problem with $A^+ \neq \emptyset$ and $A^- \neq \emptyset$ is NP-complete and W[2]-hard. To do this, we first apply the reduction for the constructive case, and then insert two additional individuals to obtain an instance where the destructive target set is also nonempty:

\begin{Corollary} \label{cor:ic_gcdi_npc}
$f^{\text{IC}}$-\textsc{Group Control by Deleting Individuals} where both $A^+ \neq \emptyset$ and $A^- \neq \emptyset$ is NP-complete and W[2]-hard with respect to $\ell$.
\end{Corollary}

\begin{Proof}
We first apply the same reduction from \textsc{Set Cover} as in \thmref{thm:ic_cgcdi_npc}. We then make the following minor modifications to obtain an instance of GCDI with $A^- \neq \emptyset$:

\begin{itemize}[nolistsep]
\item
We create two additional individuals $d_1$ and $d_2$.
\item
The individual $d_1$ qualifies only themselves, $a_\ast$ and $d_2$, and is qualified only by $z_\ast$ and the individuals in $N_X$.
\item
The individual $d_2$ qualifies only themselves and $a_\ast$, and is qualified only by $d_1$.
\item
We define $A^- = \{d_2\}$ and keep the definition of $A^+ = N_X \cup N_\mathcal{F} \cup \{ a_\ast, z_\ast \}$.
\item
We increase the budget $\ell$ by $1$, i.e.\@ we let $\ell = k+1$.
\end{itemize}

Initially, both $d_1$ and $d_2$ are socially qualified. This is because there is a consensus among the individuals in $Q^{\text{IC}}_{m+1}(N, \varphi) = \{z_\ast\}$ that $d_1$ should be qualified, i.e.\@ $Q^{\text{IC}}_{m+2}(N, \varphi) = \{d_1\}$. Everyone in $Q^{\text{IC}}_{m+2}(N, \varphi)$ then qualifies $d_2$, so we obtain $Q^{\text{IC}}_{m+3}(N, \varphi) = \{d_2\}$.

Unless $d_1$ is deleted, there will always be a consensus among the individuals in $Q^{\text{IC}}_{m+1}(N, \varphi)$ that $d_1$ should be qualified. Therefore, to turn $d_2$ socially disqualified, the attacker is forced to delete $d_1$. It is easy to see that the deletion of $d_1$ makes $d_2$ socially disqualified since $d_1$ is the only individual who qualifies $d_2$ (besides $d_2$ themselves).

The deletion of $d_1$ decreases the budget by $1$, so solving the remaining instance is again equivalent to solving the constructive case.
\end{Proof}

For the 2-stage rules $f^{\text{2IC}}$ and $f^{\text{2LIC}}$, the CGCDI problem is polynomial-time solvable. We first show this for the $f^{\text{2IC}}$ rule:

\begin{Theorem} \label{thm:2ic_cgcdi_p}
The problem $f^{\text{2IC}}$-\textsc{Constructive Group Control by Deleting Individuals} can be solved in time $\mathcal{O}(n^3)$.
\end{Theorem}

\begin{Proof}
To solve a given instance $(N, \varphi, A^+, \ell)$ of $f^{\text{2IC}}$-CGCDI, we do the following:

If any individual in $A^+$ disqualifies themselves, we immediately conclude that the instance is a NO-instance. This is because under the $f^{\text{2IC}}$ rule, individuals who disqualify themselves cannot become socially qualified.

To make all individuals in $A^+$ socially qualified, there needs to be at least one individual who is qualified by everyone (since otherwise no one is socially qualified). Thus, we guess an individual $a_\ast \in N$ to be qualified by everyone. Let $U_{a_\ast} = \{ a^\prime \in N : \varphi(a^\prime, a_\ast) = -1 \}$ denote the subset of individuals who disqualify $a_\ast$. Assuming $a_\ast$ was guessed correctly, we obviously need to delete the individuals in $U_{a_\ast}$. We then compute the set $K^{\text{2IC}}_0(N \setminus U_{a_\ast}, \varphi)$ of individuals who are qualified by everyone in $N \setminus U_{a_\ast}$. If there is a consensus among the individuals in $K^{\text{2IC}}_0(N \setminus U_{a_\ast}, \varphi)$ that everyone from $A^+$ should be qualified, we are done (everyone from $A^+$ is now socially qualified after the first iterative step). Else, we try to delete every individual from $K^{\text{2IC}}_0(N \setminus U_{a_\ast}, \varphi)$ who disqualifies someone in $A^+$ (to create a consensus among the remaining individuals). If this succeeds, we are done. Otherwise, it means that the individual $a_\ast$ was guessed wrong.

The instance is a YES-instance if and only if there exists an individual $a_\ast \in N$ for whom the above approach leads to a solution.

The running time of the algorithm is bounded by $\mathcal{O}(n \cdot n^2)$ where the additional factor of $n$ comes from guessing an individual $a_\ast \in N$ to be qualified by everyone.
\end{Proof}

By extending the algorithm from the above proof, we can also obtain a polynomial-time algorithm for $f^{\text{2LIC}}$-CGCDI:

\begin{Corollary} \label{cor:2lic_cgcdi_p}
The problem $f^{\text{2LIC}}$-\textsc{Constructive Group Control by Deleting Individuals} can be solved in time $\mathcal{O}(n^3)$.
\end{Corollary}

\begin{Proof}
To solve a given instance $(N, \varphi, A^+, \ell)$ of $f^{\text{2LIC}}$-CGCDI, we first try the same approach as in \thmref{thm:2ic_cgcdi_p}, i.e.\@ we guess an individual $a_\ast \in N$ to be qualified by everyone and see if this leads to a solution. Clearly, this takes time $\mathcal{O}(n^3)$. If this succeeds, the given instance is a YES-instance.

If the above approach does not lead to a solution, we try to delete all individuals who are qualified by everyone. This way, the set $K^{\text{2LIC}}_0(N \setminus U, \varphi)$ becomes empty, and all individuals who qualify themselves are automatically considered socially qualified. As we have shown in the proof of \thmref{thm:csr_regcdi_p}, deleting all individuals who are qualified by everyone can be done in time $\mathcal{O}(n^2)$ by using a queue. If this succeeds, the given instance is a YES-instance. Otherwise, it must be a NO-instance.
\end{Proof}

We now turn to the destructive problem. By a reduction from \textsc{Set Cover}, we show that DGCDI is NP-complete and W[2]-hard for all three considered social rules:

\begin{Theorem} \label{thm:ic_dgcdi_npc}
For all $f \in \{ f^{\text{IC}}, f^{\text{2IC}}, f^{\text{2LIC}} \}$, the problem $f$-\textsc{Destructive Group Control by Deleting Individuals} is NP-complete and W[2]-hard with respect to $\ell$.
\end{Theorem}

\begin{Proof}
Let $(X, \mathcal{F}, k)$ be an instance of \textsc{Set Cover}. Without loss of generality, we can assume that $|\mathcal{F}| > k$ (otherwise the instance would be a trivial YES-instance). Moreover, we can assume that every element from $X$ appears in at least one set from $\mathcal{F}$ (otherwise the instance would be a trivial NO-instance). We now construct an equivalent instance of DGCDI as follows:

For each element $x \in X$, we introduce one individual $a_x$. Let $N_X = \{ a_x : x \in X \}$. For each set $F \in \mathcal{F}$, we introduce two individuals $a_F$ and $\tilde{a}_F$. Let $N_\mathcal{F} = \{ a_F : F \in \mathcal{F} \}$ and $\tilde{N}_\mathcal{F} = \{ \tilde{a}_F : F \in \mathcal{F} \}$. We then set $N = N_X \cup N_\mathcal{F} \cup \tilde{N}_\mathcal{F}$ and define the profile $\varphi$ over $N$ as follows:

\begin{itemize}[nolistsep]
\item
All individuals qualify themselves.
\item
For each $x \in X$, the individual $a_x$ also qualifies all other individuals.
\item
For each $F \in \mathcal{F}$, the individual $a_F$ qualifies everyone except the individuals $a_x \in N_X$ with $x \in F$.
\item
For each $F \in \mathcal{F}$, the individual $\tilde{a}_F$ qualifies everyone except $a_F$.
\end{itemize}

Finally, we set $A^- = N_X$ and $\ell = k$. See \figref{fig:ic_dgcdi_npc} for an example illustration of the DGCDI instance created in this reduction. Clearly, the reduction can be completed in polynomial time. Before we show the correctness of the reduction, it will be useful to observe a few simple facts:

\begin{figure}[!tbh]
\centering
\raisebox{0.04\height}{
\begin{tikzpicture}[
    > = Stealth,
    shorten > = 1pt,
    auto,
    node distance = 2.5cm,
    main node/.style = {circle, draw, minimum size = 1.45cm}
]

\node[main node] (ft1)                                          {$\tilde{a}_{F_1}$};
\node[main node] (f1)   [below of = ft1]                        {$a_{F_1}$};
\node[main node] (ft2)  [right of = ft1]                        {$\tilde{a}_{F_2}$};
\node[main node] (f2)   [right of = f1]                         {$a_{F_2}$};
\node[main node] (ft3)  [right of = ft2]                        {$\tilde{a}_{F_3}$};
\node[main node] (f3)   [right of = f2]                         {$a_{F_3}$};
\node            (ft4)  [right of = ft3, node distance = 1.7cm] {$\ldots$};
\node            (f4)   [right of = f3, node distance = 1.7cm]  {$\ldots$};
\node[main node] (ftm1) [right of=ft4, node distance=1.7cm] {$\tilde{a}_{F_{m-1}}$};
\node[main node] (fm1)  [right of = f4, node distance = 1.7cm]  {$a_{F_{m-1}}$};
\node[main node] (ftm)  [right of = ftm1]                       {$\tilde{a}_{F_m}$};
\node[main node] (fm)   [right of = fm1]                        {$a_{F_m}$};

\node (x1)      [below of = f1]                         {};
\node (x2)      [right of = x1, node distance = 1cm]    {};
\node (x3)      [right of = x2, node distance = 1cm]    {};
\node (x4)      [right of = x3, node distance = 1cm]    {};
\node (x5)      [right of = x4, node distance = 1cm]    {};
\node (x6)      [right of = x5, node distance = 1cm]    {};
\node (x7)      [right of = x6, node distance = 1cm]    {};
\node (x8)      [right of = x7, node distance = 1cm]    {};
\node (x9)      [right of = x8, node distance = 1cm]    {};
\node (x10)     [right of = x9, node distance = 1cm]    {};
\node (x11)     [right of = x10, node distance = 1cm]   {};
\node (x12)     [right of = x11, node distance = 1cm]   {};
\node (nxn)     [right of = x6, node distance = 0.5cm, yshift = 0.25cm] {};
\node (nxs)     [right of = x6, node distance = 0.5cm, yshift = -0.5cm] {$N_X$};
\node[ellipse, draw, fit={(x3)(nxn)(nxs)(x10)}, inner sep=1mm] (x) {};

\path[->, dashed]
(ft1)   edge                    node {} (f1)
(ft2)   edge                    node {} (f2)
(ft3)   edge                    node {} (f3)
(ftm1)  edge                    node {} (fm1)
(ftm)   edge                    node {} (fm)
;

\path[->, dashed]
(f1)    edge                    node {} (x2.east)
        edge                    node {} (x3.west)
(f2)    edge                    node {} (x3.east)
        edge                    node {} (x4.east)
        edge                    node {} (x5.west)
        edge                    node {} (x6.west)
(f3)    edge                    node {} (x6.east)
        edge                    node {} (x7.east)
        edge                    node {} (x8.west)
(fm1)   edge                    node {} (x8.east)
        edge                    node {} (x9.east)
        edge                    node {} (x10.west)
(fm)    edge                    node {} (x10.east)
        edge                    node {} (x11.west)
;

\draw [very thick, decorate, decoration = {brace, raise=5pt, amplitude=5pt}]
(-0.5, -6.4) -- (-0.5, -3.8)
node[pos = 0.5, left = 12pt, black] {$A^-$};

\end{tikzpicture}
}
\caption{An example illustration of the DGCDI instance created in the reduction in \thmref{thm:ic_dgcdi_npc}. In the picture, we assume that $\mathcal{F}$ is of the form $\{ F_1, F_2, \ldots, F_m \}$. Every individual qualifies themselves and all other individuals, except for the drawn dashed arcs which represent disqualifications.}
\label{fig:ic_dgcdi_npc}
\end{figure}

\begin{enumerate}[
    topsep=-4pt,
    leftmargin=*,
    itemindent=4.96em,
    label={\textit{Fact \ref*{thm:ic_dgcdi_npc}.\arabic*}:},
    ref={\ref*{thm:ic_dgcdi_npc}.\arabic*}
]
\item \label{fact:ic_dgcdi_initial_situation}
Initially, all individuals in $\tilde{N}_\mathcal{F}$ are qualified by everyone, i.e.\@ $K_0(N, \varphi) = \tilde{N}_\mathcal{F}$. Among the individuals in $K_0(N, \varphi)$, there is a consensus that the individuals from $N_X$ should be qualified, and there is no consensus that anyone from $N_\mathcal{F}$ should be qualified. Thus, $K_1(N, \varphi) = \tilde{N}_\mathcal{F} \cup N_X$. Notably, this means that initially all individuals from $N_X$ are socially qualified after the first iterative step.

\item \label{fact:ic_dgcdi_no_delete_from_n_f}
To turn the individuals in $N_X$ socially disqualified, the attacker obviously has to delete some individuals from $N \setminus N_X = N_\mathcal{F} \cup \tilde{N}_\mathcal{F}$. However, in any minimal solution, the attacker only deletes individuals from $\tilde{N}_\mathcal{F}$. This is because deleting an individual from $N_\mathcal{F}$ means taking away disqualifications from certain individuals in $N_X$. Clearly, this does not help to make the individuals in $N_X$ socially disqualified.

\item \label{fact:ic_dgcdi_delete_from_n_f}
When an individual $\tilde{a}_F$ where $F \in \mathcal{F}$ is deleted from $\tilde{N}_\mathcal{F}$, the corresponding individual $a_F$ is qualified by all remaining individuals. Hence, there no longer exists a consensus among the individuals who are qualified by everyone that the individuals in $\{ a_x : x \in F \}$ should be qualified (since $a_F$ disqualifies them). This way, the individuals in $\{ a_x : x \in F \}$ are no longer socially qualified after the first iterative step.

\item \label{fact:ic_dgcdi_socially_disqualified}
Since $A^- = N_X$, the attacker must ensure that none of the individuals from $N_X$ is socially qualified. In particular, this means that none of them can be socially qualified after the first iterative step. Assume that there is a subset $U \subseteq \tilde{N}_\mathcal{F}$ such that $N_X \cap K_1(N \setminus U, \varphi) = \emptyset$. For the social rules $f^{\text{2IC}}$ and $f^{\text{2LIC}}$, this directly implies that none of the individuals from $N_X$ is socially qualified (because these rules terminate after the first iterative step). But even for the $f^{\text{IC}}$ rule, if the individuals in $U$ are deleted, all individuals in $N_X$ become socially disqualified. This is because among the individuals in $Q^{\text{IC}}_0(N \setminus U, \varphi)$, there is no consensus that anyone should be qualified, i.e.\@ $Q^{\text{IC}}_1(N \setminus U, \varphi) = \emptyset$. Therefore, the $f^{\text{IC}}$ rule also terminates after the first iterative step.
\end{enumerate}

Equipped with these facts, we now show that the constructed instance is a YES-instance if and only if there exists a set cover for $X$ in $\mathcal{F}$ of cardinality at most $k$.

($\Rightarrow$) Assume that $\mathcal{F}^\prime \subseteq \mathcal{F}$ is a set cover for $X$ with $|\mathcal{F}^\prime| \leq k$. We let $U = \{ \tilde{a}_F : F \in \mathcal{F}^\prime \}$. Clearly, $|U| \leq k = \ell$. We delete the individuals in $U$. From \factref{fact:ic_dgcdi_delete_from_n_f} and the fact that $\mathcal{F}^\prime$ is a set cover for $X$, we know that none of the individuals from $N_X$ is socially qualified after the first iterative step. In conjunction with \factref{fact:ic_dgcdi_socially_disqualified}, this implies that everyone in $N_X = A^-$ is now socially disqualified.

($\Leftarrow$) Assume we are given a minimal DGCDI solution $U \subseteq N \setminus A^-$ with $|U| \leq \ell$ such that all individuals from $A^-$ are socially disqualified after deleting the individuals in $U$. Without loss of generality, we can assume that $U \subseteq \tilde{N}_\mathcal{F}$ (see \factref{fact:ic_dgcdi_no_delete_from_n_f}). Also, from \factref{fact:ic_dgcdi_delete_from_n_f} and the fact that $N_X \cap f(N \setminus U, \varphi) = \emptyset$, we know that for each $x \in X$, there exists a set $F \in \mathcal{F}$ with $x \in F$ for which the individual $\tilde{a}_F$ is in $U$. From this, it follows that $\mathcal{F}^\prime = \{ F \in \mathcal{F} : \tilde{a}_F \in U \}$ is a set cover for $X$ of size at most $\ell = k$.
\end{Proof}

By extending the above reduction, we can also show that the general $f^{\text{2IC}}$-GCDI and $f^{\text{2LIC}}$-GCDI problems with $A^+ \neq \emptyset$ and $A^- \neq \emptyset$ are NP-complete and W[2]-hard. To do this, we first apply the reduction for the destructive case, and then insert two additional individuals to obtain an instance where the constructive target set is also nonempty:

\begin{Corollary} \label{cor:2ic_gcdi_npc}
For $f \in \{ f^{\text{2IC}}, f^{\text{2LIC}} \}$, the problem $f$-\textsc{Group Control by Deleting Individuals} with $A^+ \neq \emptyset$ and $A^- \neq \emptyset$ is NP-complete and W[2]-hard with respect to $\ell$.
\end{Corollary}

\begin{Proof}
We first apply the same reduction from \textsc{Set Cover} as in \thmref{thm:ic_dgcdi_npc}. We then make the following minor modifications to obtain an instance of GCDI with $A^+ \neq \emptyset$:

\begin{itemize}[nolistsep]
\item
We create two additional individuals $q_1$ and $q_2$.
\item
The individual $q_1$ qualifies everyone (including themselves).
\item
The individual $q_2$ disqualifies $q_1$ and qualifies everyone else (including themselves).
\item
All individuals from the original instance qualify $q_1$ and $q_2$.
\item
We define $A^+ = \{q_1\}$ and keep the definition of $A^- = N_X$.
\item
We increase the budget $\ell$ by $1$, i.e.\@ we let $\ell = k+1$.
\end{itemize}

Initially, the individual $q_1$ is socially disqualified. This is because the individual $q_2$ (who is qualified by everyone) disqualifies $q_1$; thus, $q_1$ is neither qualified by everyone nor qualified by all individuals who are qualified by everyone.

To turn $q_1$ socially qualified, the attacker is always forced to delete the individual $q_2$. This decreases the budget by $1$, so solving the remaining instance is again equivalent to solving the destructive case.
\end{Proof}

\clearpage

\subsection{Group bribery}

We now consider the group bribery problems. We begin by showing that the unpriced CGB problem is NP-complete when using the $f^{\text{IC}}$ rule. Obviously, this implies that the priced version is also NP-complete. The proof is based on a reduction from \textsc{Set Cover}. Because \textsc{Set Cover} is W[2]-hard with respect to $k$, this also implies that $f^{\text{IC}}$-CGB is W[2]-hard with respect to $\ell$:

\begin{Theorem} \label{thm:ic_cgb_npc}
$f^{\text{IC}}$-\textsc{Constructive Group Bribery} is NP-complete and W[2]-hard with respect to $\ell$.
\end{Theorem}

\begin{Proof}
Let $(X, \mathcal{F}, k)$ be an instance of \textsc{Set Cover}. Without loss of generality, we can assume that the set family $\mathcal{F}$ is of the form $\{ F_1, F_2, \ldots, F_m \}$ where $m = |\mathcal{F}| > k$. Also, we can assume that $\mathcal{F}$ contains the empty set (if not, then we simply add it; this obviously does not change the solution) and that $F_m = \emptyset$. We now construct an equivalent instance of $f^{\text{IC}}$-CGCDI as follows:

For each element $x \in X$, we introduce one individual $a_x$. Let $N_X = \{ a_x : x \in X \}$. For each set $F_i$ with $i \in \{ 1, \ldots, m \}$, we introduce $k+1$ individuals $\{ a^1_{F_i}, a^2_{F_i}, \ldots, a^{k+1}_{F_i} \}$. Let $N_\mathcal{F} = \big\{ a^j_{F_i} : i \in \{ 1, \ldots, m \} \text{ and } j \in \{ 1, \ldots, k \} \big\}$ and let $\tilde{N}_\mathcal{F} = \big\{ a^{k+1}_{F_i} : i \in \{ 1, \ldots, m \} \big\}$. Finally, we introduce $k+1$ additional individuals $\{ a^1_\ast, \ldots, a^{k+1}_\ast \}$. We then define the set $N = N_X \cup N_\mathcal{F} \cup \tilde{N}_\mathcal{F} \cup \{ a^1_\ast, \ldots, a^{k+1}_\ast \}$ and define the profile $\varphi$ over $N$ as follows:

\begin{itemize}[nolistsep]
\item
All individuals qualify themselves and the individuals in $\{ a^1_\ast, \ldots, a^{k+1}_\ast \}$.
\item
Each individual $a_x$ where $x \in X$ also qualifies all other individuals.
\item
Every individual in $\{ a^1_\ast, \ldots, a^{k+1}_\ast \}$ qualifies the individuals in $\{ a^1_{F_1}, \ldots, a^{k+1}_{F_1} \}$.
\item
For each $i \in \{ 1, \ldots, m-1 \}$, every individual in $\{ a^1_{F_i}, \ldots, a^{k+1}_{F_i} \}$ qualifies the individuals in $\{ a^1_{F_{i+1}}, \ldots, a^{k+1}_{F_{i+1}} \}$.
\item
For each $i \in \{ 1, \ldots, m \}$, the individuals in $\{ a^1_{F_i}, \ldots, a^k_{F_i} \}$ additionally qualify all individuals $a_x \in N_X$ with $x \in F_i$.
\item
All remaining entries of $\varphi$ not explicitly defined above are set to $-1$.
\end{itemize}

Finally, we set $A^+ = N$ and $\ell = k$. See \figref{fig:ic_cgb_npc} for an example illustration of the CGB instance created in this reduction. Clearly, the reduction can be completed in polynomial time. Before we show the correctness of the reduction, it will be useful to observe a few simple facts:

\begin{figure}[!tbh]
\centering
\raisebox{0.04\height}{
\begin{tikzpicture}[
    > = Stealth,
    shorten > = 1pt,
    auto,
    node distance = 2.5cm,
    main node/.style = {circle, draw, minimum size = 1.45cm}
]

\setcoverbriberyreductiongrid

\end{tikzpicture}
}
\caption{An example illustration of the CGB instance created in the reduction in \thmref{thm:ic_cgb_npc}. In addition to the drawn qualifications, all individuals qualify themselves and the individuals in the first column, i.e.\@ $\{ a^1_\ast, \ldots, a^{k+1}_\ast \}$. Also, every individual from $N_X$ qualifies all other individuals.}
\label{fig:ic_cgb_npc}
\end{figure}
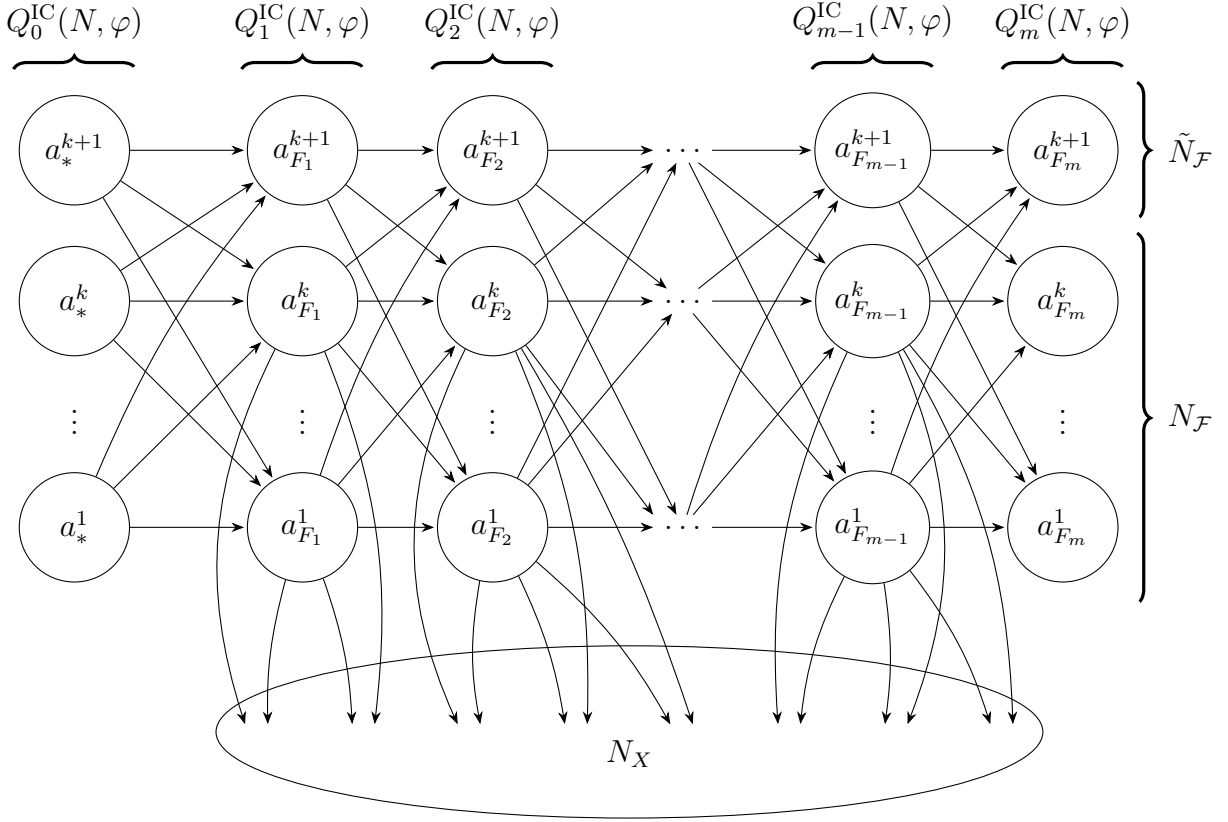

\begin{enumerate}[
    topsep=-4pt,
    leftmargin=*,
    itemindent=4.96em,
    label={\textit{Fact \ref*{thm:ic_cgb_npc}.\arabic*}:},
    ref={\ref*{thm:ic_cgb_npc}.\arabic*}
]
\item \label{fact:ic_cgb_first_column}
The individuals in $\{ a^1_\ast, \ldots, a^{k+1}_\ast \}$ are the only ones qualified by everyone. By construction of $\varphi$, every individual outside of $\{ a^1_\ast, \ldots, a^{k+1}_\ast \}$ is disqualified by more than $k$ individuals. Because the attacker can only bribe at most $\ell = k$ individuals in total, no individuals outside of $\{ a^1_\ast, \ldots, a^{k+1}_\ast \}$ can be made qualified by everyone through bribery.

\item \label{fact:ic_cgb_chain}
Among the individuals in $Q^{\text{IC}}_0(N, \varphi) = \{ a^1_\ast, \ldots, a^{k+1}_\ast \}$, there is a consensus that the individuals from $\{ a^1_{F_1}, \ldots, a^{k+1}_{F_1} \}$ should be qualified. For any $i \in \{ 1, \ldots, m-1 \}$, there is a consensus among the individuals in $\{ a^1_{F_i}, \ldots, a^{k+1}_{F_i} \}$ that the individuals from $\{ a^1_{F_{i+1}}, \ldots, a^{k+1}_{F_{i+1}} \}$ should be qualified. This way, all individuals from $N_\mathcal{F} \cup \tilde{N}_\mathcal{F}$ are socially qualified. Meanwhile, for any $i \in \{ 1, \ldots, m \}$, there is no consensus among the individuals in $\{ a^1_{F_i}, \ldots, a^{k+1}_{F_i} \}$ that someone in $N_X$ should be qualified (the individuals in $N_\mathcal{F}$ qualify some individuals in $N_X$, but the individuals in $\tilde{N}_\mathcal{F}$ do not). Therefore, we initially have $f^{\text{IC}}(N, \varphi) = \{ a^1_\ast, \ldots, a^{k+1}_\ast \} \cup N_\mathcal{F} \cup \tilde{N}_\mathcal{F}$.

\item \label{fact:ic_cgb_a_fi_must_be_in_q_i}
Let $\varphi^\prime$ be a binary profile obtained from $\varphi$ by bribing at most $k$ individuals. For any individual $a^j_{F_i}$ where $i \in \{ 1, \ldots, m \}$ and $j \in \{ 1, \ldots, k+1 \}$, it must hold that $a^j_{F_i} \in Q^{\text{IC}}_i(N, \varphi^\prime)$, i.e.\@ $a^j_{F_i}$ must become socially qualified in the $i$-th iteration. Assume for the sake of contradiction that this does not hold, and let $a^j_{F_i}$ be the first individual (i.e.\@ with the smallest $i$) from $N_\mathcal{F} \cup \tilde{N}_\mathcal{F}$ who does not become socially qualified in the $i$-th iteration. Because $a^j_{F_i} \in A^+$, they must become socially qualified at some point. But to create a consensus among the individuals in an iteration (before the $(i-1)$-th iteration where such a consensus already exists) that $a^j_{F_i}$ should be qualified, the attacker would need to bribe at least $k+1$ individuals which would exceed the budget of $\ell = k$.

\item \label{fact:ic_cgb_chain_ends}
Because $\{ a^1_{F_m}, \ldots, a^{k+1}_{F_m} \} \subseteq Q^{\text{IC}}_m(N, \varphi)$ and each individual in $\{ a^1_{F_m}, \ldots, a^{k+1}_{F_m} \}$ only qualifies themselves and the individuals from $\{ a^1_\ast, \ldots, a^{k+1}_\ast \}$, the iterative process always ends after the $m$-th iteration, regardless of who is bribed.

\item \label{fact:ic_cgb_must_bribe_in_n_f}
To turn the individuals from $N_X \subseteq A^+$ socially qualified, the attacker needs to ensure that, for each $a_x \in N_X$, there is at some point a consensus among the individuals in a set $Q^{\text{IC}}_i(N, \varphi^\prime)$ that $a_x$ should be qualified. From \factref{fact:ic_cgb_chain_ends}, we know that the iterative process goes through $m$ iterations. For every $i \in \{ 1, \ldots, m \}$, we know from \factref{fact:ic_cgb_a_fi_must_be_in_q_i} that $a^{k+1}_{F_i} \in Q^{\text{IC}}_i(N, \varphi^\prime)$. The individual $a^{k+1}_{F_i}$ is in $\tilde{N}_\mathcal{F}$ and, thus, disqualifies everyone in $N_X$. Therefore, the only way to achieve the constructive objective is by bribing certain individuals from $\tilde{N}_\mathcal{F}$ to qualify the individuals in $N_X$.

\item \label{fact:ic_cgb_bribe_some_n_f}
Once the attacker bribes an individual $a^{k+1}_{F_i} \in \tilde{N}_\mathcal{F}$ where $i \in \{ 1, \ldots, m \}$ to qualify the individuals in $N_X$, there is a consensus among everyone in $Q^{\text{IC}}_i(N, \varphi^\prime)$ that the individuals from $\{ a_x : x \in F_i \}$ should be qualified, i.e.\@ $\{ a_x : x \in F_i \} \subseteq Q^{\text{IC}}_{i+1}(N, \varphi^\prime)$. Because the individuals from $N_X$ qualify everyone, this does not cause any changes in subsequent iterations: Among the individuals in $Q^{\text{IC}}_{i+1}(N, \varphi^\prime)$, there still exists a consensus that the individuals in $\{ a^1_{F_{i+2}}, \ldots, a^{k+1}_{F_{i+2}} \}$ should be qualified. Thus, the argument also applies if multiple individuals from $\tilde{N}_\mathcal{F}$ are bribed: For each bribed individual $a^{k+1}_{F_i}$, the individuals from $\{ a_x : x \in F_i \}$ become socially qualified.
\end{enumerate}

Equipped with these facts, we now show that the constructed instance is a YES-instance if and only if there exists a set cover for $X$ in $\mathcal{F}$ of cardinality at most $k$.

($\Rightarrow$) Assume that $\mathcal{F}^\prime \subseteq \mathcal{F}$ is a set cover for $X$ with $|\mathcal{F}^\prime| \leq k$. We bribe each individual in $\{ a^{k+1}_{F_i} : {F_i} \in \mathcal{F}^\prime \}$ to qualify everyone from $N_X$. From \factref{fact:ic_cgb_bribe_some_n_f} and the fact that $\mathcal{F}^\prime$ is a set cover for $X$, we know that all individuals from $N_X$ are now socially qualified. From \factref{fact:ic_cgb_chain}, we know that the individuals in $\{ a^1_\ast, \ldots, a^{k+1}_\ast \} \cup N_\mathcal{F} \cup \tilde{N}_\mathcal{F}$ are socially qualified too. Thus, all individuals in $A^+ = N$ are now socially qualified.

($\Leftarrow$) Assume we are given a CGB solution $U \subseteq N$ with $|U| \leq \ell$ such that all individuals in $A^+$ are socially qualified after bribing the individuals in $U$. From \factref{fact:ic_cgb_must_bribe_in_n_f}, we know that $U \subseteq \tilde{N}_\mathcal{F}$. Moreover, from \factref{fact:ic_cgb_bribe_some_n_f} and the fact that $N_X \subseteq f^{\text{IC}}(N, \varphi^\prime)$, we know that for each $x \in X$, there exists a set $F_i \in \mathcal{F}$ with $x \in F_i$ for which the individual $a^{k+1}_{F_i}$ is in $U$. From this, it follows that $\mathcal{F}^\prime = \{ F_i \in \mathcal{F} : a^{k+1}_{F_i} \in U \}$ is a set cover for $X$ of size at most $\ell = k$.
\end{Proof}

By extending the above reduction, we can also show that the general and exact group bribery problems with $A^+ \neq \emptyset$ and $A^- \neq \emptyset$ are NP-complete and W[2]-hard:

\begin{Corollary} \label{cor:ic_gb_npc}
$f^{\text{IC}}$-\textsc{Exact Group Bribery} and $f^{\text{IC}}$-\textsc{Group Bribery} with $A^+ \neq \emptyset$ and $A^- \neq \emptyset$ are NP-complete and W[2]-hard with respect to $\ell$.
\end{Corollary}

\begin{Proof}
We first apply the same reduction from \textsc{Set Cover} as in \thmref{thm:ic_cgb_npc}. We then make the following minor modifications to obtain an EGB/GB instance with $A^- \neq \emptyset$:

\begin{itemize}[nolistsep]
\item
We create one additional individual $d$.
\item
The individual $d$ qualifies only themselves and everyone in $\{ a^1_\ast, \ldots, a^{k+1}_\ast \}$, and is qualified only by the individuals in $\{ a^1_{F_m}, \ldots, a^{k+1}_{F_m} \} \cup N_X$.
\item
We define $A^- = \{d\}$ and keep the definition of $A^+ = N \setminus \{d\}$.
\item
We increase the budget $\ell$ by $1$, i.e.\@ we let $\ell = k+1$.
\end{itemize}

Initially, there is a consensus among the individuals from $Q^{\text{IC}}_m(N, \varphi) = \{ a^1_{F_m}, \ldots, a^{k+1}_{F_m} \}$ that $d$ should be qualified. Since $d$ also qualifies themselves, $d$ becomes socially qualified in the $(m+1)$-th iteration.

Since $\{ a^1_{F_m}, \ldots, a^{k+1}_{F_m} \} \subseteq A^+$, it always holds that the individuals in $\{ a^1_{F_m}, \ldots, a^{k+1}_{F_m} \}$ are socially qualified in the $m$-th iteration (see \factref{fact:ic_cgb_a_fi_must_be_in_q_i}). Therefore, to turn $d$ socially disqualified, the attacker must either bribe an individual from $Q^{\text{IC}}_m(N, \varphi)$ to disqualify $d$, or bribe $d$ to disqualify themselves. However, in both scenarios, the attacker does not make any gains with regards to the constructive objective (i.e.\@ turning the individuals from $N_X$ socially qualified). This is because we assumed $F_m$ to be the empty set. Hence, the individuals in $\{ a^1_{F_m}, \ldots, a^{k+1}_{F_m} \}$ do not qualify anyone from $N_X$, so the attacker cannot use them to create a consensus that someone from $N_X$ should be qualified.

The bribery of $d$ or someone from $Q^{\text{IC}}_m(N, \varphi)$ decreases the budget by $1$, so solving the remaining instance is again equivalent to solving the constructive case.
\end{Proof}

For the social rules $f^{\text{2IC}}$ and $f^{\text{2LIC}}$, the constructive group bribery problem (including the priced version) is polynomial-time solvable. We first show this for the $f^{\text{2IC}}$ rule:

\begin{Theorem} \label{thm:2ic_cgb_p}
$f^{\text{2IC}}$-\textsc{\$Constructive Group Bribery} can be solved in time $\mathcal{O}(n^3)$.
\end{Theorem}

\begin{Proof}
Let $(N, \varphi, A^+, \rho, \ell)$ be an instance of $f^{\text{2IC}}$-\$CGB. We begin by bribing each individual in $A^+$ who disqualifies themselves to qualify themselves. This is always necessary because under the $f^{\text{2IC}}$ rule, individuals who disqualify themselves cannot become socially qualified.

For all individuals from $A^+$ to be socially qualified, at least one individual must be qualified by everyone (since otherwise $K^{\text{2IC}}_0(N, \varphi^\prime) = \emptyset$). Thus, we guess an individual $a_\ast \in N$ to be qualified by everyone. If it is not already the case, we bribe $a_\ast$ to qualify themselves and to qualify all individuals from $A^+$ (assuming $a_\ast$ was guessed correctly, this is always necessary). Moreover, if $a_\ast$ is not yet qualified by everyone, we bribe every individual who disqualifies $a_\ast$ to qualify $a_\ast$ (again, this is always necessary). Let $\varphi^\prime$ denote the resulting profile. We now compute the set $K^{\text{2IC}}_0(N, \varphi^\prime)$ of individuals who are qualified by everyone. If there is a consensus among the individuals in $K^{\text{2IC}}_0(N, \varphi^\prime)$ that everyone from $A^+$ should be qualified, we are done. Otherwise, we create such a consensus by making some additional bribes. To do this, we consider two cases:

\begin{enumerate}[
    leftmargin=*,
    label={\textit{Case \ref*{thm:2ic_cgb_p}.\arabic*}:}
]
\item
There exists a nonempty subset $A^+_\ast \subseteq A^+ \cap K^{\text{2IC}}_0(N, \varphi^\prime)$ of individuals in $A^+ \cap K^{\text{2IC}}_0(N, \varphi^\prime)$ who disqualify someone from $A^+$. \\
In this case, we can try to bribe the individuals in $A^+_\ast$ to qualify everyone from $A^+$, and check if this creates a consensus among all individuals in $K^{\text{2IC}}_0(N, \varphi^\prime)$ that the individuals in $A^+$ should be qualified. Alternatively, if $|A^+_\ast| \geq 2$, we can bribe the two individuals with the lowest bribery prices in $A^+_\ast$ to qualify only themselves. This makes $a_\ast$ the only individual who is qualified by everyone, so all individuals from $A^+$ are socially qualified after the first iterative step. However, if these approaches do not work, then the optimal solution is to bribe someone to disqualify all individuals in $K^{\text{2IC}}_0(N, \varphi^\prime)$ except $a_\ast$. This way, $a_\ast$ becomes the only individual who is qualified by everyone, making all individuals from $A^+$ socially qualified. However, note that we cannot bribe $a_\ast$ to do this because $a_\ast$ must qualify all individuals from $A^+$ (which includes the individuals in $A^+_\ast$). Likewise, we cannot bribe an individual from $A^+_\ast$ to do this since they must qualify themselves. Hence, we must bribe someone from $N \setminus (\{a_\ast\} \cup A^+_\ast)$. If we have already bribed some individuals from $N \setminus (\{a_\ast\} \cup A^+_\ast)$ up to this point, we can simply bribe them to also disqualify the individuals in $K^{\text{2IC}}_0(N, \varphi^\prime) \setminus \{a_\ast\}$ for no additional cost. Else, we bribe the individual with the lowest bribery price among all individuals in $N \setminus (\{a_\ast\} \cup A^+_\ast)$.

\item
There is no individual in $A^+ \cap K^{\text{2IC}}_0(N, \varphi^\prime)$ who disqualifies someone from $A^+$. \\
In this case, we bribe someone to disqualify all individuals in $K^{\text{2IC}}_0(N, \varphi^\prime) \setminus (A^+ \cup \{a_\ast\})$. After that, the only individuals who are qualified by everyone are $a_\ast$ and possibly some individuals from $A^+$. Among them, there exists a consensus that all individuals from $A^+$ should be qualified, thus we are done. If we have already bribed some individuals up to this point, we can simply bribe them to also disqualify the individuals in $K^{\text{2IC}}_0(N, \varphi^\prime) \setminus (A^+ \cup \{a_\ast\})$ for no additional cost. Otherwise, we bribe the individual with the lowest bribery price among all individuals.
\end{enumerate}

If there exists an individual $a_\ast \in N$ for whom the total cost of the bribery described above is at most $\ell$, the given instance is a YES-instance. Otherwise, it must be a NO-instance because it is easy to verify that all the bribes described are necessary and cost-optimal.

The running time of the algorithm is bounded by $\mathcal{O}(n \cdot n^2)$ where the additional factor of $n$ comes from guessing an individual $a_\ast \in N$ to be qualified by everyone.
\end{Proof}

When the $f^{\text{2LIC}}$ social rule is used, the constructive group bribery problem can be solved even faster:

\begin{Theorem} \label{thm:2lic_cgb_p}
$f^{\text{2LIC}}$-\textsc{\$Constructive Group Bribery} can be solved in time $\mathcal{O}(n^2)$.
\end{Theorem}

\begin{Proof}
To solve a given instance $(N, \varphi, A^+, \rho, \ell)$ of $f^{\text{2LIC}}$-\$CGB, we do the following:

Let $A^+_{-1} = \{ a \in A^+ : \varphi(a, a) = -1 \}$ denote the subset of individuals in $A^+$ who initially disqualify themselves. Clearly, it is always necessary to bribe the individuals from $A^+_{-1}$ because only individuals who qualify themselves can become socially qualified. We distinguish between two cases:

\begin{enumerate}[
    leftmargin=*,
    label={\textit{Case \ref*{thm:2lic_cgb_p}.\arabic*}:}
]
\item
$|A^+_{-1}| \geq 2$, i.e.\@ at least two individuals in $A^+$ disqualify themselves. \\
In this case, we bribe each individual $a \in A^+_{-1}$ to qualify themselves and to disqualify all other individuals. After that, no individual is qualified by everyone, and all individuals in $A^+$ qualify themselves. Thus, we are done.

\item
$|A^+_{-1}| \leq 1$, i.e.\@ at most one individual in $A^+$ disqualifies themselves. \\
If $A^+_{-1} = \emptyset$, we let $c_1$ and $c_2$ be the individuals with the lowest and second lowest bribery price among all individuals, respectively. If $A^+_{-1} = \{a\}$, we instead let $c_1 = a$ and let $c_2$ be the individual with the lowest bribery price among all individuals in $N \setminus \{a\}$; and we then bribe $a$ to qualify themselves (this is always necessary).

In the next step, we compute the set $K^{\text{2LIC}}_0(N, \varphi^\prime)$ of individuals who are now qualified by everyone. If this set is empty or if there is a consensus among everyone in $K^{\text{2LIC}}_0(N, \varphi^\prime)$ that the individuals in $A^+$ should be qualified, we are done. Otherwise, there are multiple ways how we might achieve the objective: We could bribe $c_1$ to disqualify all individuals (this way, no one is qualified by everyone, and all individuals who qualify themselves become socially qualified). However, if the individual $c_1$ is in $A^+$, then we cannot bribe $c_1$ to disqualify themselves. Thus, if $c_1$ is also qualified by all other individuals, then $c_1$ would still be qualified by everyone after the bribery. In that case, we could bribe $c_1$ to qualify only the individuals in $A^+$, and check if this creates a consensus in $K^{\text{2LIC}}_0(N, \varphi^\prime)$ that all individuals in $A^+$ should be qualified. But if this does not work, we may choose to bribe the individual $c_2$ instead of $c_1$. However, if $c_2$ is also in $A^+$ and qualified by all other individuals, we would have the same problem as with $c_1$. We could fix this by bribing each of $c_1$ and $c_2$ to disqualify all other individuals (this way, no one is qualified by everyone). But instead of doing that, we could also bribe an individual who is not in $A^+$ or who is not qualified by all other individuals to disqualify everyone. Let $c_3$ be the cheapest such individual. If among the individuals who are qualified by everyone, there is only one individual $c_4$ who does not yet qualify all individuals from $A^+$, then we could also try to bribe this individual $c_4$ to qualify everyone from $A^+$. In conclusion, we solve the instance by picking the cheapest among all option: (a) bribing only $c_1$ (if possible), or (b) bribing only $c_2$ (if possible), or (c) bribing both $c_1$ and $c_2$, or (d) bribing $c_3$, or (e) bribing $c_4$ (if applicable).

If the instance is a YES-instance, then the above approach ensures that we find an optimal solution. To see this, let $U \subseteq N$ be any optimal solution. If $|U| \geq 2$, it is obvious that bribing each of $c_1$ and $c_2$ is at most as expensive as bribing the individuals in $U$ (by definition of $c_1$ and $c_2$). Hence, our solution in this case is optimal. If $|U| \leq 1$, it implies that we can achieve the objective by bribing only one individual. If it suffices to bribe $c_1$ or $c_2$, then doing so is obviously optimal (by definition of $c_1$ and $c_2$). Otherwise, the only way to solve the instance with a single bribe is by bribing an individual who is not in $A^+$ or who is not qualified by everyone (we cover this by trying $c_3$), or alternatively by bribing an individual who is in $A^+$ and qualified by everyone, provided that there is only one such individual who does not yet qualify everyone from $A^+$ (we cover this by trying $c_4$). Either way, our solution is optimal.
\end{enumerate}

It is easy to verify that all of the above steps can be completed in time $\mathcal{O}(n^2)$. The given instance is a YES-instance if and only if the budget suffices for these bribes.
\end{Proof}

Next, we turn to destructive group bribery. For the $f^{\text{IC}}$ and $f^{\text{2IC}}$ social rules, there is a simple polynomial-time algorithm for solving \$DGB:

\begin{Theorem} \label{thm:2ic_dgb_p}
For $f \in \{ f^{\text{IC}}, f^{\text{2IC}} \}$, the problem $f$-\textsc{\$Destructive Group Bribery} can be solved in time $\mathcal{O}(n)$.
\end{Theorem}

\begin{Proof}
Let $(N, \varphi, A^-, \rho, \ell)$ be an instance of $f^{\text{IC}} / f^{\text{2IC}}$-\$DGB. To make the individuals in $A^-$ socially disqualified, we obviously need to bribe at least one individual (otherwise we would be done already). Let $a$ be the individual with the lowest bribery price among all individuals. Clearly, $a$ can be determined in time $\mathcal{O}(n)$. We bribe $a$ to disqualify everyone, including themselves. Afterwards, no individual is qualified by everyone; thus, everyone is socially disqualified. The instance is a YES-instance if and only if $\rho(a) \leq \ell$.
\end{Proof}

When the $f^{\text{2LIC}}$ social rule is used, the above algorithm does not work. This is because if no individual is qualified by everyone, the $f^{\text{2LIC}}$ rule considers all individuals who qualify themselves to be socially qualified. However, we will later show in \thmref{thm:2lic_gb_p} that the unpriced group bribery problem for the $f^{\text{2LIC}}$ rule is still polynomial-time solvable. But before that, we show that the priced destructive problem $f^{\text{2LIC}}$-\$DGB is NP-complete and W[2]-hard with respect to $\ell$. The proof is based on a reduction from \textsc{Set Cover}, similar to the one for DGCDI in \thmref{thm:ic_dgcdi_npc}:

\begin{Theorem} \label{thm:2lic_dgb_npc}
$f^{\text{2LIC}}$-\textsc{\$Destructive Group Bribery} is NP-complete and W[2]-hard with respect to $\ell$.
\end{Theorem}

\begin{Proof}
Let $(X, \mathcal{F}, k)$ be an instance of \textsc{Set Cover}. Without loss of generality, we can assume that $|\mathcal{F}| > k$ (otherwise the instance would be a trivial YES-instance). Moreover, we can assume that every element from $X$ appears in at least one set from $\mathcal{F}$ (otherwise the instance would be a trivial NO-instance). We now construct an equivalent instance of $f^{\text{2LIC}}$-\$DGB as follows:

For each element $x \in X$, we introduce one individual $a_x$. Let $N_X = \{ a_x : x \in X \}$. For each set $F \in \mathcal{F}$, we introduce two individuals $a_F$ and $\tilde{a}_F$. Let $N_\mathcal{F} = \{ a_F : F \in \mathcal{F} \}$ and $\tilde{N}_\mathcal{F} = \{ \tilde{a}_F : F \in \mathcal{F} \}$. We then set $N = N_X \cup N_\mathcal{F} \cup \tilde{N}_\mathcal{F}$ and define the profile $\varphi$ over $N$ as follows:

\begin{itemize}[nolistsep]
\item
All individuals in $N_X \cup N_\mathcal{F}$ qualify themselves.
\item
All individuals in $\tilde{N}_\mathcal{F}$ disqualify themselves.
\item
For each $x \in X$, the individual $a_x$ qualifies all other individuals.
\item
For each $F \in \mathcal{F}$, the individual $a_F$ qualifies all other individuals except the individuals $a_x \in N_X$ with $x \in F$.
\item
For each $F \in \mathcal{F}$, the individual $\tilde{a}_F$ qualifies all other individuals except $a_F$.
\end{itemize}

To conclude the reduction, we set $A^- = \tilde{N}_\mathcal{F} \cup N_X$ and $\ell = k$; and we define the bribery price function $\rho : N \rightarrow \mathbb{N}$ as follows:

\begin{itemize}[nolistsep]
\item
For each $x \in X$, we let $\rho(a_x) = \ell + 1$.
\item
For each $F \in \mathcal{F}$, we let $\rho(a_F) = \ell + 1$ and $\rho(\tilde{a}_F) = 1$.
\end{itemize}

See \figref{fig:2lic_dgb_npc} for an example illustration of the \$DGB instance created in this reduction. Clearly, the reduction can be completed in polynomial time. Before we show the correctness of the reduction, it will be useful to observe a few simple facts:

\begin{figure}[!tbh]
\centering
\raisebox{0.04\height}{
\begin{tikzpicture}[
    > = Stealth,
    shorten > = 1pt,
    auto,
    node distance = 2.5cm,
    main node/.style = {circle, draw, minimum size = 1.45cm}
]

\node[main node] (ft1)                                          {$\tilde{a}_{F_1}$};
\node[main node] (f1)   [below of = ft1]                        {$a_{F_1}$};
\node[main node] (ft2)  [right of = ft1]                        {$\tilde{a}_{F_2}$};
\node[main node] (f2)   [right of = f1]                         {$a_{F_2}$};
\node[main node] (ft3)  [right of = ft2]                        {$\tilde{a}_{F_3}$};
\node[main node] (f3)   [right of = f2]                         {$a_{F_3}$};
\node            (ft4)  [right of = ft3, node distance = 1.7cm] {$\ldots$};
\node            (f4)   [right of = f3, node distance = 1.7cm]  {$\ldots$};
\node[main node] (ftm1) [right of=ft4, node distance=1.7cm] {$\tilde{a}_{F_{m-1}}$};
\node[main node] (fm1)  [right of = f4, node distance = 1.7cm]  {$a_{F_{m-1}}$};
\node[main node] (ftm)  [right of = ftm1]                       {$\tilde{a}_{F_m}$};
\node[main node] (fm)   [right of = fm1]                        {$a_{F_m}$};

\node (x1)      [below of = f1]                         {};
\node (x2)      [right of = x1, node distance = 1cm]    {};
\node (x3)      [right of = x2, node distance = 1cm]    {};
\node (x4)      [right of = x3, node distance = 1cm]    {};
\node (x5)      [right of = x4, node distance = 1cm]    {};
\node (x6)      [right of = x5, node distance = 1cm]    {};
\node (x7)      [right of = x6, node distance = 1cm]    {};
\node (x8)      [right of = x7, node distance = 1cm]    {};
\node (x9)      [right of = x8, node distance = 1cm]    {};
\node (x10)     [right of = x9, node distance = 1cm]    {};
\node (x11)     [right of = x10, node distance = 1cm]   {};
\node (x12)     [right of = x11, node distance = 1cm]   {};
\node (nxn)     [right of = x6, node distance = 0.5cm, yshift = 0.25cm] {};
\node (nxs)     [right of = x6, node distance = 0.5cm, yshift = -0.5cm] {$N_X$};
\node[ellipse, draw, fit={(x3)(nxn)(nxs)(x10)}, inner sep=1mm] (x) {};

\path[->, dashed]
(ft1)   edge[loop above]        node {} (ft1)
        edge                    node {} (f1)
(ft2)   edge[loop above]        node {} (ft2)
        edge                    node {} (f2)
(ft3)   edge[loop above]        node {} (ft3)
        edge                    node {} (f3)
(ftm1)  edge[loop above]        node {} (ftm1)
        edge                    node {} (fm1)
(ftm)   edge[loop above]        node {} (ftm)
        edge                    node {} (fm)
;

\path[->, dashed]
(f1)    edge                    node {} (x2.east)
        edge                    node {} (x3.west)
(f2)    edge                    node {} (x3.east)
        edge                    node {} (x4.east)
        edge                    node {} (x5.west)
        edge                    node {} (x6.west)
(f3)    edge                    node {} (x6.east)
        edge                    node {} (x7.east)
        edge                    node {} (x8.west)
(fm1)   edge                    node {} (x8.east)
        edge                    node {} (x9.east)
        edge                    node {} (x10.west)
(fm)    edge                    node {} (x10.east)
        edge                    node {} (x11.west)
;

\draw [very thick, decorate, decoration = {brace, raise=5pt, amplitude=5pt}]
(-1, -6.4) -- (-1, -3.8)
node[pos = 0.5, left = 12pt, black] {$A^-$};

\draw [very thick, decorate, decoration = {brace, raise=5pt, amplitude=5pt}]
(-1, -1.1) -- (-1, 1.1)
node[pos = 0.5, left = 12pt, black] {$A^-$};

\end{tikzpicture}
}
\caption{An example illustration of the \$DGB instance created in the reduction in \thmref{thm:2lic_dgb_npc}. In the picture, we assume that $\mathcal{F}$ is of the form $\{ F_1, F_2, \ldots, F_m \}$. Every individual qualifies themselves and all other individuals, except for the drawn dashed arcs which represent disqualifications.}
\label{fig:2lic_dgb_npc}
\end{figure}

\begin{enumerate}[
    topsep=-4pt,
    leftmargin=*,
    itemindent=4.96em,
    label={\textit{Fact \ref*{thm:2lic_dgb_npc}.\arabic*}:},
    ref={\ref*{thm:2lic_dgb_npc}.\arabic*}
]
\item \label{fact:2lic_dgb_initial_situation}
Initially, no individual is qualified by everyone, i.e.\@ $K^{\text{2LIC}}_0(N, \varphi) = \emptyset$. Therefore, the socially qualified individuals are exactly the ones who qualify themselves, i.e.\@ $f^{\text{2LIC}}(N, \varphi) = N_X \cup N_\mathcal{F}$. In particular, this means that all individuals from $\tilde{N}_\mathcal{F} \subseteq A^-$ are already socially disqualified initially because they disqualify themselves.

\item \label{fact:2lic_dgb_cannot_bribe_in_n_x_or_n_f}
All individuals in $N_X$ and $N_\mathcal{F}$ have a bribery price of $\ell + 1$, so the attacker cannot afford to bribe any of them. In particular, this implies that all individuals from $N_X$ still qualify themselves after the bribery.

\item \label{fact:2lic_dgb_need_individuals_qualified_by_everyone}
From \factref{fact:2lic_dgb_cannot_bribe_in_n_x_or_n_f}, it follows that the only way to turn the individuals in $N_X$ socially disqualified is by having some individual(s) qualified by everyone, but having no consensus among them that anyone in $N_X$ should be qualified. Clearly, only individuals from $N_\mathcal{F}$ can be qualified by everyone because all individuals outside of $N_\mathcal{F}$ are part of the destructive target set $A^-$.

\item \label{fact:2lic_dgb_bribe_in_n_f_how}
We know from \factref{fact:2lic_dgb_cannot_bribe_in_n_x_or_n_f} that, in any solution for the created instance, all bribed individuals are from $\tilde{N}_\mathcal{F}$. If some individual $\tilde{a}_F$ where $F \in \mathcal{F}$ is bribed, it is always optimal to let them qualify the individual $a_F$. Doing so creates a consensus among everyone that $a_F$ should be qualified. Any further changes to the outgoing qualifications of $\tilde{a}_F$ do not contribute to achieving the objective and are therefore not necessary. This is because the individual $\tilde{a}_F$ is in $A^-$, so they cannot be among the individuals who are qualified by everyone. Hence, the opinions of $\tilde{a}_F$ about the individuals in $N_X$ are irrelevant.

\item \label{fact:2lic_dgb_bribe_in_n_f_what}
After bribing an individual $\tilde{a}_F$ where $F \in \mathcal{F}$ to qualify $a_F$, the individual $a_F$ becomes qualified by everyone. Thus, there is no consensus among the individuals who are qualified by everyone that the individuals in $\{ a_x : x \in F \}$ should be qualified (since $a_F$ disqualifies them). This way, the individuals in $\{ a_x : x \in F \}$ turn socially disqualified.
\end{enumerate}

Equipped with these facts, we now show that the constructed instance is a YES-instance if and only if there exists a set cover for $X$ in $\mathcal{F}$ of cardinality at most $k$.

($\Rightarrow$) Assume that $\mathcal{F}^\prime \subseteq \mathcal{F}$ is a set cover for $X$ with $|\mathcal{F}^\prime| \leq k$. For each $F \in \mathcal{F}^\prime$, we bribe the individual $\tilde{a}_F$ to qualify $a_F$. Clearly, the cost of this bribery is at most $k = \ell$. From \factref{fact:2lic_dgb_bribe_in_n_f_what} and the fact that $\mathcal{F}^\prime$ is a set cover for $X$, we know that all individuals from $N_X$ are now socially disqualified. Furthermore, from \factref{fact:2lic_dgb_initial_situation}, we know that the individuals from $\tilde{N}_\mathcal{F}$ are socially disqualified too.

($\Leftarrow$) Assume we are given a minimal \$DGB solution $U \subseteq N$ with $\rho(U) \leq \ell$ such that all individuals from $A^-$ are socially disqualified after bribing the individuals in $U$. From \factref{fact:2lic_dgb_cannot_bribe_in_n_x_or_n_f}, we know that $U \subseteq \tilde{N}_\mathcal{F}$. By definition of the bribery price function $\rho$, this implies that $|U| \leq \ell$. From \factref{fact:2lic_dgb_bribe_in_n_f_how}, we know that the only individuals who are qualified by everyone after the bribery are the ones in $\{ a_F : \tilde{a}_F \in U \}$. In conjunction with \factref{fact:2lic_dgb_bribe_in_n_f_what} and the fact that $N_X \cap f^{\text{2LIC}}(N, \varphi^\prime) = \emptyset$, this implies that for each $x \in X$, there exists a set $F \in \mathcal{F}$ with $x \in F$ for which the individual $\tilde{a}_F$ is in $U$. From this, it follows that $\mathcal{F}^\prime = \{ F \in \mathcal{F} : \tilde{a}_F \in U \}$ is a set cover for $X$ of size at most $\ell = k$.
\end{Proof}

It is possible to alter the above reduction in such a way that all bribery prices are at most $\ell$, i.e.\@ there are no individuals whom the attacker cannot afford to bribe outright. However, this requires a much more involved argumentation in the proof, which is why we refrain from doing so in this thesis.

By extending the above reduction, we can also show that \$EGB and \$GB are NP-complete and W[2]-hard for the social rules $f^{\text{2IC}}$ and $f^{\text{2LIC}}$:

\begin{Corollary} \label{cor:2ic_egb_gb_npc}
For $f \in \{ f^{\text{2IC}}, f^{\text{2LIC}} \}$, the problems $f$-\textsc{\$Exact Group Bribery} and $f$-\textsc{\$Group Bribery} with $A^+ \neq \emptyset$ and $A^- \neq \emptyset$ are NP-complete and W[2]-hard with respect to $\ell$.
\end{Corollary}

\begin{Proof}
We first apply the same reduction from \textsc{Set Cover} as in \thmref{thm:2lic_dgb_npc}. We then make the following minor modifications to obtain an \$EGB/\$GB instance with $A^+ \neq \emptyset$:

\begin{itemize}[nolistsep]
\item
We create two additional individuals $a_\ast$ and $q$.
\item
The individual $a_\ast$ qualifies everyone, and is qualified by everyone.
\item
The individual $q$ qualifies everyone, except themselves. All individuals in $\tilde{N}_\mathcal{F}$ disqualify $q$, and all other individuals qualify $q$.
\item
We define $A^+ = \{ a_\ast, q \} \cup N_\mathcal{F}$ and keep the definition of $A^- = \tilde{N}_\mathcal{F} \cup N_X = N \setminus A^+$.
\item
We increase the budget $\ell$ by $1$, i.e.\@ we let $\ell = k+1$.
\item
We set the bribery prices for $a_\ast$ and $q$ to $\rho(a_\ast) = \ell$ and $\rho(q) = 1$.
\end{itemize}

Because the individual $a_\ast$ is qualified by everyone, we now have $K_0(N, \varphi) = \{a_\ast\}$. Since $a_\ast$ qualifies all other individuals, the individuals from $N_X \cup N_\mathcal{F}$ are also socially qualified. However, the individuals from $\tilde{N}_\mathcal{F} \cup \{q\}$ are socially disqualified because they disqualify themselves. We therefore initially have the same set of socially qualified individuals as in \thmref{thm:2lic_dgb_npc}, with the addition of $a_\ast$, i.e.\@ $f(N, \varphi) = N_X \cup N_\mathcal{F} \cup \{a_\ast\}$. The attacker now has to turn the individual $q$ socially qualified and turn everyone in $N_X$ socially disqualified.

Clearly, to turn $q$ socially qualified, the attacker is always forced to bribe $q$ to qualify themselves (because only individuals who qualify themselves can be socially qualified). After that, $q$ automatically becomes socially qualified since all individuals who are qualified by everyone already qualify $q$ (including the individuals from $N_\mathcal{F}$ who might become qualified by everyone through the bribery). Any additional changes to the outgoing qualifications of $q$ do not contribute to achieving the objective and are therefore not necessary. This is because $q$ is disqualified by all individuals from $\tilde{N}_\mathcal{F}$, and we assumed that $|\mathcal{F}| > k$. Since each individual in $\tilde{N}_\mathcal{F}$ has a bribery price of $1$, it is therefore impossible to make $q$ qualified by everyone through bribery with the remaining budget. Hence, the opinions of $q$ about the individuals in $N_X$ are irrelevant.

The bribery of $q$ decreases the budget by $\rho(q) = 1$, so the available budget for achieving the destructive objective is only $\ell - 1 = k$. Obviously, the remaining budget does not suffice for bribing any individuals in $N_X \cup N_\mathcal{F} \cup \{a_\ast\}$ because they all have a bribery price of $k+1$. Therefore, the attacker can only bribe individuals in $\tilde{N}_\mathcal{F}$.

Clearly, to achieve the objective, there must be at least one individual who is qualified by everyone. For the $f^{\text{2IC}}$ rule, this is because we have $A^+ \neq \emptyset$ (and if no individual is qualified by everyone, then no one can be socially qualified). For the $f^{\text{2LIC}}$ rule, this is because all individuals in $N_X \subseteq A^-$ qualify themselves (and they cannot be bribed to disqualify themselves). Therefore, the only way to achieve the objective is to have some individuals qualified by everyone (in addition to $a_\ast$ who is already qualified by everyone), but having no consensus among them that anyone from $N_X$ should be qualified. Clearly, this is equivalent to solving the destructive case as described in \thmref{thm:2lic_dgb_npc}.
\end{Proof}

Unlike the priced group bribery problem, the unpriced version is polynomial-time solvable for both the $f^{\text{2IC}}$ rule and the $f^{\text{2LIC}}$ rule. Obviously, this implies that the unpriced exact and unpriced destructive problems are also polynomial-time solvable. We first show this for the $f^{\text{2LIC}}$ rule:

\begin{Theorem} \label{thm:2lic_gb_p}
$f^{\text{2LIC}}$-\textsc{Group Bribery} can be solved in time $\mathcal{O}(n^3)$.
\end{Theorem}

\begin{Proof}
To solve a given instance $(N, \varphi, A^+, A^-, \ell)$ of $f^{\text{2LIC}}$-GB, we do the following:

If $A^+ = N$, we solve the instance with the algorithm for $f^{\text{2LIC}}$-CGB from \thmref{thm:2lic_cgb_p}. Below, we assume that $A^+ \neq N$ and thus $N \setminus A^+ \neq \emptyset$.

Let $A^+_{-1} = \{ a \in A^+ : \varphi(a, a) = -1 \}$ denote the subset of individuals in $A^+$ who initially disqualify themselves. Clearly, it is always necessary to bribe the individuals from $A^+_{-1}$ because only individuals who qualify themselves can become socially qualified. Hence, if $|A^+_{-1}| > \ell$, we immediately conclude that the instance is a NO-instance.

Let $A^-_{1} = \{ a \in A^- : \varphi(a, a) = 1 \}$ denote the subset of individuals in $A^-$ who initially qualify themselves. If $|A^+_{-1}| + |A^-_{1}| \leq \ell$, we conclude that the instance is a YES-instance. To solve the instance, we bribe the individuals in $A^+_{-1}$ to qualify only themselves, and we bribe the individuals in $A^-_{1}$ to disqualify everyone, including themselves. If $A^+_{-1} = A^-_{1} = \emptyset$, we instead bribe an arbitrary individual from $N \setminus A^+$ to disqualify everyone. Afterwards, no individual is qualified by everyone; hence, the socially qualified individuals are exactly the ones who qualify themselves. Since all individuals in $A^+$ now qualify themselves, and all individuals in $A^-$ now disqualify themselves, that means we are done.

If $|A^+_{-1}| + |A^-_{1}| > \ell$, we know that after making all individuals in $A^+$ qualify themselves (which is always necessary), the remaining budget will not suffice to also make all individuals in $A^-$ disqualify themselves. Thus, the only way to achieve the objective is by making some individual(s) qualified by everyone, and having a consensus among them that the individuals in $A^+$ should be qualified, but no consensus that any individual in $A^-$ who qualifies themselves should be qualified. To do this, we guess an individual $a_\ast \in N \setminus A^-$ to be qualified by everyone. Let $U_{a_\ast} = \{ a^\prime \in N : \varphi(a^\prime, a_\ast) = -1 \}$ denote the subset of individuals who disqualify $a_\ast$. To solve the instance, we distinguish between the following cases:

\begin{enumerate}[
    leftmargin=*,
    label={\textit{Case \ref*{thm:2lic_gb_p}.\arabic*}:},
    ref={\ref*{thm:2lic_gb_p}.\arabic*}
]

\item
$|A^+_{-1} \cup U_{a_\ast}| > \ell$. \\
In this case, we immediately know that the individual $a_\ast$ was guessed wrong because we cannot afford to make all individuals in $A^+$ qualify themselves and simultaneously make $a_\ast$ qualified by everyone.

\item
$|A^+_{-1} \cup U_{a_\ast}| < \ell$. \label{case:2lic_gb_p_easy} \\
In this case, we solve the instance by bribing all individuals in $A^+_{-1}$ to qualify themselves, and bribing all individuals in $U_{a_\ast}$ to qualify $a_\ast$. Finally, we bribe $a_\ast$ to qualify only themselves and the individuals in $A^+$. If $a_\ast$ is now the only individual qualified by everyone, we are done (since $a_\ast$ qualifies all individuals from $A^+$ and disqualifies all individuals from $A^-$). But if some of the individuals in $A^+$ are also qualified by everyone, then there might not yet be a consensus among the individuals qualified by everyone that the individuals in $A^+$ should be qualified. In that case, we need to create such a consensus by making some additional bribes:

If no individuals have been bribed up to this point, or if $a_\ast$ is the only individual who has been bribed, we bribe an arbitrary individual from $N \setminus (A^+ \cup \{a_\ast\})$ to disqualify everyone except $a_\ast$ (which makes $a_\ast$ the only individual qualified by everyone). If we have already bribed at least one individual from $N \setminus (A^+ \cup \{a_\ast\})$ up to this point, we can bribe them to disqualify everyone except $a_\ast$ for no additional cost (which again makes $a_\ast$ the only individual qualified by everyone). If the only individuals who have been bribed up to this point are from $A^+$, we bribe them to qualify only themselves and $a_\ast$. However, if only a single individual in $A^+$ is bribed like this, that might not be sufficient because this individual could then still be qualified by everyone. In this case, we can bribe this individual to qualify only $a_\ast$ and the individuals in $A^+$, and check if this creates a consensus among the individuals who are qualified by everyone that the individuals from $A^+$ should be qualified. But if this does not work, we need to bribe one more arbitrary individual from $N \setminus (A^+ \cup \{a_\ast\})$ to disqualify everyone except $a_\ast$.

If the instance is a YES-instance and the individual $a_\ast$ was guessed correctly, the above approach ensures that we find a minimal solution. To see this, note that it is always necessary to bribe the individuals from $A^+_{-1} \cup U_{a_\ast}$. Clearly, if there is a solution where all bribed individuals are from $A^+_{-1} \cup U_{a_\ast}$, then we find it. Otherwise, we have seen that it suffices to bribe just one more arbitrary individual from $N \setminus (A^+ \cup \{a_\ast\})$. Either way, our solution is minimal.

\item
$|A^+_{-1} \cup U_{a_\ast}| = \ell$. \\
If $a_\ast \in A^+_{-1} \cup U_{a_\ast}$, we can solve the instance the same way as in \caseref{case:2lic_gb_p_easy}. Otherwise, we know that after bribing the individuals in $A^+_{-1} \cup U_{a_\ast}$ (which is always necessary), we cannot afford to also bribe the individual $a_\ast$. Hence, if $a_\ast$ disqualifies someone from $A^+$, then $a_\ast$ was guessed wrong.

If $a_\ast$ qualifies everyone from $A^+$, we check for each $a^\prime \in A^+_{-1} \cup U_{a_\ast}$ whether $N^{-1}_\varphi(a^\prime) \subseteq A^+_{-1} \cup U_{a_\ast}$, i.e.\@ whether all individuals who disqualify $a^\prime$ are in $A^+_{-1} \cup U_{a_\ast}$. If we find an $a^\prime \in A^+_{-1} \cup U_{a_\ast}$ where this is the case, we can solve the instance by first bribing all individuals in $A^+_{-1}$ to qualify themselves, and bribing all individuals in $U_{a_\ast}$ to qualify $a_\ast$ (i.e.\@ the same bribes as in \caseref{case:2lic_gb_p_easy}). Additionally, we bribe all individuals in $A^+_{-1} \cup U_{a_\ast}$ to qualify $a^\prime$, and we bribe $a^\prime$ to qualify only themselves, $a_\ast$, and the individuals in $A^+$. If $a_\ast$ and $a^\prime$ are now the only individuals qualified by everyone, we are done since they both qualify everyone from $A^+$, and $a^\prime$ disqualifies everyone from $A^-$. But if some of the individuals in $A^+$ are also qualified by everyone, then there might not yet be a consensus among the individuals qualified by everyone that the individuals in $A^+$ should be qualified. In that case, we create such a consensus by using the same approach as in \caseref{case:2lic_gb_p_easy} (with $a^\prime$ taking the role of $a_\ast$). After that, we are done.

If we do not find an individual $a^\prime$ who is disqualified only by individuals from $A^+_{-1} \cup U_{a_\ast}$, then we cannot be certain that each individual in $A^-$ will be disqualified by at least one individual from $K^{\text{2LIC}}_0(N, \varphi^\prime)$ after the bribery. In this case, the optimal strategy is to determine the set $U^+ = \{ a^\prime \in N \setminus A^- : \varphi(a^\prime, a^{\prime\prime}) \text{ for all } a^{\prime\prime} \in A^+ \}$ of individuals in $N \setminus A^-$ who qualify everyone from $A^+$. Any individual in $U^+$ can safely be among the individuals who are qualified by everyone (without destroying the consensus among them that the individuals from $A^+$ should be qualified). Hence, to solve the instance, we bribe the individuals in $A^+_{-1}$ to qualify only themselves and everyone from $U^+$, and we bribe the individuals in $U_{a_\ast} \setminus A^+_{-1}$ to qualify only the individuals from $U^+$. This way, we get the largest possible number of individuals in $N \setminus A^-$ to be qualified by everyone, while ensuring that there is a consensus among them that the individuals from $A^+$ should be qualified.
\end{enumerate}

The instance is a YES-instance if and only if there exists an individual $a_\ast \in N \setminus A^-$ for whom the above bribery is successful.

The running time of the algorithm is bounded by $\mathcal{O}(n \cdot n^2)$ where the additional factor of $n$ comes from guessing an individual $a_\ast \in N \setminus A^-$ to be qualified by everyone.
\end{Proof}

By slightly adapting the above algorithm, we can also make it work for the $f^{\text{2IC}}$ rule:

\begin{Corollary} \label{cor:2ic_gb_p}
$f^{\text{2IC}}$-\textsc{Group Bribery} can be solved in time $\mathcal{O}(n^3)$.
\end{Corollary}

\begin{Proof}
To solve a given instance $(N, \varphi, A^+, A^-, \ell)$ of $f^{\text{2IC}}$-GB, we do the following:

If $A^+ = N$, we can solve the instance with the algorithm for $f^{\text{2IC}}$-CGB from \thmref{thm:2ic_cgb_p}. Below, we assume that $A^+ \neq N$ and thus $N \setminus A^+ \neq \emptyset$.

If $A^+ = \emptyset$, we can solve the instance with the algorithm for $f^{\text{2IC}}$-DGB from \thmref{thm:2ic_dgb_p}. Below, we assume that $A^+ \neq \emptyset$.

Let $A^+_{-1} = \{ a \in A^+ : \varphi(a, a) = -1 \}$ denote the subset of individuals in $A^+$ who initially disqualify themselves. Clearly, it is always necessary to bribe the individuals from $A^+_{-1}$ because only individuals who qualify themselves can become socially qualified. Hence, if $|A^+_{-1}| > \ell$, we immediately conclude that the instance is a NO-instance.

Because $A^+ \neq \emptyset$, the only way to achieve the objective is by making some individual(s) qualified by everyone, and having a consensus among them that the individuals in $A^+$ should be qualified, but no consensus that any individual in $A^-$ who qualifies themselves should be qualified. To do this, we guess an individual $a_\ast \in N \setminus A^-$ to be qualified by everyone. Let $U_{a_\ast} = \{ a^\prime \in N : \varphi(a^\prime, a_\ast) = -1 \}$ denote the subset of individuals who disqualify $a_\ast$. We can now use the same case distinction as in \thmref{thm:2lic_gb_p} to solve the instance. The instance is a YES-instance if and only if there exists an individual $a_\ast \in N \setminus A^-$ for whom the described bribery is successful.

The running time of the algorithm is bounded by $\mathcal{O}(n \cdot n^2)$ where the additional factor of $n$ comes from guessing an individual $a_\ast \in N \setminus A^-$ to be qualified by everyone.
\end{Proof}

\clearpage

\subsection{Group microbribery}

To conclude this section, we now consider the group microbribery problems. We begin by showing that $f^{\text{IC}}$-CGMB is NP-complete and W[2]-hard with respect to $\ell$. Obviously, this implies that the priced version is also NP-complete and W[2]-hard. The proof is based on a reduction from \textsc{Set Cover}, very similar to the one for $f^{\text{IC}}$-CGB in \thmref{thm:ic_cgb_npc}.

\begin{Theorem} \label{thm:ic_cgmb_npc}
$f^{\text{IC}}$-\textsc{Constructive Group Microbribery} is NP-complete and W[2]-hard with respect to $\ell$.
\end{Theorem}

\begin{Proof}
Let $(X, \mathcal{F}, k)$ be an instance of \textsc{Set Cover}. We first apply the same reduction as in \thmref{thm:ic_cgb_npc}: We assume that $\mathcal{F}$ is of the form $\{ F_1, F_2, \ldots, F_m \}$ where $m = |\mathcal{F}| > k$, and that $F_m = \emptyset$. For each element $x \in X$, we introduce one individual $a_x$, and for each set $F_i$ with $i \in \{ 1, \ldots, m \}$, we introduce $k+1$ individuals $\{ a^1_{F_i}, \ldots, a^{k+1}_{F_i} \}$. We use $N_X$, $N_\mathcal{F}$ and $\tilde{N}_\mathcal{F}$ to denote these sets of individuals. Finally, we introduce further $k+1$ individuals $\{ a^1_\ast, \ldots, a^{k+1}_\ast \}$. See \figref{fig:ic_cgmb_npc} for an example illustration of the created instance. The only difference between this CGMB instance and the CGB instance from \thmref{thm:ic_cgb_npc} is that we define $A^+ = N \setminus \tilde{N}_\mathcal{F}$ (instead of $A^+ = N$). We keep the definition of $\ell = k$.

\begin{figure}[!tbh]
\centering
\raisebox{0.04\height}{
\begin{tikzpicture}[
    > = Stealth,
    shorten > = 1pt,
    auto,
    node distance = 2.5cm,
    main node/.style = {circle, draw, minimum size = 1.45cm}
]

\setcoverbriberyreductiongrid

\end{tikzpicture}
}
\caption{An example illustration of the CGMB instance created in the reduction in \thmref{thm:ic_cgmb_npc}. In addition to the drawn qualifications, all individuals qualify themselves and the individuals in the first column, i.e.\@ $\{ a^1_\ast, \ldots, a^{k+1}_\ast \}$. Also, every individual from $N_X$ qualifies all other individuals.}
\label{fig:ic_cgmb_npc}
\end{figure}
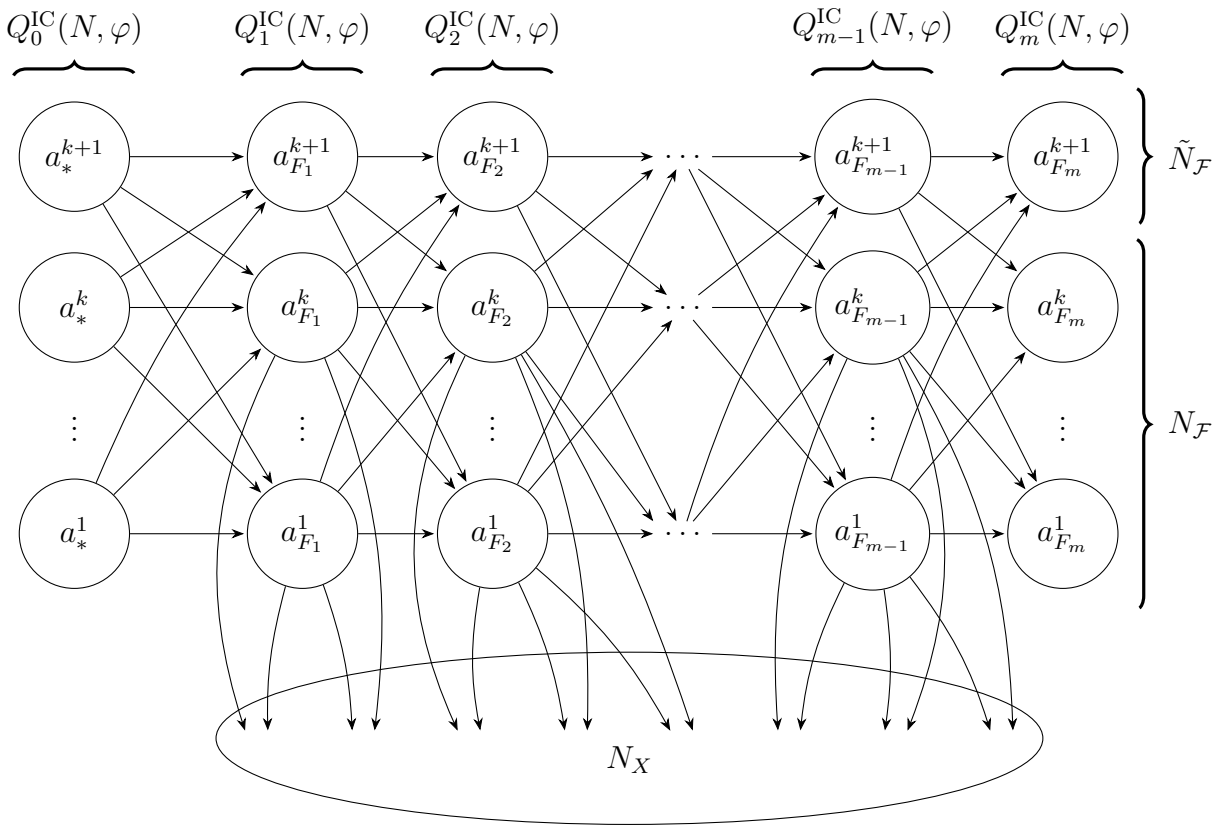

Before we show the correctness of the reduction, it will be useful to observe a few simple facts:

\begin{enumerate}[
    topsep=-4pt,
    leftmargin=*,
    itemindent=4.96em,
    label={\textit{Fact \ref*{thm:ic_cgmb_npc}.\arabic*}:},
    ref={\ref*{thm:ic_cgmb_npc}.\arabic*}
]
\item \label{fact:ic_cgmb_first_column}
The individuals in $\{ a^1_\ast, \ldots, a^{k+1}_\ast \}$ are the only ones qualified by everyone. By construction of $\varphi$, every individual outside of $\{ a^1_\ast, \ldots, a^{k+1}_\ast \}$ is disqualified by more than $k$ individuals. Because the attacker can only change at most $\ell = k$ entries of $\varphi$ in total, no individuals outside of $\{ a^1_\ast, \ldots, a^{k+1}_\ast \}$ can be made qualified by everyone through bribery.

\item \label{fact:ic_cgmb_chain}
Among the individuals in $Q^{\text{IC}}_0(N, \varphi) = \{ a^1_\ast, \ldots, a^{k+1}_\ast \}$, there is a consensus that the individuals from $\{ a^1_{F_1}, \ldots, a^{k+1}_{F_1} \}$ should be qualified. For any $i \in \{ 1, \ldots, m-1 \}$, there is a consensus among the individuals in $\{ a^1_{F_i}, \ldots, a^{k+1}_{F_i} \}$ that the individuals from $\{ a^1_{F_{i+1}}, \ldots, a^{k+1}_{F_{i+1}} \}$ should be qualified. This way, all individuals from $N_\mathcal{F} \cup \tilde{N}_\mathcal{F}$ are socially qualified. Meanwhile, for any $i \in \{ 1, \ldots, m \}$, there is no consensus among the individuals in $\{ a^1_{F_i}, \ldots, a^{k+1}_{F_i} \}$ that someone in $N_X$ should be qualified. Therefore, we initially have $f^{\text{IC}}(N, \varphi) = \{ a^1_\ast, \ldots, a^{k+1}_\ast \} \cup N_\mathcal{F} \cup \tilde{N}_\mathcal{F}$.

\item \label{fact:ic_cgmb_a_fi_must_be_in_q_i}
Let $\varphi^\prime$ be a binary profile obtained from $\varphi$ by changing at most $k$ entries. For any individual $a^j_{F_i} \in N_\mathcal{F}$ where $i \in \{ 1, \ldots, m \}$ and $j \in \{ 1, \ldots, k \}$, it must hold that $a^j_{F_i} \in Q^{\text{IC}}_i(N, \varphi^\prime)$, i.e.\@ $a^j_{F_i}$ must become socially qualified in the $i$-th iteration. Assume for the sake of contradiction that this does not hold, and let $a^j_{F_i}$ be the first individual (i.e.\@ with the smallest $i$) from $N_\mathcal{F}$ who does not become socially qualified in the $i$-th iteration. Because $a^j_{F_i} \in A^+$, they must become socially qualified at some point. But to create a consensus in an iteration (before the $(i-1)$-th iteration where such a consensus already exists) that $a^j_{F_i}$ should be qualified, the attacker would have to change at least $k+1$ entries of $\varphi$ which would exceed the budget of $\ell = k$.

\item \label{fact:ic_cgmb_can_only_create_consensus_for_n_x}
For every $i \in \{ 1, \ldots, m \}$, the only individuals in $N_X$ for whom the attacker can create a consensus among everyone in the $i$-th iteration that they should be qualified are the ones from $\{ a_x : x \in F_i \}$. The reason for this is as follows: We know from \factref{fact:ic_cgmb_a_fi_must_be_in_q_i} that $\{ a^1_{F_i}, \ldots, a^k_{F_i} \} \subseteq Q^{\text{IC}}_i(N, \varphi^\prime)$. Initially, the only individuals in $N_X$ who are qualified by everyone in $\{ a^1_{F_i}, \ldots, a^k_{F_i} \}$ are the ones from $\{ a_x : x \in F_i \}$. Clearly, making everyone in $\{ a^1_{F_i}, \ldots, a^k_{F_i} \}$ qualify some further individual $a \in N_X \setminus \{ a_x : x \in F_i \}$ would require at least $k$ changes in $\varphi$. But that alone would not be sufficient to create a consensus among everyone in the $i$-th iteration that the individual $a$ should be qualified. This is because there would still be the individual $a^{k+1}_{F_i} \in Q^{\text{IC}}_i(N, \varphi^\prime)$ who initially disqualifies everyone from $N_X$, and there would be no more budget left to bribe $a^{k+1}_{F_i}$.

\item \label{fact:ic_cgmb_chain_ends}
From \factref{fact:ic_cgmb_a_fi_must_be_in_q_i}, we know that $\{ a^1_{F_m}, \ldots, a^k_{F_m} \} \subseteq Q^{\text{IC}}_m(N, \varphi^\prime)$. Initially, the individuals from $\{ a^1_{F_m}, \ldots, a^k_{F_m} \}$ only qualify themselves and the individuals in $\{ a^1_\ast, \ldots, a^{k+1}_\ast \}$. Notably, they do not qualify any individuals from $N_X$ since we assumed that $F_m = \emptyset$. In conjunction with \factref{fact:ic_cgmb_can_only_create_consensus_for_n_x}, this implies that it is impossible to create a consensus among the individuals in the $m$-th iteration that some further individual should be qualified. Therefore, the iterative process always ends after the $m$-th iteration.

\item \label{fact:ic_cgmb_optimal_to_bribe_in_n_f}
To turn the individuals from $N_X \subseteq A^+$ socially qualified, the attacker needs to ensure that, for each $a_x \in N_X$, there is at some point a consensus among the individuals in a set $Q^{\text{IC}}_i(N, \varphi^\prime)$ that $a_x$ should be qualified. From \factref{fact:ic_cgmb_chain_ends}, we know that the iterative process goes through $m$ iterations. For every $i \in \{ 1, \ldots, m \}$, we know from \factref{fact:ic_cgmb_can_only_create_consensus_for_n_x} that the only individuals for whom the attacker could create a consensus among everyone in $Q^{\text{IC}}_i(N, \varphi^\prime)$ that they should be qualified are the ones from $\{ a_x : x \in F_i \}$. Clearly, the cheapest way to create this consensus is to ensure that the individual $a^{k+1}_{F_i}$ (who initially disqualifies the individuals in $N_X$) is not in the set $Q^{\text{IC}}_i(N, \varphi^\prime)$. To do this, the attacker could either bribe $a^{k+1}_{F_i}$ to disqualify themselves, or bribe some individual from the previous iteration $Q^{\text{IC}}_{i-1}(N, \varphi^\prime)$ to disqualify $a^{k+1}_{F_i}$. Both of these options lead to the desired result. Hence, without loss of generality, we can assume that in any minimal solution, the attacker only bribes individuals from $\tilde{N}_\mathcal{F}$ to disqualify themselves.

\item \label{fact:ic_cgmb_bribe_some_n_f}
Once the attacker bribes an individual $a^{k+1}_{F_i}$ with $i \in \{ 1, \ldots, m \}$ to disqualify themselves, there is a consensus among everyone in $Q^{\text{IC}}_i(N, \varphi^\prime)$ that the individuals from $\{ a_x : x \in F_i \}$ should be qualified, i.e.\@ $\{ a_x : x \in F_i \} \subseteq Q^{\text{IC}}_{i+1}(N, \varphi^\prime)$. Because the individuals from $N_X$ qualify everyone, this does not cause any changes in subsequent iterations: Among the individuals in $Q^{\text{IC}}_{i+1}(N, \varphi^\prime)$, there still exists a consensus that the individuals in $\{ a^1_{F_{i+2}}, \ldots, a^{k+1}_{F_{i+2}} \}$ should be qualified. Thus, the argument also applies if multiple individuals from $\tilde{N}_\mathcal{F}$ are bribed: For each bribed individual $a^{k+1}_{F_i}$, the individuals from $\{ a_x : x \in F_i \}$ become socially qualified.
\end{enumerate}

Equipped with these facts, we now show that the constructed instance is a YES-instance if and only if there exists a set cover for $X$ in $\mathcal{F}$ of cardinality at most $k$.

($\Rightarrow$) Assume that $\mathcal{F}^\prime \subseteq \mathcal{F}$ is a set cover for $X$ with $|\mathcal{F}^\prime| \leq k$. We bribe each individual in $\{ a^{k+1}_{F_i} : {F_i} \in \mathcal{F}^\prime \}$ to disqualify themselves. From \factref{fact:ic_cgmb_bribe_some_n_f} and the fact that $\mathcal{F}^\prime$ is a set cover for $X$, we know that all individuals from $N_X$ are now socially qualified. From \factref{fact:ic_cgmb_chain}, we know that the individuals in $\{ a^1_\ast, \ldots, a^{k+1}_\ast \} \cup N_\mathcal{F}$ are socially qualified too. Thus, all individuals in $A^+$ are now socially qualified.

($\Leftarrow$) Assume we are given a minimal CGMB solution $M \subseteq N \times N$ with $|M| \leq \ell$ such that all individuals in $A^+$ are socially qualified after changing the entries of $\varphi$ for the pairs of individuals in $M$. Without loss of generality, we can assume that $M \subseteq \tilde{N}_\mathcal{F} \times \tilde{N}_\mathcal{F}$, i.e.\@ the attacker has only bribed individuals from $\tilde{N}_\mathcal{F}$ to disqualify themselves (see \factref{fact:ic_cgmb_optimal_to_bribe_in_n_f}). Let $U = \{ a \in \tilde{N}_\mathcal{F} : (a, a) \in M \}$. Clearly, it holds that $|U| \leq \ell$. From \factref{fact:ic_cgmb_bribe_some_n_f} and the fact that $N_X \subseteq f^{\text{IC}}(N, \varphi^\prime)$, we know that for each $x \in X$, there exists a set $F_i \in \mathcal{F}$ with $x \in F_i$ for which the individual $a^{k+1}_{F_i}$ is in $U$. From this, it follows that $\mathcal{F}^\prime = \{ F_i \in \mathcal{F} : a^{k+1}_{F_i} \in U \}$ is a set cover for $X$ of size at most $\ell = k$.
\end{Proof}

For the social rules $f^{\text{2IC}}$ and $f^{\text{2LIC}}$, the unpriced CGMB problem can be solved in polynomial time. We first show this for the $f^{\text{2IC}}$ rule:

\begin{Theorem} \label{thm:2ic_cgmb_p}
The problem $f^{\text{2IC}}$-\textsc{Constructive Group Microbribery} can be solved in time $\mathcal{O}(n^3)$.
\end{Theorem}

\begin{Proof}
Let $(N, \varphi, A^+, \ell)$ be an instance of $f^{\text{2IC}}$-CGMB. We begin by bribing each individual in $A^+$ who disqualifies themselves to qualify themselves. This is always necessary because under the $f^{\text{2IC}}$ rule, only individuals who qualify themselves can become socially qualified.

To achieve the constructive objective, we need at least one individual who is qualified by everyone (since otherwise no one is socially qualified). Hence, we guess an individual $a_\ast \in N$ to be qualified by everyone. Assuming $a_\ast$ was guessed correctly, we can solve the instance as follows:

For each individual $a^\prime \in N$ who disqualifies $a_\ast$, we bribe $a^\prime$ to qualify $a_\ast$ (clearly, this is always necessary). Furthermore, for each individual $a^+ \in A^+$ who is disqualified by $a_\ast$, we bribe $a_\ast$ to qualify $a^+$ (again, this is always necessary). Let $\varphi^\prime$ denote the resulting profile. We now compute the set $K^{\text{2IC}}_0(N, \varphi^\prime)$ of individuals who are qualified by everyone. If there is a consensus among the individuals in $K^{\text{2IC}}_0(N, \varphi^\prime)$ that everyone from $A^+$ should be qualified, we are done. Otherwise, we iterate over every $a_{\ast\ast} \in K^{\text{2IC}}_0(N, \varphi^\prime)$ who does not yet qualify everyone from $A^+$, and bribe someone to disqualify $a_{\ast\ast}$. This way, we create a consensus among the remaining individuals in $K^{\text{2IC}}_0(N, \varphi^\prime)$ that everyone from $A^+$ should be qualified.

When deciding which individual we should bribe to disqualify a given $a_{\ast\ast} \in K^{\text{2IC}}_0(N, \varphi^\prime)$, we must keep in mind that the individuals in $A^+$ must qualify themselves and must be qualified by all individuals who are qualified by everyone. Therefore, if $a_{\ast\ast} \in A^+$, we let $a_{\ast\ast}$ be disqualified by someone from $N \setminus K^{\text{2IC}}_0(N, \varphi^\prime)$. Otherwise, we can let $a_{\ast\ast}$ be disqualified by an arbitrary individual from $N$.

If there exists an individual $a_\ast \in N$ for whom the approach described above requires at most $\ell$ bribes, the given instance is a YES-instance. Otherwise, it must be a NO-instance.

The running time of the algorithm is bounded by $\mathcal{O}(n \cdot n^2)$ where the additional factor of $n$ comes from guessing an individual $a_\ast \in N$ to be qualified by everyone.
\end{Proof}

When the $f^{\text{2LIC}}$ social rule is used, the CGMB problem can be solved even faster:

\begin{Theorem} \label{thm:2lic_cgmb_p}
The problem $f^{\text{2LIC}}$-\textsc{Constructive Group Microbribery} can be solved in time $\mathcal{O}(n^2)$.
\end{Theorem}

\begin{Proof}
To solve a given instance $(N, \varphi, A^+, \ell)$ of $f^{\text{2LIC}}$-CGMB, we do the following:

We again begin by bribing each individual in $A^+$ who disqualifies themselves to qualify themselves (this is always necessary). Let $\varphi^\prime$ denote the resulting profile. We now compute the set $K^{\text{2LIC}}_0(N, \varphi^\prime)$ of individuals who are qualified by everyone. If this set is empty or there exists a consensus among the individuals in $K^{\text{2LIC}}_0(N, \varphi^\prime)$ that everyone from $A^+$ should be qualified, we are done. Otherwise, we iterate over every $a_{\ast\ast} \in K^{\text{2LIC}}_0(N, \varphi^\prime)$ who does not yet qualify everyone from $A^+$, and bribe someone to disqualify $a_{\ast\ast}$. This way, we either empty the set of individuals who are qualified by everyone, or we create a consensus among the remaining individuals in this set that everyone from $A^+$ should be qualified. Either way, all individuals from $A^+$ are socially qualified after this bribery. If $a_{\ast\ast} \in A^+$, we let $a_{\ast\ast}$ be disqualified by someone from $N \setminus K^{\text{2LIC}}_0(N, \varphi^\prime)$ to ensure that $a_{\ast\ast}$ is still qualified by themselves and by all individuals who are qualified by everyone after the bribery. Otherwise, we can let $a_{\ast\ast}$ be disqualified by an arbitrary individual from $N$.

The instance is a YES-instance if and only if the total number of bribes is at most $\ell$. It is easy to see that the running time of the algorithm is bounded by $\mathcal{O}(n^2)$.
\end{Proof}

The priced version of constructive group microbribery is NP-complete for both the $f^{\text{2IC}}$ rule and the $f^{\text{2LIC}}$ rule. We show this via a reduction from \textsc{Independent Set}:

\begin{Theorem} \label{thm:2ic_cgmb_npc}
For $f \in \{ f^{\text{2IC}}, f^{\text{2LIC}} \}$, the problem $f$-\textsc{\$Constructive Group Microbribery} is NP-complete.
\end{Theorem}

\begin{Proof}
Let $(G, k)$ with $G = (V, E)$ be an instance of \textsc{Independent Set}. Without loss of generality, we can assume that $|E| \geq 1$ (otherwise solving the instance is trivial). Also, we can assume that for all $v \in V$, it holds that $\operatorname{deg}(v) \geq 1$ (if this is not the case, we can simply delete all vertices of degree $0$ and decrease the value of $k$ accordingly). We now construct an equivalent instance of \$CGMB as follows:

For each vertex $v \in V$, we introduce one individual $a_v$. Let $N_V = \{ a_v : v \in V \}$. For each edge $\{u, v\} \in E$, we introduce one individual $a_{\{u, v\}}$. Let $N_E = \{ a_e : e \in E \}$. In addition, we introduce two more individuals $a_\ast$ and $q$. We set $N = N_V \cup N_E \cup \{ a_\ast, q \}$, $A^+ = N_E \cup \{q\}$, and define the profile $\varphi$ over $N$ as follows:

\begin{itemize}[nolistsep]
\item
All individuals qualify themselves.
\item
The individuals $a_\ast$ and $q$ also qualify all other individuals.
\item
For each vertex $v \in V$, the individual $a_v$ qualifies everyone except the individuals $a_e \in N_E$ with $e \in E$ and $v \in e$.
\item
For each edge $e \in E$, the individual $a_e$ qualifies everyone except $q$.
\end{itemize}

In the next step, we assign the bribery prices. Let $D$ be the smallest positive integer such that, for every $v \in V$, $D$ is divisible by $\operatorname{deg}(v)$. We set the budget to $\ell = |V| \cdot (D + 1) - k$ and define the microbribery price function $\rho : N \times N \rightarrow \mathbb{N}$ as follows:

\begin{itemize}[nolistsep]
\item
For each $v \in V$, we let $\rho\big((a_v, a_v)\big) = D + 1$.
\item
For each $v \in V$ and $e \in E$ with $v \in e$, we let $\rho\big((a_v, a_e)\big) = \frac{D}{\operatorname{deg}(v)}$.
\item
For all pairs of individuals $(a, b) \in N \times N$ not specified above, we let $\rho\big((a, b)\big) = \ell + 1$.
\end{itemize}

See \figref{fig:2ic_cgmb_npc} for an example illustration of the \$CGMB instance created in this reduction. Clearly, the reduction can be completed in polynomial time.

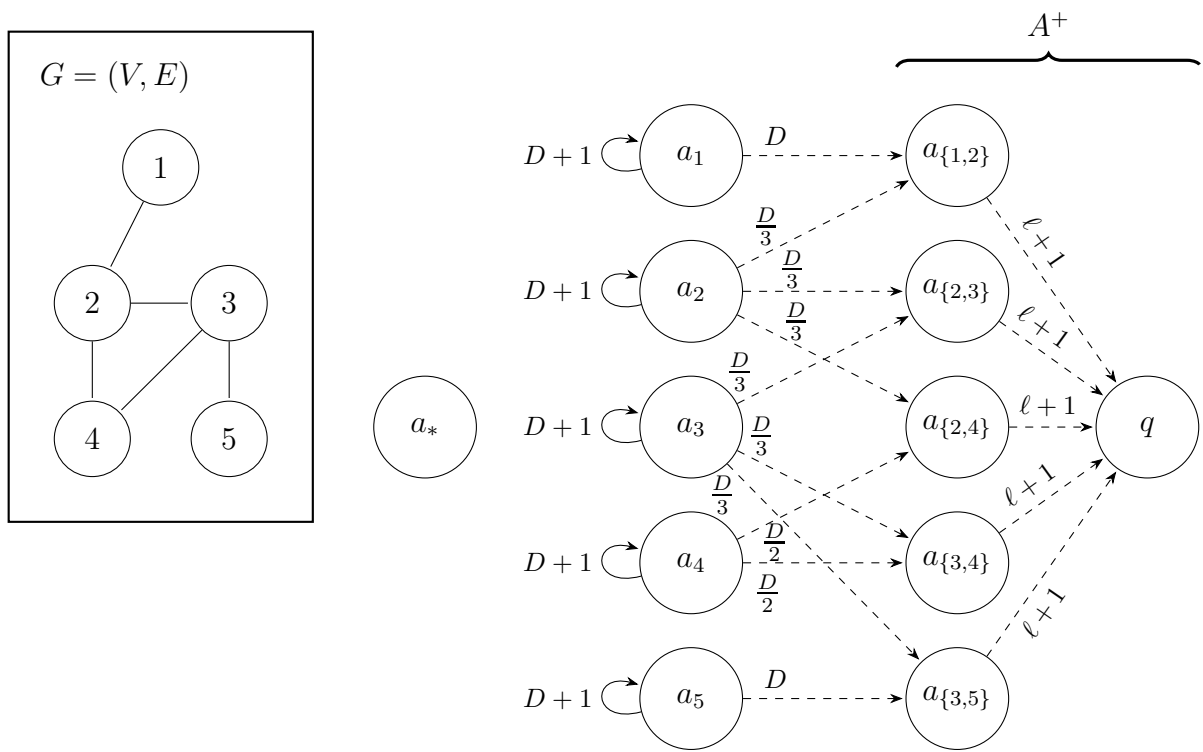
\begin{figure}[!tbh]
\begin{minipage}[b]{.25\textwidth}
\centering

\raisebox{0.04\height}{
\begin{tikzpicture}[
    > = Stealth,
    shorten > = 1pt,
    auto,
    node distance = 1.8cm,
    main node/.style = {circle, draw, minimum size = 1cm}
]

\node[main node] (1)                                            {$1$};
\node[main node] (2)    [below of = 1, xshift = -0.9cm]         {$2$};
\node[main node] (3)    [below of = 1, xshift = 0.9cm]          {$3$};
\node[main node] (4)    [below of = 2]                          {$4$};
\node[main node] (5)    [below of = 3]                          {$5$};

\node (g) [above of = 1, node distance = 1.2cm, xshift=-0.6cm]  {$G = (V, E)$};

\node (i) [below of = 5, node distance = 4cm]                   {};

\draw[thick] (-2,-4.7) rectangle (2,1.8);

\path[-]
(1)     edge                    node {} (2)
(2)     edge                    node {} (3)
        edge                    node {} (4)
(3)     edge                    node {} (4)
        edge                    node {} (5)
;

\end{tikzpicture}
}

\end{minipage}\hfill
\begin{minipage}[b]{.7\textwidth}
\centering

\raisebox{0.04\height}{
\begin{tikzpicture}[
    > = Stealth,
    shorten > = 1pt,
    auto,
    node distance = 1.8cm,
    main node/.style = {circle, draw, minimum size = 1.35cm},
    every loop/.style={looseness = 5}
]

\node[main node] (a1)                                           {$a_1$};
\node[main node] (a2)   [below of = a1]                         {$a_2$};
\node[main node] (a3)   [below of = a2]                         {$a_3$};
\node[main node] (a4)   [below of = a3]                         {$a_4$};
\node[main node] (a5)   [below of = a4]                         {$a_5$};

\node[main node] (aa)   [left of = a3, node distance = 3.5cm]   {$a_\ast$};

\node[main node] (a12)  [right of = a1, node distance = 3.5cm]  {$a_{\{1, 2\}}$};
\node[main node] (a23)  [below of = a12]                        {$a_{\{2, 3\}}$};
\node[main node] (a24)  [below of = a23]                        {$a_{\{2, 4\}}$};
\node[main node] (a34)  [below of = a24]                        {$a_{\{3, 4\}}$};
\node[main node] (a35)  [below of = a34]                        {$a_{\{3, 5\}}$};

\node[main node] (q)    [right of = a24, node distance = 2.5cm] {$q$};

\path[->]
(a1)    edge[loop left]     node [left]     {\footnotesize $D + 1$}     (a1)
(a2)    edge[loop left]     node [left]     {\footnotesize $D + 1$}     (a2)
(a3)    edge[loop left]     node [left]     {\footnotesize $D + 1$}     (a3)
(a4)    edge[loop left]     node [left]     {\footnotesize $D + 1$}     (a4)
(a5)    edge[loop left]     node [left]     {\footnotesize $D + 1$}     (a5)
;

\path[->, dashed]
(a1)    edge    node [pos=0.2]                  {\footnotesize $D$}     (a12)
(a2)    edge    node [pos=0.3, yshift=-0.2cm]   {$\frac{D}{3}$}         (a12)
        edge    node [pos=0.3, yshift=-0.1cm]   {$\frac{D}{3}$}         (a23)
        edge    node [pos=0.2, yshift=-0.25cm]  {$\frac{D}{3}$}         (a24)
(a3)    edge    node [pos=0.15, yshift=-0.15cm] {$\frac{D}{3}$}         (a23)
        edge    node [pos=0, yshift=-0.2cm]     {$\frac{D}{3}$}         (a34)
        edge    node [pos=-0.15, yshift=-1.2cm] {$\frac{D}{3}$}         (a35)
(a4)    edge    node [pos=0.35, yshift=-0.85cm] {$\frac{D}{2}$}         (a24)
        edge    node [pos=0.15, yshift=-0.8cm]  {$\frac{D}{2}$}         (a34)
(a5)    edge    node [pos=0.2]                  {\footnotesize $D$}     (a35)
;

\path[->, dashed]
(a12)   edge    node [pos=0.3, sloped]          {\footnotesize $\ell + 1$}  (q)
(a23)   edge    node [pos=0.3, sloped]          {\footnotesize $\ell + 1$}  (q)
(a24)   edge    node [pos=0.45, sloped]         {\footnotesize $\ell + 1$}  (q)
(a34)   edge    node [pos=0.4, sloped]          {\footnotesize $\ell + 1$}  (q)
(a35)   edge    node [pos=0.3, sloped, below]   {\footnotesize $\ell + 1$}  (q)
;

\draw [very thick, decorate, decoration = {brace, raise=5pt, amplitude=5pt}]
(2.7, 1) -- (6.7, 1)
node[pos = 0.5, above = 12pt, black] {$A^+$};

\end{tikzpicture}
}

\end{minipage}
\caption{An example illustration of the \$CGMB instance created in the reduction in \thmref{thm:2ic_cgmb_npc}. In the picture, every individual qualifies themselves and all other individuals, and the price of changing this qualification is $\ell + 1$. The only exceptions are the drawn arcs: Each continuous arc represents a qualification with the specified bribery price, and each dashed arc represents a disqualification with the specified bribery price.}
\label{fig:2ic_cgmb_npc}
\end{figure}

Before we show the correctness of the reduction, it will be useful to observe a few simple facts:

\begin{enumerate}[
    topsep=-4pt,
    leftmargin=*,
    itemindent=4.96em,
    label={\textit{Fact \ref*{thm:2ic_cgmb_npc}.\arabic*}:},
    ref={\ref*{thm:2ic_cgmb_npc}.\arabic*}
]
\item \label{fact:2ic_cgmb_initial_situation}
Initially, the individuals in $\{a_\ast\} \cup N_V$ are the only ones qualified by everyone. Among them, there exists a consensus that the individual $q$ should be qualified, but no consensus that anyone from $N_E$ should be qualified (since each individual in $N_E$ is disqualified by exactly two individuals from $N_V$). Therefore, we initially have $f(N, \varphi) = \{a_\ast\} \cup N_V \cup \{q\}$.

\item \label{fact:2ic_cgmb_n_e_not_qualified_by_everyone}
In any solution for the created instance, none of the individuals from $N_E$ can be qualified by everyone after the bribery. This is because all individuals from $N_E$ disqualify the individual $q$ (who is in $A^+$), and the attacker cannot afford to change this through the bribery. Hence, if some individual from $N_E$ were qualified by everyone, the individual $q$ would no longer be socially qualified.

\item \label{fact:2ic_cgmb_make_n_e_socially_qualified}
To turn the individuals from $N_E$ socially qualified, the attacker must create a consensus among the individuals who are qualified by everyone that the individuals from $N_E$ should be qualified. This also holds when the $f^{\text{2LIC}}$ rule is used because the individual $a_\ast$ is initially qualified by everyone, and the attacker cannot afford to bribe someone to disqualify $a_\ast$. Hence, it is impossible to empty the set of individuals who are qualified by everyone.

\item \label{fact:2ic_cgmb_all_bribed_are_from_n_v}
By definition of the microbribery price function $\rho$, the only entries of $\varphi$ which the attacker can afford to change are certain outgoing qualifications of the individuals in $N_V$. Hence, in any solution for the created instance, all bribed individuals are from $N_V$.

\item \label{fact:2ic_cgmb_all_from_n_v_are_bribed}
In any solution for the created instance, every individual $a_v$ where $v \in V$ must be bribed in one of two ways: Either $a_v$ must be bribed to disqualify themselves, or $a_v$ must be bribed to qualify the individuals in $\{ a_e \in N_E : e \in E \text{ and } v \in e \}$. Assume for the sake of contradiction that there is an individual $a_v \in N_V$ such that it is possible to make everyone from $N_E$ socially qualified without bribing $a_v$ in one of the ways described above. The fact that $a_v$ was not bribed to disqualify themselves implies that $a_v$ is still qualified by everyone after the bribery. And the fact that $a_v$ was not bribed to qualify the individuals in $N_E$ implies that there is at least one individual in $N_E \subseteq A^+$ who is not socially qualified after the bribery, contradicting our assumption.

\item \label{fact:2ic_cgmb_cannot_bribe_both_from_a_v}
For each individual $a_e \in N_E$, at most one of the two individuals from $N_V$ who initially disqualify $a_e$ can be bribed to qualify $a_e$. If the attacker would bribe both individuals from $N_V$ to qualify $a_e$, then $a_e$ would be qualified by everyone, contradicting \factref{fact:2ic_cgmb_n_e_not_qualified_by_everyone}.
\end{enumerate}

Equipped with these facts, we now show that the constructed instance is a YES-instance if and only if there exists an independent set $V^\prime \in V$ in $G$ of size at least $k$.

($\Rightarrow$) Assume that $V^\prime \subseteq V$ is an independent set in $G$ with $|V^\prime| \geq k$. For each $v \in V^\prime$, we bribe the individual $a_v$ to qualify the individuals in $\{ a_e \in N_E : e \in E \text{ and } v \in e \}$. This requires a total bribery cost of $|V^\prime| \cdot D$. For each $v \in V \setminus V^\prime$, we bribe the individual $a_v$ to disqualify themselves. This requires an additional bribery cost of $|V \setminus V^\prime| \cdot (D + 1)$. Thus, the total bribery cost is $|V^\prime| \cdot D + (|V| - |V^\prime|) \cdot (D + 1) = |V| \cdot (D + 1) - |V^\prime| \leq |V| \cdot (D + 1) - k = \ell$. After this bribery, the only individuals who are qualified by everyone are $a_\ast$ and the individuals from $\{ a_v : v \in V^\prime \}$. Since $V^\prime$ is an independent set in $G$, each individual from $N_E$ is still disqualified by at least one individual from $N_V$, and therefore not qualified by everyone. Among the individuals who are qualified by everyone, there now exists a consensus that all individuals from $N_E \cup \{q\} = A^+$ should be qualified.

($\Leftarrow$) Assume we are given a \$CGMB solution $M \subseteq N \times N$ with $\rho(M) \leq \ell$ such that all individuals in $A^+$ are socially qualified after changing the entries of $\varphi$ for the pairs of individuals in $M$. From \factref{fact:2ic_cgmb_all_bribed_are_from_n_v}, we know that all bribed individuals are from $N_V$. Furthermore, we know from \factref{fact:2ic_cgmb_all_from_n_v_are_bribed} that for each individual $a_v \in N_V$, the attacker either bribed $a_v$ to disqualify themselves, or bribed them to qualify the individuals from $\{ a_e \in N_E : e \in E \text{ and } v \in E \}$. Let $N_V^\prime = \{ a_v \in N_V : (a_v, a_v) \in M \}$ denote the subset of individuals from $N_V$ who were bribed to disqualify themselves. The total cost required to bribe these individuals is $|N_V^\prime| \cdot (D + 1)$. Since the available budget is $\ell = |V| \cdot (D + 1) - k$, this leaves a budget of $(|N_V| - |N_V^\prime|) \cdot (D + 1) - k = |N_V \setminus N_V^\prime| \cdot D + |N_V \setminus N_V^\prime| - k$ for the remaining bribes. The cost required to bribe the remaining individuals from $N_V \setminus N_V^\prime$ is $|N_V \setminus N_V^\prime| \cdot D$. This implies that $|N_V \setminus N_V^\prime| \geq k$ (since otherwise the budget would not suffice). We know from \factref{fact:2ic_cgmb_cannot_bribe_both_from_a_v} that each individual in $N_E$ is qualified by at most one individual from $N_V \setminus N_V^\prime$. By construction of $\varphi$, it follows that $V^\prime = \{ v \in V : a_v \in N_V \setminus N_V^\prime \}$ is an independent set in $G$ of size at least $k$.
\end{Proof}

It is possible to alter the above reduction in such a way that all bribery prices are at most $\ell$, i.e.\@ there are no bribes which the attacker cannot afford outright. However, this requires a much more involved argumentation in the proof, which is why we refrain from doing so in this thesis.

We now consider the destructive group microbribery problems. For all three considered social rules, the DGMB problem is NP-complete. Obviously, this implies that the priced version is also NP-complete. The proof is based on a reduction from \textsc{Restricted Exact Cover by 3-sets}.

\begin{Theorem} \label{thm:ic_dgmb_npc}
For all $f \in \{ f^{\text{IC}}, f^{\text{2IC}}, f^{\text{2LIC}} \}$, the problem $f$-\textsc{Destructive Group Microbribery} is NP-complete.
\end{Theorem}

\begin{Proof}
Given a RX3C instance $(X, \mathcal{F})$ with $|X|=3m$ (and thus $|\mathcal{F}|=3m$), we construct an equivalent instance of DGMB as follows:

For each element $x \in X$, we introduce one individual $a_x$. Let $N_X = \{ a_x : x \in X \}$. For each triplet $F \in \mathcal{F}$, we introduce one individual $a_F$. Let $N_\mathcal{F} = \{ a_F : F \in \mathcal{F} \}$. In addition, we introduce $m+1$ dummy individuals $\{ d_1, \ldots, d_{m+1} \} = N_D$. Finally, we introduce one more individual $d$. We set $N = N_X \cup N_\mathcal{F} \cup N_D \cup \{d\}$ and define the profile $\varphi$ over $N$ as follows:

\begin{itemize}[nolistsep]
\item
The individual $d$ disqualifies themselves. All other individuals qualify themselves.
\item
For each $x \in X$, the individual $a_x$ qualifies all other individuals.
\item
For each $F \in \mathcal{F}$, the individual $a_F$ qualifies all other individuals except the three individuals $a_x \in N_X$ with $x \in F$.
\item
For every $i \in \{ 1, \ldots, m+1 \}$, the individual $d_i$ qualifies all other individuals.
\item
The individual $d$ qualifies all other individuals except the ones from $N_\mathcal{F}$.
\end{itemize}

To conclude the reduction, we set $A^- = N_X$ and $\ell = m$. See \figref{fig:ic_dgmb_npc} for an example illustration of the DGMB instance created in this reduction. Clearly, the reduction can be completed in polynomial time.

\begin{figure}[!tbh]
\centering
\raisebox{0.04\height}{
\begin{tikzpicture}[
    > = Stealth,
    shorten > = 1pt,
    auto,
    node distance = 2.5cm,
    main node/.style = {circle, draw, minimum size = 1.45cm}
]

\node[main node] (f1)                                           {$a_{F_1}$};
\node[main node] (f2)   [right of = f1]                         {$a_{F_2}$};
\node[main node] (f3)   [right of = f2]                         {$a_{F_3}$};
\node            (f4)   [right of = f3, node distance = 1.7cm]  {$\ldots$};
\node[main node] (fm1)  [right of = f4, node distance = 1.7cm]  {$a_{F_{3m-1}}$};
\node[main node] (fm)   [right of = fm1]                        {$a_{F_{3m}}$};

\node[main node] (d)    [above of = f1, xshift = 2.5cm, yshift = 1cm] {$d$};

\node[main node] (d1)   [right of = d, node distance = 2.8cm]   {$d_1$};
\node[main node] (d2)   [right of = d1, node distance = 1.8cm]  {$d_2$};
\node            (d3)   [right of = d2, node distance = 1.3cm]  {$\ldots$};
\node[main node] (dm1)  [right of = d3, node distance = 1.3cm]  {$d_{m+1}$};

\node (x1)      [below of = f1, xshift = -1cm]          {};
\node (x2)      [right of = x1, node distance = 1cm]    {};
\node (x3)      [right of = x2, node distance = 1cm]    {};
\node (x4)      [right of = x3, node distance = 1cm]    {};
\node (x5)      [right of = x4, node distance = 1cm]    {};
\node (x6)      [right of = x5, node distance = 1cm]    {};
\node (x7)      [right of = x6, node distance = 1cm]    {};
\node (x8)      [right of = x7, node distance = 1cm]    {};
\node (x9)      [right of = x8, node distance = 1cm]    {};
\node (x10)     [right of = x9, node distance = 1cm]    {};
\node (x11)     [right of = x10, node distance = 1cm]   {};
\node (x12)     [right of = x11, node distance = 1cm]   {};
\node (x13)     [right of = x12, node distance = 1cm]   {};
\node (x14)     [right of = x13, node distance = 1cm]   {};
\node (nxn)     [right of = x7, node distance = 0.5cm, yshift = 0.25cm] {};
\node (nxs)     [right of = x7, node distance = 0.5cm, yshift = -0.5cm] {$N_X$};
\node[ellipse, draw, fit={(x3)(nxn)(nxs)(x12)}, inner sep=1mm] (x) {};

\path[->, dashed]
(d)     edge[loop left]         node {} (d)
        edge                    node {} (f1)
        edge                    node {} (f2)
        edge                    node {} (f3)
        edge                    node {} (fm1)
        edge                    node {} (fm)
;

\path[->, dashed]
(f1)    edge                    node {} (x2.east)
        edge                    node {} (x3.west)
        edge                    node {} (x4.west)
(f2)    edge                    node {} (x4.east)
        edge                    node {} (x5.west)
        edge                    node {} (x6.west)
(f3)    edge                    node {} (x6.east)
        edge                    node {} (x7.east)
        edge                    node {} (x8.west)
(fm1)   edge                    node {} (x9.east)
        edge                    node {} (x10.east)
        edge                    node {} (x11.west)
(fm)    edge                    node {} (x11.east)
        edge                    node {} (x12.west)
        edge                    node {} (x13.west)
;

\draw [very thick, decorate, decoration = {brace, raise=5pt, amplitude=5pt}]
(-1.5, -3.8) -- (-1.5, -1.4)
node[pos = 0.5, left = 12pt, black] {$A^-$};

\end{tikzpicture}
}
\caption{An example illustration of the DGMB instance created in the reduction in \thmref{thm:ic_dgmb_npc}. In the picture, we assume that $\mathcal{F}$ is of the form $\{ F_1, F_2, \ldots, F_{3m} \}$. Every individual qualifies themselves and all other individuals, except for the drawn dashed arcs which represent disqualifications.}
\label{fig:ic_dgmb_npc}
\end{figure}

Before we show the correctness of the reduction, it will be useful to observe a few simple facts:

\begin{enumerate}[
    topsep=-4pt,
    leftmargin=*,
    itemindent=4.96em,
    label={\textit{Fact \ref*{thm:ic_dgmb_npc}.\arabic*}:},
    ref={\ref*{thm:ic_dgmb_npc}.\arabic*}
]
\item \label{fact:ic_dgmb_initial_situation}
Initially, the individuals in $N_D$ are the only ones qualified by everyone. Since the individuals in $N_D$ qualify all other individuals, the individuals from $N_\mathcal{F} \cup N_X$ are socially qualified after the first iterative step. Only the individual $d$ is socially disqualified because they disqualify themselves. Hence, we initially have $f(N, \varphi) = N_D \cup N_\mathcal{F} \cup N_X$.

\item \label{fact:ic_dgmb_only_bribe_d}
Because $|N_D| = m+1 > \ell$, the attacker cannot bribe the individuals in such a way that no individual is qualified by everyone. Therefore, the only way to achieve the objective is to ensure that there is no consensus among the individuals who are qualified by everyone that any individual from $A^-$ who qualifies themselves should be qualified. However, because $|N_X| = 3m > \ell$, the attacker cannot make every individual in $N_X$ disqualified by themselves or by someone from $N_D$. Hence, the attacker must utilize the individuals from $N_\mathcal{F}$, i.e.\@ they must make some individuals in $N_\mathcal{F}$ qualified by everyone. Obviously, for every $a_F \in N_\mathcal{F}$, to make $a_F$ qualified by everyone, the attacker needs to change exactly one outgoing qualification of the individual $d$ (who initially disqualifies $a_F$). Since $|N_X| = 3m$ and each individual in $N_\mathcal{F}$ disqualifies only three individuals from $N_X$, this implies that the attacker must change at least $m$ entries of $\varphi$ to achieve the objective. Because the attacker can only change at most $\ell = m$ entries in total, it follows that any solution for the created instance consists of bribing the individual $d$ to qualify exactly $m$ individuals from $N_\mathcal{F}$.

\item \label{fact:ic_dgmb_bribe_d}
When the individual $d$ is bribed to qualify some individual $a_F$ with $F \in \mathcal{F}$, the individual $a_F$ becomes qualified by everyone. Thus, there no longer exists a consensus among the individuals who are qualified by everyone that the individuals in $\{ a_x : x \in F \}$ should be qualified (since $a_F$ disqualifies them). This way, the individuals in $\{ a_x : x \in F \}$ are no longer socially qualified after the first iterative step.

\item \label{fact:ic_dgmb_exact_cover}
For any $U \subseteq N_\mathcal{F}$, let $\varphi^U$ denote the profile obtained from $\varphi$ by bribing the individual $d$ to qualify the individuals in $U$. Let $\mathcal{F}^\prime \subseteq \mathcal{F}$ be an exact 3-set cover for $X$, and let $U = \{ a_F : F \in \mathcal{F}^\prime \}$. From \factref{fact:ic_dgmb_bribe_d}, we know that $N_X \cap K_1(N, \varphi^U) = \emptyset$, i.e.\@ if we bribe $d$ to qualify the individuals in $U$, none of the individuals in $N_X$ is socially qualified after the first iterative step. For the social rules $f^{\text{2IC}}$ and $f^{\text{2LIC}}$, this directly implies that none of the individuals from $N_X$ is socially qualified (because these rules terminate after the first iterative step). But if the $f^{\text{IC}}$ rule is used, there are further iterations to consider: The initial set of socially qualified individuals is $Q^{\text{IC}}_0(N, \varphi^U) = N_D \cup U$. The set of newly socially qualified individuals after the first iteration is $Q^{\text{IC}}_1(N, \varphi^U) = N_\mathcal{F} \setminus U$. Because the individuals in $U$ represent an exact 3-set cover for $X$, each individual $a_x$ with $x \in X$ is disqualified by exactly one individual from $U$, and is disqualified by exactly two individuals from $N_\mathcal{F} \setminus U$. Hence, there is no consensus among the individuals in $Q^{\text{IC}}_1(N, \varphi^U) = N_\mathcal{F} \setminus U$ that any individual in $N_X$ should be qualified, i.e.\@ $Q^{\text{IC}}_2(N, \varphi^U) = \emptyset$. In other words, for all three considered social rules (including the $f^{\text{IC}}$ rule), the final set of socially qualified individuals is $f(N, \varphi^U) = K_1(N, \varphi^U) = N_D \cup N_\mathcal{F}$.
\end{enumerate}

Equipped with these facts, we now show that the constructed instance is a YES-instance if and only if there exists an exact 3-set cover for $X$ in $\mathcal{F}$.

($\Rightarrow$) Assume that $\mathcal{F}^\prime \subseteq \mathcal{F}$ is an exact 3-set cover for $X$. Let $U = \{ a_F : F \in \mathcal{F}^\prime \}$. From \factref{fact:ic_dgmb_exact_cover}, we know that after bribing the individual $d$ to qualify the individuals in $U$, all individuals from $A^- = N_X$ are socially disqualified. Since $|U| = |\mathcal{F}^\prime| = m$, the number of entries in $\varphi$ that need to be changed for this bribery is exactly $m = \ell$.

($\Leftarrow$) Assume we are given a DGMB solution $M \subseteq N \times N$ with $|M| \leq \ell$ such that all individuals in $A^-$ are socially disqualified after changing the entries of $\varphi$ for the pairs of individuals in $M$. We know from \factref{fact:ic_dgmb_only_bribe_d} that $M \subseteq \{d\} \times N_\mathcal{F}$ and that $|M| = m$, i.e.\@ the attacker has bribed the individual $d$ to qualify exactly $m$ individuals from $N_\mathcal{F}$. Let $U = \{ a \in N_\mathcal{F} : (d, a) \in M \}$. The fact that none of the individuals from $N_X$ is socially qualified after the bribery implies that, for each $a_x \in N_X$, there is an individual $a \in U$ who disqualifies $a_x$. From this, it follows that $\{ F \in \mathcal{F} : a_F \in U \}$ is an exact 3-set cover for $X$.
\end{Proof}

By extending the above reduction, we can also show that EGMB and GMB are NP-complete for all three considered social rules:

\begin{Corollary} \label{cor:ic_egmb_gmb_npc}
For all $f \in \{ f^{\text{IC}}, f^{\text{2IC}}, f^{\text{2LIC}} \}$, the problems $f$-\textsc{Exact Group Microbribery} and $f$-\textsc{Group Microbribery} with $A^+ \neq \emptyset$ and $A^- \neq \emptyset$ are NP-complete.
\end{Corollary}

\begin{Proof}
We first apply the same reduction from RX3C as in \thmref{thm:ic_dgmb_npc}. We then make the following minor modifications to obtain an EGMB/GMB instance with $A^+ \neq \emptyset$:

\begin{itemize}[nolistsep]
\item
We add $d$ to the destructive target set, i.e.\@ we let $A^- = N_X \cup \{d\}$.
\item
We define the constructive target set as $A^+ = N \setminus A^- = N_D \cup N_\mathcal{F}$.
\item
To fulfill the requirement that at least one individual from $A^+$ must be socially disqualified initially, we let $d_1$ disqualify themselves.
\item
We increase the budget $\ell$ by $1$, i.e.\@ we let $\ell = m+1$.
\end{itemize}

Clearly, to achieve the constructive objective, the attacker is always forced to bribe the individual $d_1$ to qualify themselves (because only individuals who qualify themselves can become socially qualified). This decreases the budget to $\ell - 1 = m$.

From Facts \ref{fact:ic_dgmb_initial_situation} and \ref{fact:ic_dgmb_only_bribe_d}, we already know that in any solution for the created instance, the individuals from $N_D \cup N_\mathcal{F}$ are socially qualified, and the individual $d$ is socially disqualified. Therefore, the only individuals left to take care of are the ones from $N_X$. In other words, solving the remaining instance (with the remaining budget of $m$) is equivalent to solving the destructive case as described in \thmref{thm:ic_dgmb_npc}.
\end{Proof}

\clearpage

\section{Conclusion}
\label{sec:conclusion}

In this thesis, we have extended the scope of the complexity-theoretic analysis of manipulative attacks in Group Identification. We studied previously unconsidered social rules and problem variants, conducting a detailed analysis of the computational complexity for all examined problems. We also closed some cases that were left open in previous works.

\subsection{Protective problem instances}
\label{sec:conclusion_protective}

In all previous publications on the computational complexity of manipulative attacks in Group Identification, it was implicitly assumed that at least one individual from the constructive target set is initially socially disqualified, and at least one individual from the destructive target set is initially socially qualified. In this thesis, for the first time, we also considered instances where the objective is already fulfilled for one of the two target sets, referring to these instances as \emph{protective}. The complexity of $f^{\text{CSR}} / f^{\text{LSR}}$-GMB on protective instances is particularly interesting because, on general instances, only the constructive case is NP-complete while the destructive case is polynomial-time solvable. Via a reduction from CNF-SAT, we were able to show that the $f^{\text{CSR}} / f^{\text{LSR}}$-GMB problem remains NP-complete even when it is restricted to protective instances (\thmref{thm:gmb_protective_instances_npc}).

For most other manipulative attack problems with a general or exact objective, we presume that restricting them to protective instances does not change their complexity. This is especially true for problems that are polynomial-time solvable, because it is unlikely that some polynomial-time algorithm would fail just because the objective is already fulfilled for one of the two target sets. However, it likely also applies to most NP-complete problems. We only found one case where the consideration of protective instances leads to a difference in complexity: On instances that are not protective, the $f^{\text{CSR}}$ rule is immune to GCDI (\obsref{obs:csr_immune_gcdi}), but on protective instances, the $f^{\text{CSR}}$-GCDI problem is NP-complete (\corref{cor:csr_gcdi_npc}). The fact that the general problem has a different complexity than the constructive and destructive cases (which both are polynomial-time solvable) is already surprising in itself. But the change in complexity on protective instances makes it even more interesting. Roughly speaking, the reason for these discrepancies is that the special nature of the CSR rule makes it impossible to simply combine the algorithms for the constructive and destructive cases into one that works for the general objective. In particular, when both target sets contain individuals for whom the objective is not initially fulfilled, then there exists no solution at all.

\subsection{Relaxed group control by deleting individuals}
\label{sec:conclusion_relaxed}

Another aspect that we focused on in our analysis is the specific rules regarding the deletion of individuals. In the classical definition of the group control by deleting individuals problem, the attacker is not allowed to delete individuals that are in $A^+$ or $A^-$. In this thesis, we relaxed that restriction by allowing the attacker to also delete individuals from the destructive target set $A^-$. For certain objectives paired with the consent rules, the CSR rule, and the LSR rule, we found that the relaxation changes the complexity of the problem. Our results are summarized in \tableref{tab:relaxed_group_control_deleting}.

For the consent rules, we obtained dichotomous results: All consent rules with $s=1$ are immune against the non-relaxed destructive problem, but the relaxed variant is polynomial-time solvable for these rules (\thmref{thm:fst_rdgcdi_p}). In other words, the relaxation helps the attacker to carry out the attack. On the other hand, the non-relaxed destructive problem is polynomial-time solvable for all consent rules with $s=2$, but through the relaxation it becomes NP-complete (\thmref{thm:fst_rdgcdi_npc}). Here, the relaxation makes it more difficult for the attacker to carry out the attack. In both of these cases, the change in complexity can be explained by the fact that the relaxation gives the attacker more options, i.e.\@ some attacks that were previously impossible become feasible through the relaxation, while other cases that were previously easy to solve become more complex.

For the CSR and LSR rules, the complexity of the relaxed problems is mostly the same as for the non-relaxed variants. However, as discussed in \secref{sec:conclusion_protective}, for the $f^{\text{CSR}}$-GCDI problem, we found that it is only NP-complete when one considers protective instances, and immune otherwise. Meanwhile, the relaxed problem $f^{\text{CSR}}$-R-GCDI is always NP-complete, regardless of whether the given instance is protective or not. This leads to the interesting situation where the relaxed problems with a purely constructive, a purely destructive, or an exact objective are all polynomial-time solvable (\obsref{obs:f_rcgcdi}, \corref{cor:csr_rdgcdi_p}, \thmref{thm:csr_regcdi_p}), but the general problem is NP-complete (\thmref{thm:csr_rgcdi_npc}). Again, the reason for this is that the algorithms for the constructive and destructive objective cannot simply be combined into one that works for the general objective. Only when the attacker has an exact objective, it is possible to handle both target sets separately because in this special case there are no individuals ``in between''.

\subsection{Iterative consensus rules}
\label{sec:conclusion_iterative_consensus}

All previous works on the complexity of manipulative attacks in Group Identification only considered the consent rules, the CSR rule, and the LSR rule. We extended the analysis to three new social rules, called iterative-consensus rule $f^{\text{IC}}$, 2-stage-iterative-consensus rule $f^{\text{2IC}}$, and 2-stage-liberal-iterative-consensus rule $f^{\text{2LIC}}$. The three rules are based on proposals by \textcite{KR97}. Our results are summarized in \tableref{tab:manipulative_attacks_iterative_consensus}.

Our results show that the three new social rules are susceptible to all considered manipulative attacks, i.e.\@ we did not find any immunity results. However, most of the problems are NP-complete, so carrying out the attacks is rather difficult. In particular, when the $f^{\text{IC}}$ rule is used, almost all problems are NP-complete, the only exception being \$DGB. This suggests that the $f^{\text{IC}}$ rule is a good candidate for scenarios where resilience against manipulative attacks is needed.

When using the 2-stage rules $f^{\text{2IC}}$ and $f^{\text{2LIC}}$, several of the problems that are NP-complete for the $f^{\text{IC}}$ rule become polynomial-time solvable. This is not very surprising, since the functioning of the 2-stage rules is simpler than that of the $f^{\text{IC}}$ rule. However, there is one exception to this pattern: The $f^{\text{2LIC}}$ rule is the only one for which the \$DGB problem is NP-complete. This is because for the rules $f^{\text{IC}}$ and $f^{\text{2IC}}$, one can easily make all individuals socially disqualified by bribing someone to disqualify everyone (\thmref{thm:2ic_dgb_p}). But when the $f^{\text{2LIC}}$ rule is used, this approach does not work, and finding a solution is much harder (\thmref{thm:2lic_dgb_npc}).

It should be noted that in many real-world instances (especially ones with a large number of individuals), one can expect that no individual is initially qualified by everyone. On such instances, the $f^{\text{2LIC}}$ rule behaves exactly like the liberal rule $f^{(1, 1)}$, i.e.\@ the subsets of socially qualified individuals consists of those and only those individuals who qualify themselves. As shown in previous works, the $f^{(1, 1)}$ rule is immune against all group control problems \cite{YD18,ERY20,J22}. Meanwhile, all bribery and microbribery problems are polynomial-time solvable for the $f^{(1, 1)}$ rule \cite{BBKL20}. Therefore, the various NP-completeness results we obtained for the $f^{\text{2LIC}}$ rule only stem from the fact that the rule behaves differently when we have (or create) a consensus among all individuals.

One of our most notable findings is that there are certain bribery and microbribery problems where the unpriced version is polynomial-time solvable while the priced version is NP-complete. For the problems EGB, GB, and CGMB, this is the case for both the $f^{\text{2IC}}$ rule and the $f^{\text{2LIC}}$ rule. Meanwhile, when the $f^{\text{IC}}$ rule is used, even the unpriced versions of these problems are NP-complete. As before, this can be explained by the fact that the functioning of the 2-stage rules is simpler than that of the $f^{\text{IC}}$ rule. The only outlier is again the \$DGB problem: Here, the priced version is NP-complete only when using the $f^{\text{2LIC}}$ rule. This suggests that the difficulty of the $f^{\text{2LIC}}$-\$DGB problem is a result of both the bribery prices (the unpriced $f^{\text{2LIC}}$-DGB problem is polynomial-time solvable) and the fact that the $f^{\text{2LIC}}$ rule is liberal (for the two non-liberal rules, even the priced version \$DGB is polynomial-time solvable).

\subsection{Future work}
\label{sec:conclusion_future}

The fact that there are social rules for which the complexity of certain bribery problems differs between the priced and unpriced versions suggests that it could be worth revisiting the complexity of bribery problems for other social rules as well. Even though the priced and unpriced bribery/microbribery problems for the consent rules, the CSR rule, and the LSR rule have been shown to be in the same complexity class, it is possible that their parameterized complexity or the complexity of approximating a solution differs between the priced and unpriced versions. For problems that are polynomial-time solvable, there could also be faster algorithms for the unpriced versions than for the priced ones.

In general, the parameterized complexity of the problems studied in this thesis would be an interesting research topic. We have already seen that certain group control and bribery problems for the iterative consensus rules are W[2]-hard with respect to $\ell$. We obtained these results directly through the reduction from \textsc{Set Cover}. For the remaining NP-complete problems, one could investigate whether they are fixed-parameter-tractable with respect to $\ell$. Also, various other parameters could be considered, such as the sizes of the target sets $A^+$ and $A^-$, or the values of $s$ and $t$ (for the consent rules).

There are various possible combinations of manipulative attack problems and social rules that we did not cover in this thesis. This includes relaxed group control by deleting individuals in combinations with iterative consensus rules (we only studied the consent rules, the CSR rule, and the LSR rule). Also, for the iterative consensus rules, we did not consider the group control by partitioning individuals problem in this thesis. Thus, to get a more complete picture, one could try to analyze the complexity of these problems. However, it should be noted that at the time of writing this thesis the complexity of the partitioning problems is still open for the CSR rule. This suggests that finding results for the iterative consensus rules could be difficult.

Finally, as suggested in \secref{sec:intro_social_rules}, one could define additional social rules, e.g.\@ a variant of the iterative consensus rule that is liberal, but not 2-stage. One could also consider variants where, to make an individual socially qualified, instead of requiring a consensus among all individuals, it suffices that a majority of individuals qualify that individual. This could solve the problem that the initial set of socially qualified individuals is empty in many real-world instances: Even though there may not exist a consensus among all individuals, it is conceivable that there are some individuals for whom a majority (or a certain share) of individuals agree that they should be qualified. This majority rule could then also be applied in the iterative step. It would be interesting to see how these changes in the definition affect the complexity of the manipulative attack problems.

\clearpage
\nocite{*}
\printbibliography

@article{YD18,
  title={How hard is it to control a group?},
  author={Yang, Yongjie and Dimitrov, Dinko},
  journal={Autonomous Agents and Multi-Agent Systems},
  volume={32},
  number={5},
  pages={672--692},
  year={2018}
}

@article{ERY20,
  title={The complexity of bribery and control in group identification},
  author={Erd{\'e}lyi, G{\'a}bor and Reger, Christian and Yang, Yongjie},
  journal={Autonomous Agents and Multi-Agent Systems},
  volume={34},
  number={1},
  pages={1--31},
  year={2020}
}

@inproceedings{EY20,
  title={Microbribery in group identification},
  author={Erd{\'e}lyi, G{\'a}bor and Yang, Yongjie},
  booktitle={Proceedings of the 19th International Conference on Autonomous Agents and Multiagent Systems (AAMAS)},
  pages={1840--1842},
  year={2020}
}

@inproceedings{BBKL20,
  title={Fine-grained view on bribery for group identification},
  author={Boehmer, Niclas and Bredereck, Robert and Knop, Du{\v{s}}an and Luo, Junjie},
  booktitle={Proceedings of the 29th International Joint Conference on Artificial Intelligence (IJCAI)},
  pages={67--73},
  year={2020},
  addendum={Long version appeared as: \textit{arXiv preprint arXiv:2105.08376 [cs.GT]}}
}

@article{YD22,
  title={Group control for procedural rules: Parameterized complexity and consecutive domains},
  author={Yang, Yongjie and Dimitrov, Dinko},
  journal={arXiv preprint arXiv:2203.16872 [cs.GT]},
  year={2022}
}

@article{J22,
  title={Manipulative attacks and group identification},
  author={Junker, Emil},
  journal={arXiv preprint arXiv:2203.16151 [cs.GT]},
  year={2022},
  titleaddon={Supervised student publication}
}

@inproceedings{ERY17,
  title={Complexity of group identification with partial information},
  author={Erd{\'e}lyi, G{\'a}bor and Reger, Christian and Yang, Yongjie},
  booktitle={Proceedings of the 5th International Conference on Algorithmic Decision Theory (ADT)},
  pages={182--196},
  year={2017},
  organization={Springer}
}

@phdthesis{R18,
    title={Complexity of strategic influences in elections and group identification with a main focus on incomplete information},
    school={Universit{\"a}t Siegen},
    author={Reger, Christian},
    pages={187--206},
    year={2018}
}

@article{G85,
  title={Clustering to minimize the maximum intercluster distance},
  author={Gonzalez, Teofilo F.},
  journal={Theoretical Computer Science},
  volume={38},
  pages={293--306},
  year={1985},
  publisher={Elsevier}
}

@incollection{K72,
  title={Reducibility among combinatorial problems},
  author={Karp, Richard M.},
  booktitle={Complexity of Computer Computations},
  pages={85--103},
  year={1972}
}

@article{DF95,
  title={Fixed-parameter tractability and completeness I: Basic results},
  author={Downey, Rod G. and Fellows, Michael R.},
  journal={SIAM Journal on Computing},
  volume={24},
  number={4},
  pages={873--921},
  year={1995},
  publisher={SIAM}
}

@article{E75,
  title={An algorithm for determining whether the connectivity of a graph is at least k},
  author={Even, Shimon},
  journal={SIAM Journal on Computing},
  volume={4},
  number={3},
  pages={393--396},
  year={1975},
  publisher={SIAM}
}

@article{ET75,
  title={Network flow and testing graph connectivity},
  author={Even, Shimon and Tarjan, R. Endre},
  journal={SIAM Journal on Computing},
  volume={4},
  number={4},
  pages={507--518},
  year={1975},
  publisher={SIAM}
}

@article{BBKL21,
  title={Finding small multi-demand set covers with ubiquitous elements and large sets is fixed-parameter tractable},
  author={Boehmer, Niclas and Bredereck, Robert and Knop, Du{\v{s}}an and Luo, Junjie},
  journal={arXiv preprint arXiv:2104.10124 [cs.DS]},
  year={2021}
}

@article{K93,
  title={Jewish collective identity},
  author={Kasher, Asa},
  journal={Jewish Identity, Temple University Press, Philadelphia},
  volume={2},
  pages={56--78},
  year={1993}
}

@article{KR97,
  title={On the question "Who is a J?": A social choice approach},
  author={Kasher, Asa and Rubinstein, Ariel},
  journal={Logique et Analyse},
  volume={160},
  pages={385--395},
  year={1997},
  publisher={JSTOR}
}

@article{DSX07,
  title={Procedural group identification},
  author={Dimitrov, Dinko and Sung, Shao Chin and Xu, Yongsheng},
  journal={Mathematical Social Sciences},
  volume={54},
  number={2},
  pages={137--146},
  year={2007},
  publisher={Elsevier}
}

@article{N07,
  title={"I want to be a J!": Liberalism in group identification problems},
  author={Nicolas, Houy},
  journal={Mathematical Social Sciences},
  volume={54},
  number={1},
  pages={59--70},
  year={2007},
  publisher={Elsevier}
}

@article{M08,
  title={Group identification},
  author={Miller, Alan D.},
  journal={Games and Economic Behavior},
  volume={63},
  number={1},
  pages={188--202},
  year={2008},
  publisher={Elsevier}
}

@article{SS03,
  title={Between liberalism and democracy},
  author={Samet, Dov and Schmeidler, David},
  journal={Journal of Economic Theory},
  volume={110},
  number={2},
  pages={213--233},
  year={2003},
  publisher={Elsevier}
}

@incollection{D11,
  title={The social choice approach to group identification},
  author={Dimitrov, Dinko},
  booktitle={Consensual Processes},
  pages={123--134},
  year={2011},
  publisher={Springer}
}

@article{FT20,
  title={Asking infinite voters ‘Who is a J?’: Group identification problems in $\mathbb{N}$},
  author={Fioravanti, Federico and Tohm{\'e}, Fernando},
  journal={Journal of Classification},
  volume={37},
  number={1},
  pages={58--65},
  year={2020},
  publisher={Springer}
}

@article{FT22,
  title={Fuzzy group identification problems},
  author={Fioravanti, Federico and Tohm{\'e}, Fernando},
  journal={Fuzzy Sets and Systems},
  volume={434},
  pages={159--171},
  year={2022},
  publisher={Elsevier},
  relatedstring={arXiv preprint arXiv:1912.05540 [econ.TH]}
}

@article{FHH09,
  title={How hard is bribery in elections?},
  author={Faliszewski, Piotr and Hemaspaandra, Edith and Hemaspaandra, Lane A.},
  journal={Journal of Artificial Intelligence Research},
  volume={35},
  pages={485--532},
  year={2009}
}

@article{FHHR09,
  title={Llull and Copeland voting computationally resist bribery and constructive control},
  author={Faliszewski, Piotr and Hemaspaandra, Edith and Hemaspaandra, Lane A. and Rothe, J{\"o}rg},
  journal={Journal of Artificial Intelligence Research},
  volume={35},
  pages={275--341},
  year={2009}
}

@article{BEFGMR14,
  title={The complexity of probabilistic lobbying},
  author={Binkele-Raible, Daniel and Erd{\'e}lyi, G{\'a}bor and Fernau, Henning and Goldsmith, Judy and Mattei, Nicholas and Rothe, J{\"o}rg},
  journal={Discrete Optimization},
  volume={11},
  pages={1--21},
  year={2014},
  publisher={Elsevier}
}

@article{CFRS07,
  title={On complexity of lobbying in multiple referenda},
  author={Christian, Robin and Fellows, Mike and Rosamond, Frances and Slinko, Arkadii},
  journal={Review of Economic Design},
  volume={11},
  number={3},
  pages={217--224},
  year={2007},
  publisher={Springer}
}

@inproceedings{C11,
  title={Estimating the margin of victory for instant-runoff voting},
  author={Cary, David},
  booktitle={Proceedings of the 2011 conference on Electronic Voting Technology / Workshop on Trustworthy Elections (EVT/WOTE)},
  year={2011}
}

@inproceedings{MRSW11,
  title={Computing the margin of victory in IRV elections},
  author={Magrino, Thomas R. and Rivest, Ronald L. and Shen, Emily and Wagner, David},
  booktitle={Proceedings of the 2011 conference on Electronic Voting Technology / Workshop on Trustworthy Elections (EVT/WOTE)},
  year={2011}
}

@inproceedings{BBFN21,
  title={Winner robustness via swap- and shift-bribery: Parameterized counting complexity and experiments},
  author={Boehmer, Niclas and Bredereck, Robert and Faliszewski, Piotr and Niedermeier, Rolf},
  booktitle={Proceedings of the 30th International Joint Conference on Artificial Intelligence (IJCAI)},
  pages={52--58},
  year={2021},
  relatedstring={arXiv preprint arXiv:2010.09678 [cs.GT]}
}

@inproceedings{FST17,
  title={Bribery as a measure of candidate success: Complexity results for approval-based multiwinner rules},
  author={Faliszewski, Piotr and Skowron, Piotr and Talmon, Nimrod},
  booktitle={Proceedings of the 16th International Conference on Autonomous Agents and Multiagent Systems (AAMAS)},
  pages={6--14},
  year={2017},
  relatedstring={arXiv preprint arXiv:2104.09130 [cs.GT]}
}

@inproceedings{X12,
  title={Computing the margin of victory for various voting rules},
  author={Xia, Lirong},
  booktitle={Proceedings of the 13th ACM Conference on Electronic Commerce (EC)},
  pages={982--999},
  year={2012}
}

@article{FSST17,
  title={Multiwinner voting: A new challenge for social choice theory},
  author={Faliszewski, Piotr and Skowron, Piotr and Slinko, Arkadii and Talmon, Nimrod},
  journal={Trends in Computational Social Choice},
  volume={74},
  pages={27--47},
  year={2017},
  publisher={AI Access Foundation}
}

@article{EFSS17,
  title={Properties of multiwinner voting rules},
  author={Elkind, Edith and Faliszewski, Piotr and Skowron, Piotr and Slinko, Arkadii},
  journal={Social Choice and Welfare},
  volume={48},
  number={3},
  pages={599--632},
  year={2017},
  publisher={Springer}
}

@article{BTT92,
  title={How hard is it to control an election?},
  author={Bartholdi III, John J. and Tovey, Craig A. and Trick, Michael A.},
  journal={Mathematical and Computer Modelling},
  volume={16},
  number={8-9},
  pages={27--40},
  year={1992},
  publisher={Elsevier}
}

@article{MPRZ08,
  title={Complexity of strategic behavior in multi-winner elections},
  author={Meir, Reshef and Procaccia, Ariel D. and Rosenschein, Jeffrey S. and Zohar, Aviv},
  journal={Journal of Artificial Intelligence Research},
  volume={33},
  pages={149--178},
  year={2008}
}

@inproceedings{YG14,
  title={Controlling elections with bounded single-peaked width},
  author={Yang, Yongjie and Guo, Jiong},
  booktitle={Proceedings of the 13th International Conference on Autonomous Agents and Multiagent Systems (AAMAS)},
  pages={629--636},
  year={2014}
}

@inproceedings{YG15,
  title={How hard is control in multi-peaked elections: A parameterized study},
  author={Yang, Yongjie and Guo, Jiong},
  booktitle={Proceedings of the 14th International Conference on Autonomous Agents and Multiagent Systems (AAMAS)},
  pages={1729--1730},
  year={2015}
}

@misc{FR16,
  title={Control and bribery in voting},
  author={Faliszewski, Piotr and Rothe, J{\"o}rg},
  year={2016}
}

@incollection{BEHHR10,
  title={Computational aspects of approval voting},
  author={Baumeister, Dorothea and Erd{\'e}lyi, G{\'a}bor and Hemaspaandra, Edith and Hemaspaandra, Lane A. and Rothe, J{\"o}rg},
  booktitle={Handbook on Approval Voting},
  pages={199--251},
  year={2010},
  publisher={Springer}
}

@article{FB81,
  title={Approval voting, Condorcet's principle, and runoff elections},
  author={Fishburn, Peter C. and Brams, Steven J.},
  journal={Public Choice},
  volume={36},
  number={1},
  pages={89--114},
  year={1981},
  publisher={Springer}
}

@article{LS10,
  title={The basic approval voting game},
  author={Laslier, Jean-Fran{\c{c}}ois and Sanver, M. Remzi},
  journal={Handbook on Approval Voting},
  pages={153--163},
  year={2010},
  publisher={Springer}
}

@article{HHR07,
  title={Anyone but him: The complexity of precluding an alternative},
  author={Hemaspaandra, Edith and Hemaspaandra, Lane A. and Rothe, J{\"o}rg},
  journal={Artificial Intelligence},
  volume={171},
  number={5-6},
  pages={255--285},
  year={2007},
  publisher={Elsevier}
}

@inproceedings{L11,
  title={The complexity of manipulating $k$-approval elections},
  author={Lin, Andrew},
  booktitle={Proceedings of the 3rd International Conference on Agents and Artificial Intelligence (ICAART)},
  volume={2},
  pages={212--218},
  year={2011},
  relatedstring={arXiv preprint arXiv:1005.4159 [cs.AI]}
}

@article{YG17,
  title={The control complexity of r-approval: From the single-peaked case to the general case},
  author={Yang, Yongjie and Guo, Jiong},
  journal={Journal of Computer and System Sciences},
  volume={89},
  pages={432--449},
  year={2017},
  publisher={Elsevier}
}

@inproceedings{Y19,
  title={Complexity of manipulating and controlling approval-based multiwinner voting},
  author={Yang, Yongjie},
  booktitle={Proceedings of the 28th International Joint Conference on Artificial Intelligence (IJCAI)},
  pages={637--643},
  year={2019}
}

@inproceedings{Y20,
  title={On the complexity of destructive bribery in approval-based multi-winner voting},
  author={Yang, Yongjie},
  booktitle={Proceedings of the 19th International Conference on Autonomous Agents and Multiagent Systems (AAMAS)},
  pages={1584--1592},
  year={2020},
  relatedstring={arXiv preprint arXiv:2002.00836 [cs.GT]}
}

@article{BFNT21,
  title={Complexity of shift bribery in committee elections},
  author={Bredereck, Robert and Faliszewski, Piotr and Niedermeier, Rolf and Talmon, Nimrod},
  journal={ACM Transactions on Computation Theory (TOCT)},
  volume={13},
  number={3},
  pages={1--25},
  year={2021},
  publisher={ACM New York, NY, USA}
}

@article{BKN21,
  title={On coalitional manipulation for multiwinner elections: Shortlisting},
  author={Bredereck, Robert and Kaczmarczyk, Andrzej and Niedermeier, Rolf},
  journal={Autonomous Agents and Multi-Agent Systems},
  volume={35},
  number={2},
  pages={1--41},
  year={2021},
  publisher={Springer}
}

@article{BBMN04,
  title={Stability and voting by committees with exit},
  author={Berga, Dolors and Berganti{\~n}os, Gustavo and Mass{\'o}, Jordi and Neme, Alejandro},
  journal={Social Choice and Welfare},
  volume={23},
  number={2},
  pages={229--247},
  year={2004},
  publisher={Springer}
}

@article{K16,
  title={Approval elections with a variable number of winners},
  author={Kilgour, D. Marc},
  journal={Theory and Decision},
  volume={81},
  number={2},
  pages={199--211},
  year={2016},
  publisher={Springer}
}

@inproceedings{FST20,
  title={Multiwinner rules with variable number of winners},
  author={Faliszewski, Piotr and Slinko, Arkadii and Talmon, Nimrod},
  booktitle={Proceedings of the 24th European Conference on Artificial Intelligence (ECAI)},
  pages={67--74},
  year={2020},
  relatedstring={arXiv preprint arXiv:1711.06641 [cs.GT]}
}

@inproceedings{YW18,
  title={Multiwinner voting with restricted admissible sets: Complexity and strategyproofness},
  author={Yang, Yongjie and Wang, Jianxin},
  booktitle={Proceedings of the 27th International Joint Conference on Artificial Intelligence (IJCAI)},
  pages={576--582},
  year={2018}
}

@article{DPZ16,
  title={Aggregation of binary evaluations: A Borda-like approach},
  author={Duddy, Conal and Piggins, Ashley and Zwicker, William S.},
  journal={Social Choice and Welfare},
  volume={46},
  number={2},
  pages={301--333},
  year={2016},
  publisher={Springer}
}

@inproceedings{LM21,
  title={Approval-based shortlisting},
  author={Lackner, Martin and Maly, Jan},
  booktitle={Proceedings of the 20th International Conference on Autonomous Agents and Multiagent Systems (AAMAS)},
  pages={737--745},
  year={2021},
  relatedstring={arXiv preprint arXiv:2005.07094 [cs.GT]}
}

@inproceedings{AFPT11,
  title={Sum of us: Strategyproof selection from the selectors},
  author={Alon, Noga and Fischer, Felix and Procaccia, Ariel and Tennenholtz, Moshe},
  booktitle={Proceedings of the 13th Conference on Theoretical Aspects of Rationality and Knowledge (TARK)},
  pages={101--110},
  year={2011}
}

@inproceedings{ALMRW16,
  title={Strategyproof peer selection: Mechanisms, analyses, and experiments},
  author={Aziz, Haris and Lev, Omer and Mattei, Nicholas and Rosenschein, Jeffrey and Walsh, Toby},
  booktitle={Proceedings of the 30th AAAI Conference on Artificial Intelligence (AAAI)},
  pages={390--396},
  year={2016}
}

@book{W01,
  title={Introduction to graph theory},
  author={West, Douglas Brent},
  volume={2},
  year={2001},
  publisher={Prentice Hall}
}

\end{document}